\documentclass[a4paper, 11pt, oneside]{article}
\pdfoutput=1
\usepackage{aapreprint}
\usepackage{lmodern}
\usepackage{breqn}
\usepackage{enumitem}

\usepackage{mathrsfs}
\usepackage[english]{babel}
\usepackage{slashed}

\usepackage{graphicx,wrapfig,float,slashed}
\usepackage{mathtools,amsmath,amssymb,empheq,dsfont,epsfig}
\usepackage{bbm}
\usepackage{epstopdf}
\usepackage{ragged2e}
\usepackage[utf8]{inputenc}
\usepackage[usenames,dvipsnames,table]{xcolor}
\usepackage{tabu}
\usepackage[toc]{appendix}
\usepackage[normalem]{ulem}
\usepackage[font={small},skip=0pt]{caption}
\allowdisplaybreaks 
\usepackage{pifont}
\usepackage{subcaption}

\newcommand{\nc}{\newcommand}
\nc{\non}{\nonumber}
\nc{\hc}{\hbox {H.c.}}
\nc{\noi}{\noindent}
\nc{\barx}{\bar{x}}
\nc{\pbarn}{\;\hbox {pb}}
\nc{\fbarn}{\;\hbox {fb}}

\newcommand{\arh}{a_{\text{\tiny{RH}}}}

\newcommand{\Hrh}{H_{\text{\tiny{RH}}}}
\nc{\hsp}{\hspace{0.5cm}}
\nc{\lsp}{\hspace{1cm}}
\nc{\Lsp}{\hspace{2cm}}
\nc{\LLsp}{\lsp\lsp}
\nc{\lra}{\longrightarrow}
\nc{\p}{\prime}
\nc{\sgn}{\text{sgn}}
\nc{\ph}{\varphi}
\nc{\op}{{\cal O}}
\nc{\cL}{{\cal L}}
\nc{\tr}{{\text{Tr}}}
\nc{\eq}{\text{Eq.~}}
\nc{\cg}{{\cal G}}
\nc{\ch}{{\cal H}}
\nc{\cZ}{\mathbb Z}
\nc{\cw}{\cos\theta_{\textsc w}}
\nc{\sw}{\sin\theta_{\textsc w}}
\nc{\cwsq}{\cos^2\theta_{\textsc w}}
\nc{\swsq}{\sin^2\theta_{\textsc w}}
\def\zBB{{\mathbbm Z}}
\def\z2{\zBB_2}
\nc{\beq}{\begin{equation}}  \nc{\eeq}{\end{equation}}
\nc{\bea}{\begin{eqnarray}}  \nc{\eea}{\end{eqnarray}}
\nc{\baa}{\begin{array}}     \nc{\eaa}{\end{array}}
\nc{\bit}{\begin{itemize}}   \nc{\eit}{\end{itemize}}
\nc{\ben}{\begin{enumerate}} \nc{\een}{\end{enumerate}}
\nc{\bce}{\begin{center}}    \nc{\ece}{\end{center}}
\nc{\bpm}{\begin{pmatrix}}   \nc{\epm}{\end{pmatrix}}
\nc{\bvt}{\begin{verbatim}}  \nc{\evt}{\end{verbatim}}
\def\lsim{\mathrel{\raise.3ex\hbox{$<$\kern-.75em\lower1ex\hbox{$\sim$}}}}
\def\gsim{\mathrel{\raise.3ex\hbox{$>$\kern-.75em\lower1ex\hbox{$\sim$}}}}

\def\udots{\mathinner{\mkern1mu\raise1pt\vbox{\kern7pt\hbox{.}}\mkern2mu\raise4pt\hbox{.}\mkern2mu\raise7pt\hbox{.}\mkern1mu}}

\def\mev{\;\hbox{MeV}}
\def\gev{\;\hbox{GeV}}

\def\mpl{M_{\rm Pl}}

\def\hi{H_{\rm I}}
\def\ai{a_{\rm I}}

\definecolor{agray}{rgb}{0.95, 0.95, 0.99}

\def\xtl{\mathcal{\widetilde{X}}_L}

\def\app#1{Appendix~\ref{#1}}
\def\eq#1{Eq.~(\ref{#1})}
\def\fig#1{Fig.~\ref{#1}}
\def\sec#1{Sec.~\ref{#1}}

\newcommand\fverb{\setbox\fverbbox=\hbox\bgroup\verb}
\newcommand\fverbdo{\egroup\medskip\noindent%
			\fbox{\unhbox\fverbbox}\ }
\newcommand\fverbit{\egroup\item[\fbox{\unhbox\fverbbox}]}
\newbox\fverbbox

\setlist[itemize]{itemsep=0em, topsep=0.3em}

\begin{document}
\title{Gravitational production of vector dark matter}
\author[1]{Aqeel Ahmed,}
\author[2]{Bohdan Grzadkowski,}
\author[2]{and Anna Socha}
\affiliation[1]{Theoretische Natuurkunde \& IIHE/ELEM,\\ Vrije Universiteit Brussel, Pleinlaan 2, 1050 Brussels, Belgium}
\affiliation[2]{Faculty of Physics, University of Warsaw, Pasteura 5, 02-093 Warsaw, Poland}
\emailAdd{aqeel.ahmed@vub.be}
\emailAdd{bohdan.grzadkowski@fuw.edu.pl}
\emailAdd{anna.socha@fuw.edu.pl}

\abstract{
A model of vector dark matter that communicates with the Standard Model only through gravitational interactions has been investigated. 
It has been shown in detail how does the canonical quantization of the vector field in varying FLRW geometry implies a tachyonic enhancement of some of its momentum modes. 
Approximate solutions of the mode equation have been found and verified against exact numerical ones. 
De~Sitter geometry has been assumed during inflation while after inflation a non-standard cosmological era of reheating with a generic equation of state has been adopted which is followed by the radiation-dominated universe. 
It has been shown that the spectrum of dark vectors produced gravitationally is centered around a characteristic comoving momentum~$k_\star$ that is determined in terms of the mass of the vector~$m_X$, the Hubble parameter during inflation~$H_{\rm I}$, the equation of state parameter~$w$ and the efficiency of reheating~$\gamma$. 
Regions in the parameter space consistent with the observed dark matter relic abundance have been determined, justifying the gravitational production as a viable mechanism for vector dark matter. 
The results obtained in this paper are applicable within various possible models of inflation/reheating with non-standard cosmology parametrized effectively by the corresponding equation of state and efficiency of reheating.  
}

\arxivnumber{2005.01766}

\maketitle
\flushbottom
\section{Introduction}
\label{Introduction}
One of the outstanding puzzles of high energy physics is to understand the nature of dark matter~(DM). 
There is overwhelming evidence that the dominant component of matter density in the universe is due to DM. 
However, all the observational pieces of evidence in favor of DM have been established by its interactions with the Standard Model (SM) only through gravity~\cite{Aghanim:2018eyx,Sofue:2000jx,Bartelmann:1999yn,Clowe:2003tk}. 
A wide range of DM models have been proposed with different DM production mechanisms depending on the DM interaction strength with the SM and/or beyond the SM new physics, for a review see~\cite{Bernal:2017kxu} and original references therein. 
Arguably the most popular DM models involve weakly interacting massive particles (WIMPs) whose mass $\sim\!\op({\rm TeV})$ and interaction strength $\sim\!\op(0.1)$ are similar to those of the SM particles. 
The production mechanism for WIMPs is a thermal freeze-out scenario where it is assumed that the SM and DM were produced in the early universe during the reheating phase. The WIMP DM was in thermal equilibrium with the SM and as the universe cooled down it went out of the equilibrium to constitute the observed relic abundance. 
Thermal freeze-out and other DM production mechanisms~\cite{Bernal:2017kxu} are theoretically appealing, however there has been no sign of DM interactions with the SM in a variety of experiments thus far, apart from its gravitational interactions. 

The lack of DM signals motivates us to envision novel mechanisms of DM production where the DM and SM interactions are absent or negligible. 
One such mechanism is the particle production due to quantum fluctuations in a rapidly expanding universe~\cite{Parker:1969au,Zeldovich:1971mw}. 
An inflationary scenario in the early universe not only explains successfully some of the puzzles of modern cosmology (for a review see~\cite{Baumann:2009ds}), it can also provide an ideal phase for DM production due to quantum fluctuations~\cite{Ford:1986sy,Lyth:1996yj}.  
Furthermore, a very heavy DM can also be produced during the (p)reheating period -- the last stage of inflation -- where the inflaton field transfers its energy density to the visible and dark sectors through coherent oscillations~\cite{Kofman:1997yn,Chung:1998zb,Chung:2001cb}. 
In particular if the reheating does not happen instantaneously the maximum temperature achieved during the reheating phase can be larger than the temperature at the end of reheating which allows supper-heavy DM production~\cite{Giudice:2000ex}.
Recently, there has been a renewed interest in models of gravitational production of DM during and after inflation due to quantum fluctuations~\cite{Graham:2015rva,Ema:2018ucl,Alonso-Alvarez:2018tus,Dror:2018pdh,Co:2018lka,Agrawal:2018vin,Bastero-Gil:2018uel,Hashiba:2018tbu,Li:2019ves,Ema:2019yrd,Tenkanen:2019aij,Nakayama:2019yjv,AlonsoAlvarez:2019cgw,Nakayama:2019rhg,Velazquez:2019mpj,Ahmed:2019mjo,Nakayama:2020rka,Nakai:2020cfw,Herring:2020cah}.

In this work, we study a minimal model of gravitationally produced DM where the DM is a massive Abelian gauge boson $X_\mu$.  
In particular, we focus on vector DM production due to the quantum fluctuations during and after inflation in an early era of non-standard cosmology parameterized effectively by the equation of state $w$. 
We assume the dark vector field couples minimally to gravity and it has no other interactions with the visible sector. 
The vector DM mass is assumed to be generated via the Stueckelberg mechanism and it is non-zero during and after inflation.

A massive vector field $X_\mu$ has three physical polarizations: two transverse $X_\pm$ and one longitudinal $X_L$, whereas $X_0$ is an auxiliary component. 
It is well known that the transverse components of a minimally coupled vector field produce scale-invariant density fluctuations during inflation~\cite{Dimopoulos:2006ms,Nelson:2011sf,Arias:2012az,Graham:2015rva} therefore they can not constitute all the observed DM relic density. 
However, if the vector DM is non-minimally coupled to gravity or the inflaton field then the transverse components of a vector field may constitute all the observed DM density, see e.g.~\cite{Dror:2018pdh,Co:2018lka,Agrawal:2018vin,Bastero-Gil:2018uel} for the latter case. 
On the other hand, the longitudinal modes in a minimally or non-minimally coupled scenarios do not lead to scale-invariant density fluctuations during and after inflation~\cite{Graham:2015rva,Ema:2019yrd,AlonsoAlvarez:2019cgw} and hence can account for all the observed DM abundance.
Since we are interested in a minimally coupled vector DM scenario, therefore our focus is confined to the production of the longitudinal modes of the vector DM during and after inflation. 

The main new feature of our work in comparison to the previous works on the gravitational production of vector DM is an early epoch of non-standard cosmology during the reheating period parameterized by a general equation of state $w$ of the inflaton/reheaton field.
During inflation, without specifying the inflationary dynamics, we assume a de Sitter phase with constant Hubble parameter $\hi$. 
After the end of inflation a reheating phase starts where a non-standard cosmological evolution is assumed with the dominant energy density scaling as $a^{-3(1+w)}$, where $a$ is the scale factor and $-1/3\!<\!w\!<\!1$. 
This specific range of $w$ during reheating period is motivated by our analytic analysis which is robustly verified against exact numerical calculations with specific values of $w\!=\!\big\{\!\!-\!1/4,\!-1/6,0,1/3,2/3\big\}$.
At the end of reheating phase, with non-standard cosmology, the standard cosmological evolution resumes with the radiation-dominated~(RD) era followed by the matter-dominated~(MD) universe.
We define the end of the reheating phase when the energy density of the non-standard cosmology $\propto\!a^{-3(1+w)}$ is dominated by the SM radiation energy density $\rho_{\text{\tiny RD}}\!\propto\! a^{-4}$.
The production of dark matter during the non-instantaneous reheating phase due to the coherent oscillations/decays of the inflaton field are studied in Refs.~\cite{Gelmini:2006pw,Harigaya:2014waa,Drees:2017iod,Kaneta:2019zgw,Harigaya:2019tzu,Maldonado:2019qmp}, however we neglect such effects here. 
Furthermore, we do not specify the dynamics of reheating mechanism, however we assume that such mechanism exists and leads to a non-standard cosmology where the dominant energy density scales as $a^{-3(1+w)}$ and the SM radiation energy density during this phase is subdominant. 
In recent years there has been many works done on DM scenarios in the presence of an early universe non-standard cosmology~\cite{Davoudiasl:2015vba,Randall:2015xza,Tenkanen:2016jic,DEramo:2017gpl,Hamdan:2017psw,Visinelli:2017qga,DEramo:2017ecx,Maity:2018dgy,Bernal:2018kcw,Arbey:2018uho,Drees:2018dsj,Poulin:2019omz,Arias:2019uol,Bramante:2017obj,Bernal:2018ins,Bernal:2019mhf,Mahanta:2019sfo,Cosme:2020mck,Garcia:2020eof,Bernal:2020bfj}.
Our focus in this paper is to study, for the first time, the production of vector DM purely due to quantum fluctuations during and after inflation employing non-standard cosmological period of reheating in the early universe. 

In \sec{grav_prod} we provide a detailed description of the gravitational production of a minimally coupled vector DM in an expanding universe. 
Focusing only on the longitudinal modes, we give approximate analytic and exact numerical solutions for the mode functions during and after inflation for different regimes of DM mass and wavelengths. 
In particular, we note that for certain mass and wavelength range the frequency squared of the longitudinal modes becomes negative leading to a tachyonic enhancement of those modes. 
We calculate the vector DM relic abundance in \sec{sec:relicDM} in two DM mass ranges, $H_{\rm rh}\!\leq\!m_X \!<\! H_{\rm I}$ and $m_X \!<\!H_{\rm rh}$, where $H_{\rm rh}$ is the Hubble scale at the end of reheating with non-standard cosmology.
In \sec{sec:con} we present our conclusions. Supplementary material including details of the quantization of a vector field on the curved background is given in \app{app:quanta}.

\section{Gravitational vector DM production in expanding universe}
\label{grav_prod}
During an epoch of rapidly expanding universe (inflation and partially reheating) the gravitational production dominates over other DM production mechanisms. In this work we consider renormalizable (dim-4) Lagrangian for the vector DM minimally coupled to gravity. The action for the vector DM in a background metric $g_{\mu\nu}$ is given by 
\begin{align}
    S_{\rm DM} = \int d^{4}x \sqrt{-g} \left( - \frac{1}{4} g^{\mu \alpha} g^{\nu \beta} X_{\mu \nu} X_{\alpha \beta} - 
        \frac{1}{2} m_X^2 g^{\mu \nu} X_{\mu} X_{\nu}  \right),    \label{eq:dmaction}
\end{align}
where the background metric $g_{\mu \nu}$ is of the FLRW form with the line element
\begin{align}
    ds^2 = dt^2 - a^2(t) d \vec{x}^2\,,    \label{eq:metric}
\end{align}
The mass for the dark vector boson $m_X$ is generated via the Stueckelberg mechanism~\footnote{Alternatively one may assume an Abelian Higgs mechanism such that the extra neutral Higgs boson (radial mode) is very heavy. Then it is possible to show that, in an appropriate limit of constant $m_X$, effectively the Higgs model reduces to the Stueckelberg mechanism. However, this Higgs mechanism has to be {\it effective} already during the inflationary period.}.
Now, using the gravitational definition of the energy-momentum tensor
\begin{align*}
    T_{\mu \nu}=  \frac{2}{\sqrt{-g}} \frac{\delta (\sqrt{-g} S_{\rm DM})}{\delta g_{\mu \nu}},
\end{align*}
one can find the energy density of the vector DM as
\begin{align}
    \rho_{X}^{} = \frac{1}{2 a^2} \Big( \lvert \dot{\vec{X}} - \vec{\nabla} X_0 \lvert^2  + \frac{1}{a^2} \lvert \vec{\nabla} \times \vec{X}  \rvert^2 + a^2 m_X^2 X_0^2 + m_X^2 \vec{X}^2\Big),
\end{align}
where $X_0$ and $\vec X\equiv X_i$ are components of the vector field,  the {\it over-dot} denotes derivative w.r.t. $t$ and $\vec{\nabla}\equiv \partial/(\partial \vec x)$.

In Appendix~\ref{app:quanta}, we provide details of the quantization of a vector field in a curved background. 
Adopting the notation from the appendix, we write the vacuum expectation values for the longitudinal ($L$) and transverse ($\pm$) components of the energy density as
\begin{align}
    \langle \rho_{L}^{} \rangle \!&=\! \frac{1}{4 \pi^2 a^4}\!\int\!\! dk k^2 \bigg\{\!\lvert \mathcal{\widetilde{X}}^{\prime}_L \rvert^2 \!-\! \frac{A^{\prime}(\tau)}{A(\tau)}\!\Big(\! \mathcal{\widetilde{X}}^{\prime}_L \mathcal{\widetilde{X}}^*_L \!+\!\mathcal{\widetilde{X}}^{\prime*}_L \mathcal{\widetilde{X}}_L\!\Big)\!+\! \bigg[\!\bigg(\!\frac{A^{\prime}(\tau)}{A(\tau)}\!\bigg)^2 \!+\! k^2 \!+\! a^2 m_X^2 \! \bigg] \lvert \mathcal{\widetilde{X}}_L \rvert^2 \bigg\},        \label{eq:rhox_long}        \\
    \langle \rho_{\pm}^{} \rangle \!&=\! \frac{1}{2 \pi^2 a^4}\!\int\!\! dk k^2 \bigg\{\lvert {\mathcal{X}}'_{\pm} \rvert^2 + \left( k^2 + a^2 m_X^2  \right) \lvert {\mathcal{X}}_{\pm} \rvert^2 \bigg\},    \label{eq:rhox_trans}    
\end{align}
where $\langle \rho_L \rangle \equiv \langle 0|\colon\hat\rho_L\colon|0 \rangle$ and $\langle \rho_{\pm} \rangle \equiv \langle 0|\colon\hat\rho_{\pm}\colon|0 \rangle$~\footnote{The hat over $\hat\rho_{L,\pm}$ reminds that we are dealing with quantum operators while the colon $\colon$ stands for the normal ordering. Note also the disappearance of $\vec x$ dependence for $\langle \rho_{L,\pm}^{} \rangle$ is a consequence of taking vacuum expectation values before integrating over momenta in (\ref{eq:rhox_long} - \ref{eq:rhox_trans}).} and
\begin{align}
  \widetilde{\cal X}_{L}(\tau,\vec k) &\equiv A(\tau) {\cal X}_{L}(\tau,\vec k) ,  &A(\tau) &\equiv \frac{a(\tau) m_X}{\sqrt{k^2 + a^2(\tau) m_X^2}}\,. \label{eq:Atau}
\end{align}
Above the ${\cal X}_{L}(\tau,\vec k)$ and ${\cal X}_{\pm}(\tau,\vec k)$ are Fourier transforms of the vector-field longitudinal $X_L(\tau,\vec x)$ and transverse $X_\pm(\tau,\vec x)$ components, respectively (see Appendix~\ref{app:quanta}). The {\it  prime} denotes derivative w.r.t. to the conformal time $\tau$, defined as $dt=a(\tau)d\tau$. 

Equations of motion for the longitudinal and transverse Fourier modes, i.e. $\widetilde{\cal X}_{L}(t,\vec k)$ and ${\cal X}_{\pm}(t,\vec k)$, are
\begin{align}
\mathcal{\widetilde{X}''}_{L} + \omega^2_L(\tau)\mathcal{\widetilde{X}}_{L}&=0,     
&{\mathcal{X}''_\pm} + \omega^2_\pm(\tau){\mathcal{X}_\pm}&=0, \label{eq:doe}
\end{align} 
where the time-dependent frequencies are given by
\begin{align}
\omega^2_L(\tau)&= k^2 + a^2 m_X^2 - \frac{k^2}{k^2 + a^2 m_X^2}\bigg(\frac{a^{\p\p}}{a} - \frac{3a^2 m_X^2}{k^2 + a^2 m_X^2}\frac{a^{\p2}}{a^2}\bigg),     \label{eq:omegaL}    \\
\omega^2_\pm(\tau)&= k^2 + a^2 m_X^2\,.     \label{eq:omegaT}
\end{align} 
Note that the transverse mode frequency $\omega^2_\pm(\tau)$ is always positive, however, the longitudinal mode frequency $\omega^2_L(\tau)$ can be negative in some regions of the parameter space. 
Therefore, for $\omega^2_L(\tau)\!<\!0$, the production of longitudinal modes would receive the tachyonic enhancement. 
Hence the production of the longitudinal modes can be parametrically larger than that of the transverse modes, see also~\cite{Graham:2015rva,Ema:2019yrd,AlonsoAlvarez:2019cgw}. 
In the following, we only focus on the gravitational production of the longitudinal modes in the expanding universe. 
In particular, we will present both numeric and approximate analytic solutions for the longitudinal mode functions 
$\widetilde{\cal X}_L$ in two phases, (i) during slow-roll inflation, and (ii) after inflation during the reheating phase with non-standard cosmology followed by the RD universe.

\paragraph{Cosmological evolution:}
First, let us focus on the background dynamics.   
We assume that during the inflationary period the evolution of the scale factor is approximated by the de~Sitter universe with an exactly exponential variation of the scale factor, $a_{\rm I}(t)\propto e^{H_{\rm I} t}$, with $H_{\rm I}$ being constant Hubble parameter defined as $H\equiv \dot a/a=a^\prime/a^2$ during inflation. 
In general, the evolution during inflation is governed dynamically by a homogeneous inflaton field that rolls slowly down in a properly adjusted potential interacting with gravity~\cite{Baumann:2009ds}. 
This more elaborate approach is not necessary here, as it would not alter the qualitative results presented in this work.
Hence, neglecting corrections from the inflaton dynamics, the inflationary stage can be well approximated by the de~Sitter scale factor $a_{\rm I}(\tau) \propto -1/(\tau+\text{const.})$ in the conformal coordinates. 
Inflation starts at conformal time $\tau_i \!\rightarrow\! - \infty $ and lasts up to $\tau_e \!\simeq\! 0^{+}$.
Then, at $\tau\!=\!\tau_e$ the universe smoothly (in the presence of the inflaton/reheaton field $\phi$) transfers from the de Sitter stage into the reheating period which employs non-standard cosmology. 
During the reheating the energy density of the universe is dominated by the energy density $\rho_{\phi}$ of the inflaton/reheaton field $\phi$. 
In this period, we assume that the inflaton equation of state is parametrized by a parameter $w$ describing the non-standard cosmology that we will vary and investigate its relevance:
\begin{align}
    p_{\phi}= w \rho_{\phi} \label{eq:eosinf},
\end{align}
where e.g. $w\!=\!0$ and $w\!=\!1/3$ correspond to an early MD and RD universe, respectively. 

Strictly speaking, we consider transfer of energy density from $\rho_{\phi}$ to the SM (radiation) density $\rho_{\text{\tiny SM}}$ during the epoch of non-standard cosmology until the very end of this period when $\rho_{\text{\tiny SM}} \approx \rho_{\text{\tiny RD}} \!\propto\!a^{-4}$ dominates and standard cosmological universe resumes with RD era. In particular, the evolution of non-standard cosmology is governed by the following set of Boltzmann equations for $\rho_\phi$ and $\rho_{\text{\tiny SM}}$,
\begin{align}
\dot \rho_\phi + 3(1+w) H\rho_\phi&=-\Gamma_\phi\, \rho_\phi,	&\dot \rho_{\text{\tiny SM}} + 4 H\rho_{\text{\tiny SM}}&=+\Gamma_\phi\, \rho_\phi,		\label{eq:rhophi_rhord}
\end{align}
where $\Gamma_\phi$ is the rate of transfer of inflaton energy to the SM. Assuming $\rho_\phi\!\gg\!\rho_{\text{\tiny SM}}$ and $\Gamma_\phi\ll 3(1+w) H$, the generic form of the equation of state for the inflaton/reheaton~\eqref{eq:eosinf} implies that during the reheating period the scale factor $\arh$ evolves as a function of the conformal time $\tau$ as
\beq
  \arh(\tau) \propto \tau^{\frac{2}{1 + 3w}}.
	\label{a_evol}
\eeq
The reheating period is followed, in turn, by the RD epoch when the total energy density is dominated by the SM radiation during which the scale factor varies as $a_{\text{\tiny{RD}}}\propto \tau$.

We require the scale factor $a(\tau)$ and the Hubble rate $H(\tau)$ to be continuous at the two transition points, i.e.
\begin{align}
  a_{\rm I}(\tau_e)&= \arh(\tau_e),  &H_{\rm I}(\tau_e)&= \Hrh(\tau_e),\\ 
  \arh(\tau_{\rm rh})&= a_{\text{\tiny{RD}}}(\tau_{\rm rh}),   &\Hrh(\tau_{\rm rh})&= H_{\text{\tiny{RD}}}(\tau_{\rm rh}),
\end{align}
where $\tau_{\rm rh}$ refers to the conformal time at which the transition from the reheating to the RD era happens~\footnote{In order to satisfy the continuity conditions we adopt the freedom of shifting the conformal time by a constant and a freedom of adjusting normalization of the scale factor separately in each considered region.}.
The continuity of the scale factor and Hubble rate imply
\begin{align}
  a(\tau) &= \begin{dcases}
  \frac{-1}{H_{\rm I} \big(\tau -\frac{3}{2} \tau_e (1+w)\big)}, & \tau \leq \tau_e \\
  \frac{2}{H_{\rm I}(1+3w)} \tau_e^{\frac{-3(1+w)}{1+3w}}\tau^{\frac{2}{1+3w}}, &\tau_e < \tau \leq \tau_{\rm rh}\\
  \frac{1}{H_{\rm I} } \bigg( \frac{2}{1+3w}\bigg)^2\tau_e^{-\frac{3 (1+w)}{1+3 w}}\tau_{\rm rh}^{\frac{1-3 w}{1+3w}}\Big(\tau - \tfrac{1}{2}\tau_{\rm rh}(1-3w) \Big), & \tau_{\rm rh}<\tau,
  \end{dcases}	\label{eq:scalefactor}	\\ 
H(\tau) &= \begin{dcases}
  H_{\rm I}, & \tau \leq \tau_e \\
  H_{\rm I}\, \tau_e^{\frac{3 (1+w)}{1+3 w}} \tau ^{-\frac{3 (1+w)}{1+3 w}}, & \tau_e < \tau \leq \tau_{\rm rh}\\
  \frac{H_{\rm I}}{4} \bigg(\frac{1+3w}{\tau- \frac{1}{2}\tau_{\rm rh}(1-3w) }\bigg)^2\, \tau_e^{\frac{3(1+w)}{1+3w}}\tau_{\rm rh}^{\frac{-1+3w}{1+3w}}, & \tau_{\rm rh}<\tau\,.
  \end{dcases}
\end{align}
Consequently, we can express the Hubble rate as a function of the scale factor $a$,
\begin{align}
H(a) &= \begin{dcases}
  H_{\rm I}, & a \leq a_e \\
  H_{\rm I} \bigg(\frac{a_e}{a}\bigg)^{\frac{3 (1+w)}{2}}, & a_e < a \leq a_{\rm rh}\\
  H_{\rm{I}}\bigg(\frac{a_e}{a_{\rm rh}}\bigg)^{\frac{3(1+w)}{2}} \bigg(\frac{a_{\rm rh}}{a}\bigg)^{2}, 	\qquad&  a_{\rm rh}<a\,,
  \end{dcases}	\label{eq:Ha}
\end{align}
where $a_e\!\equiv\! a(\tau_e)$ and $a_{\rm rh}\!\equiv\! a(\tau_{\rm rh})$ define the scale factors at the end of inflation and reheating periods, respectively.
The total energy density $\rho(a)\!=\!\rho_\phi+\rho_{\text{\tiny SM}}$ in these different epochs during the evolution of universe is, 
\beq
\rho(a)\!=\! 3\mpl^2 H^2(a),  \label{eq:rho_a}
\eeq
where $\mpl\!\equiv\!1/\sqrt{8 \pi G}\!=\!2.435 \!\times\! 10^{18}\gev$ is the reduced Planck mass. Approximate form of individual components of the energy density, $\rho_\phi$ and $\rho_{\text{\tiny SM}}$, can be obtained by solving \eq{eq:rhophi_rhord}:
\begin{align}
\rho_\phi(a) &\approx \begin{dcases}
  3\mpl^2 H_{\rm I}^2\,, & a \leq a_e \\
  3\mpl^2 H_{\rm I}^2\,\bigg(\frac{a_e}{a}\bigg)^{3 (1+w)}, & a_e < a \leq a_{\rm rh}\\
  0, 	&  a_{\rm rh}<a\,,	
  \end{dcases} 	\label{eq:rhophi}	\\
\rho_{\text{\tiny SM}}(a) &\approx \begin{dcases}
  0\,, & a \leq a_e \\
  3\mpl^2 H_{\rm I}^2\bigg(\frac{a_e}{a_{\rm rh}}\bigg)^{\frac{3 (1+w)}{2}}\bigg[\bigg(\frac{a_e}{a}\bigg)^{\frac{3 (1+w)}{2}}-\bigg(\frac{a_e}{a}\bigg)^{4}\bigg], & a_e < a \leq a_{\rm rh}\\
  3\mpl^2H_{\rm I}^2\bigg(\frac{a_e}{a_{\rm rh}}\bigg)^{3(1+w)} \bigg(\frac{a_{\rm rh}}{a}\bigg)^{4}, 	&  a_{\rm rh}<a\,,
  \end{dcases}	\label{eq:rhord}
\end{align}
where we have used $\Gamma_{\phi}\!=\!(5-3w)H_{\rm rh}/2$, such that $H_{\rm rh}\!\equiv\!H(a_{\rm rh})$ defines the end of reheating period when $\rho_\phi(a_{\rm rh})\!=\!\rho_{\text{\tiny SM}}(a_{\rm rh})$. As mentioned in Introduction we confine $w\!\in\!(\!-1/3,1)$, therefore during the non-standard reheating period the SM radiation energy density $\rho_{\text{\tiny SM}}$ approximately scales as $a^{-\frac{3 (1+w)}{2}}$ (the second term in the parenthesis in the middle line of (\ref{eq:rhord}) proportional to $a^{-4}$ is negligible). Hence the temperature during the reheating period scales as $a^{-\frac{3 (1+w)}{8}}$, this is a useful result for later use. Note that the SM energy density $\rho_{\text{\tiny SM}}$ is zero during the inflationary period and is continuous at the end of reheating since $a^{-4}$ term in the middle line of (\ref{eq:rhord}) is negligible in comparison to $a^{-3(1+w)/2}$ term for $-1/3\!<\!w\!<\!1$.

Finally, it is convenient to define {\it reheating efficiency} $\gamma$ as
\beq
  \gamma \equiv \sqrt{\frac{H_{\rm rh}}{H_{\rm I}}}\,,
    \label{reh_eff}
\eeq
which parametrizes the duration of reheating with non-standard cosmology.
By fixing $\gamma$ and $w$, we can express $a_{\rm rh}$ in terms of $a_e$, or equivalently $\tau_{\rm rh}$ in terms of $\tau_e$, as
\begin{align}
  a_{\rm rh}&= a_e \,\gamma^{-\frac{4}{3(1+w)}}, 	&\tau_{\rm rh}&= \tau_e \,\gamma^{-\frac{2(1+3w)}{3(1+w)}}.		\label{eq:a_rh}
\end{align}
The efficiency of reheating $\gamma$ is a constrained parameter. 
Requiring $H_{\rm rh}\!<\!H_{\rm I}$ sets an upper limit on $\gamma\!<\!1$. 
Whereas the lower limit on $\gamma\gtrsim 10^{-18}$ can be deduced by using the fact that curvature perturbations during inflation constrain $H_{\rm I}\!\leq\!6.6\!\times\!10^{13}\gev$ at 95\% C.L.~\cite{Akrami:2018odb} and Big Bang Nucleosynthesis (BBN) sets the lower limit on the reheating temperature $T_{\rm rh}\!\gtrsim 10 \mev$~\cite{Sarkar:1995dd}, (see \eq{eq:Trh}).

\begin{figure}[t]
\centering
  \!\!\!\includegraphics[width=0.5\textwidth]{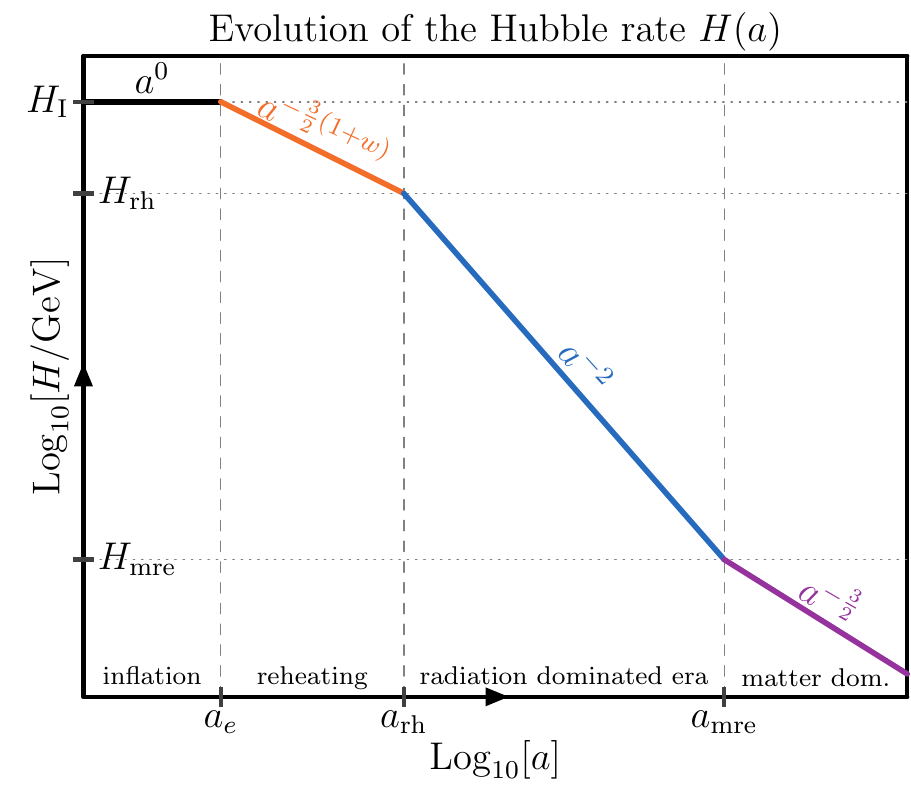}\!\!\!
  \includegraphics[width=0.5\textwidth]{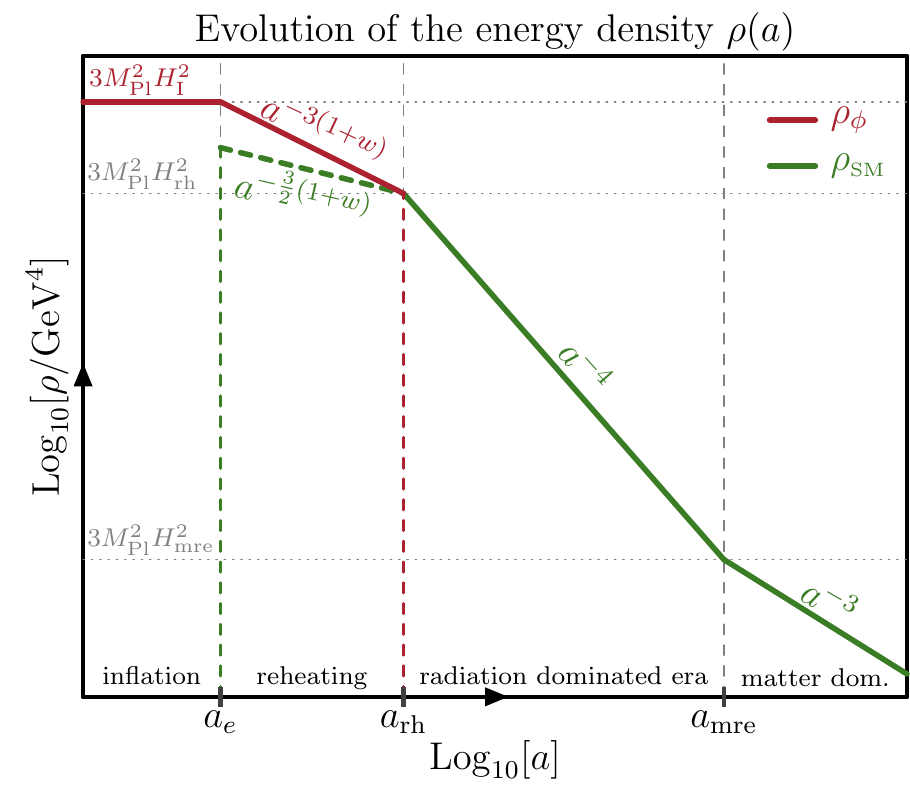}
\caption{The cosmological evolution of the Hubble rate $H(a)$ (left-panel) and energy density $\rho(a)$ (right-panel) as a function of the scale factor $a$. The scale factor at the end of inflation $a_e$, end of reheating $a_{\rm rh}$, and matter-radiation equality $a_{\rm mre}$ are represented as gray dashed vertical lines.}
\label{fig:H_and_rho}
\end{figure}

Evolution of the Hubble rate $H(a)$ and the energy density $\rho(a)$ as a function of the scale factor $a$ are illustrated in \fig{fig:H_and_rho}. 
During the reheating period, we consider non-standard cosmological evolution parameterized by $w\!\in\!(-1/3,1)$. 
The end of inflation $a_e$, end of reheating $a_{\rm rh}$, and the end of RD, i.e. the matter-radiation equality~(mre) $a_{\rm mre}$, periods are shown as gray dashed vertical lines. In the left-panel, the gray dotted horizontal lines represent the Hubble rate at the end of inflation~$H_{\rm I}$, the Hubble rate at the end of reheating~$H_{\rm rh}$ and the Hubble rate at the end of RD epoch~$H_{\rm mre}$.
In the right-panel, we show the total energy density $\rho(a)\!=\!\rho_\phi+\rho_{\text{\tiny SM}}$ as solid curve. The inflaton energy density $\rho_\phi$ is shown as the red curve, whereas the SM energy density $\rho_{\text{\tiny SM}}$ is shown as green curve. Note that the $\rho_{\text{\tiny SM}}$ is zero during the inflationary period, however, it has non-zero value proportional to $a^{-\frac32(1+w)}$ during the non-standard reheating, shown as dashed green curve.

\subsection{Longitudinal modes during inflation}
During the slow-roll period of inflation the Hubble rate $\hi$ is almost constant, so we adopt the de Sitter solution. The relation between the scale factor and conformal time is given by~\eqref{eq:scalefactor},
\begin{align}
    a(\tau\leq \tau_e) \equiv \ai(\tau) = \frac{-1}{\hi (\tau-\frac{3}{2}\tau_e(1+w))}\,.
\end{align}
Consequently, during the de Sitter stage the longitudinal frequency $\omega_L^2$~\eqref{eq:omegaL} simplifies as
\begin{align}
    \omega_L^2 (\tau\leq \tau_e) \equiv  \omega_{\rm I}^2(\tau)= k^2 + a^2_{\rm I}(\tau) m_X^2 - \frac{2 k^4-k^2 a^2_{\rm I}(\tau) m_X^2}{(k^2 + a^2_{\rm I}(\tau) m_X^2)^2} a^2_{\rm I}(\tau) H^2_{\rm I} \,,	\label{eq: omegainf}
\end{align}
where we have used the fact that during the slow-roll period of inflation 
\begin{align*}
\Big(\frac{a^{\prime }}{a}\Big)^2 = a_{\rm I}^2(\tau) H_{\rm I}^2,        \qquad   \frac{a^{\p\p}}{a} = 2 a^2_{\rm I}(\tau) H_{\rm I}^2.
\end{align*}
It is seen that for vector DM mass $m_X\!\gg\! H_{\rm I}$, $\omega_L^2$ remains positive with no chance for the tachyonic enhancement. Furthermore, another reason not to consider vector DM heavier than the inflationary scale $H_{\rm I}$ is the fact that we assume during inflation the inflaton energy density $3\mpl^2 H_{\rm I}^2$ to be the dominant energy density. Therefore, hereafter we consider only vector DM masses $m_X \!\lesssim\! H_{\rm I}$.

In the far past during inflation (subhorizon limit) all modes were deep inside the horizon, so that $\ai(\tau) m_X\! \ll\! \ai(\tau) H_{\rm I} \!\ll\! k$.  In this limit, the frequency~$\omega_L^2$ becomes constant which results in a simple harmonic oscillator equation for modes,
\begin{align}
    \xtl^{\prime \prime}(\tau) +k^2 \xtl(\tau) =0.
\end{align}
The unique solution that minimizes the energy is known as the Bunch-Davies state,
\begin{align}
    \lim_{\tau \rightarrow - \infty} \xtl(\tau) = \frac{1}{\sqrt{2k}}e^{-i k \tau}. \label{eq:bunchdavies}
\end{align}
Hereafter we will take the above solution as our initial condition.

\begin{figure}[t!]
   \includegraphics[width=0.5\textwidth]{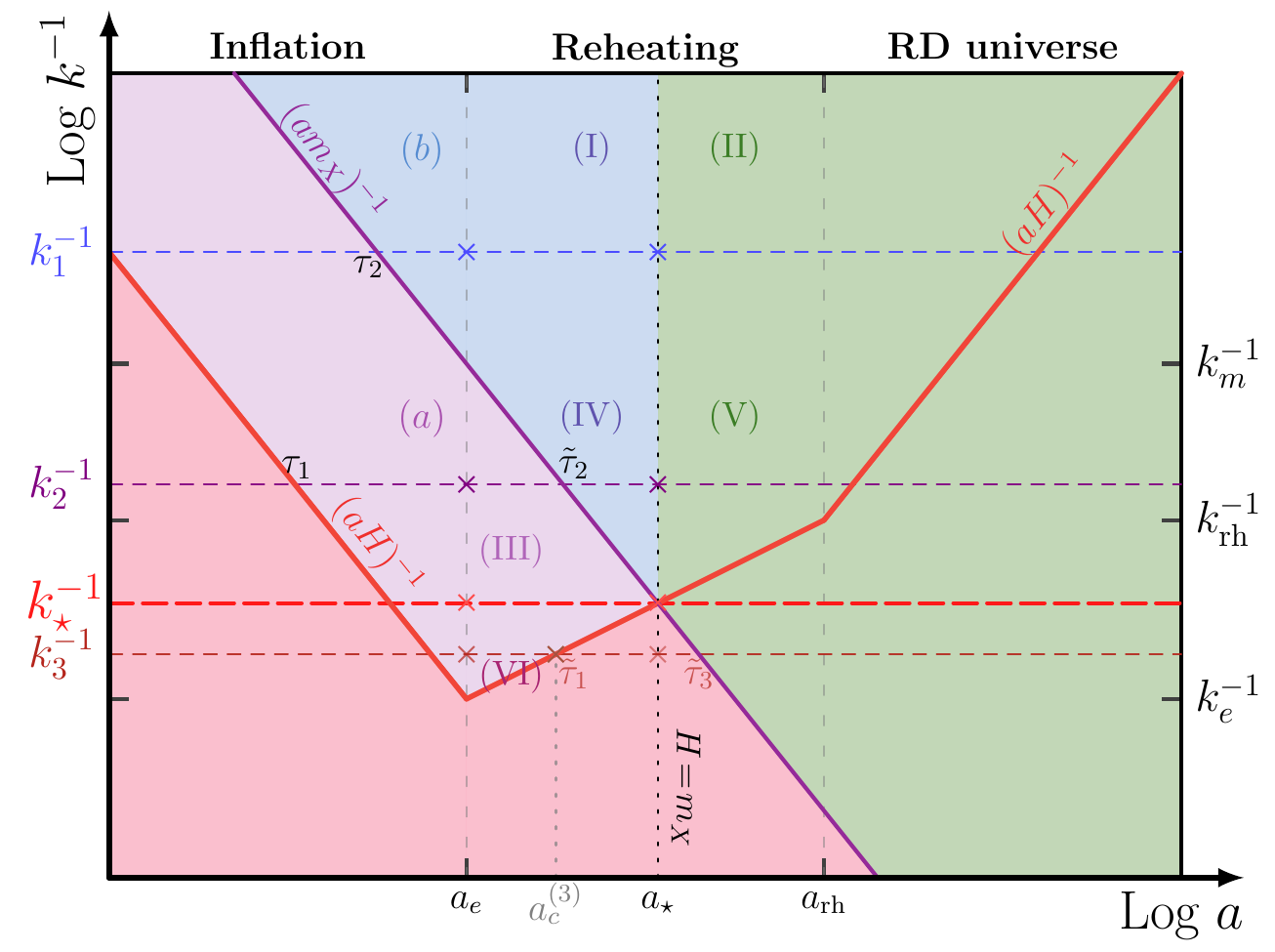}\!\!\!
    \includegraphics[width=0.5\textwidth]{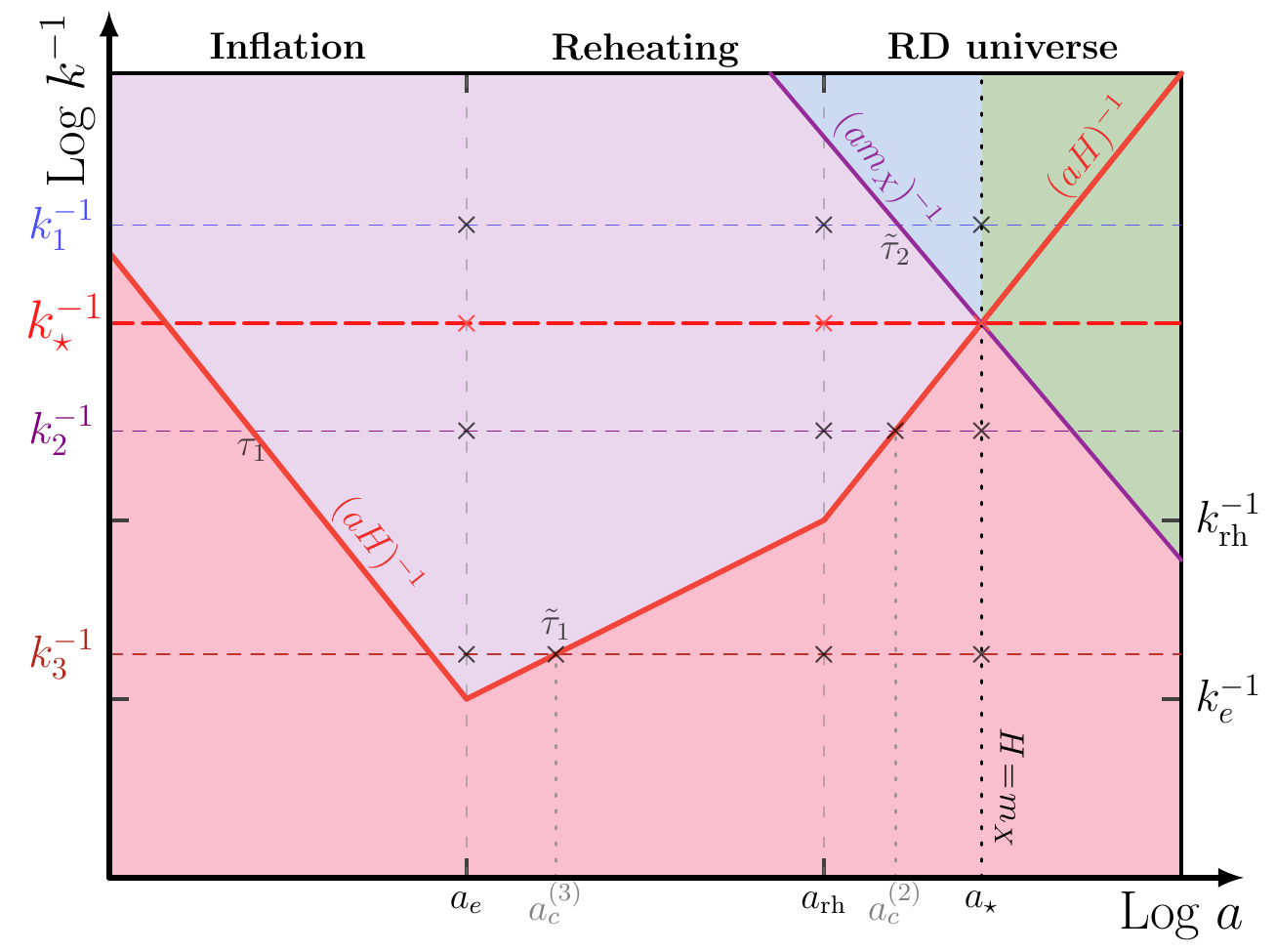}
\caption{Evolution of various cosmological distances during and after inflation for heavy vector DM i.e. $H_{\rm rh}\!\leq\!m_X\!<\!H_{\rm I}$ (left diagram) and light vector DM $m_X \!<\! H_{\rm rh}$ (right diagram).  The red region corresponds to modes with wavevector in range $am_X, aH \!\ll\! k$, purple refers to the region where $am_X\!<\!k\!<\!aH$, blue corresponds to the condition $k\!<\!am_X\!<\!aH$  and in the green region $aH, k \!\ll\! a m_X$. Here $a_c=k/H$ refers to the second \textit{horizon crossing}, $a_{\star}\equiv a(\tau_{\star})$, $k_m\equiv a_em_X$, $k_{\star}=a_{\star} m_X$, $k_e \equiv a_e H_{\rm{I}}$ and $k_{\rm rh} \equiv a_{\rm rh} H_{\rm rh}$. The plot assumes $-1/3<w<1/3$ during the reheating phase.} 
\label{fig:diagenergyscales}
\end{figure}

For the vector DM mass $m_X\! \lesssim\! H_{\rm I}$, there are two other regimes relevant during inflation: 
\begin{itemize}
 \item[({\it a})] {\it Intermediate-wavelength case} ($\ai(\tau) m_X \!\ll \!k\! \ll\! \ai(\tau) \hi$), such that  
 \beq
 \omega_{\rm I}^2 \approx k^2 - 2 a_{\rm I}^2(\tau) \hi^2.
 \eeq
 \item[({\it b})] {\it Long-wavelength case} ($k\! \ll\! \ai(\tau) m_X \!\ll\! \ai(\tau) \hi$), such that 
 \beq
 \omega_{\rm I}^2 \approx \ai^2(\tau)  m_X^2 + \frac{k^2}{\ai^2(\tau)m_X^2} \ai^2(\tau) \hi^2.
 \eeq
\end{itemize}
In \fig{fig:diagenergyscales} we sketch the evolution of various cosmological distances as functions of the scale factor~$a$ for the two vector DM mass regimes; heavy $H_{\rm rh}\!\leq\!m_X\!<\!H_{\rm I}$ (left-panel) and light $m_X \!<\! H_{\rm rh}$ (right-panel).
During inflation the intermediate- and long-wavelength regimes are shown as the {\it purple} region `$(a)$', and the {\it blue} region `$(b)$', respectively. 
The intermediate- and long-wavelength modes are represented in the left-panel of \fig{fig:diagenergyscales} by $k_{1}\!<\!k_m\!\equiv\!a_e m_X$ and $k_2$ such that $k_m\!<\!k_2\!<\!k_e\!\equiv\!a_e H_{\rm I}$~\footnote{Note that, in fact, this is the case only towards the end of inflation.}, respectively. Early enough both these modes were in the {\it red} region (sub-horizon) and they satisfied the initial condition provided by \eqref{eq:bunchdavies} known as the Bunch-Davies vacuum. Next they enter the intermediate-wavelength region ({\it purple}) and before the end of inflation at $a_e$ the mode with momentum $k_1$ crosses the Compton wavelength $(a m_X)^{-1}$ at $\tau_2$ and enters the long-wavelength regime ({\it blue}) region. 
We solve the mode equation in different regimes of $k$ until the end of inflation. After that, at $a_e$, we match solutions found during inflation with those during the reheating phase, that will be discussed in the next subsection~\ref{sec:long_rh}. 

Starting with the first ({\it purple}) region during inflation, i.e. intermediate-wavelength $\ai(\tau) m_X \!\ll\! k \!\ll\! \ai(\tau) \hi$, the equation of motion for modes takes the form
\begin{align}
    \frac{d^2}{d\eta^2}\xtl + \left(k^2 - \frac{2}{\eta^2}\right)\xtl =0,
\qquad {\rm where} \quad 
    \eta\equiv \tau -\frac{3}{2}(1+w)\tau_e.
\end{align}
The solution to the above equation is given by
\begin{align}
    \xtl^{(a)}(\eta) = \sqrt{\frac{2}{\pi k}}\, C_1^{(a)} \left[\frac{\sin (k \eta)}{k \eta }-\cos (k \eta )\right]-\sqrt{\frac{2}{\pi k}} \,C_2^{(a)} \left[\frac{\cos (k \eta)}{k \eta }+\sin (k \eta )\right],
\end{align}
where the integration constants are obtained, by imposing the Bunch-Davies initial condition~\eqref{eq:bunchdavies}, as
\begin{align}
    C_1^{(a)} = - \frac{\sqrt{\pi}}{2}e^{i k \left(\frac{3}{2}(1+w) \tau_e\right)}, \qquad\qquad
    C_2^{(a)} = - i c_1^{(a)}\,.
\end{align}
Hence the mode solution takes the form
\begin{align}
    \xtl^{(a)}(\tau) &= \frac{1}{\sqrt{2k}}\bigg( 1- \frac{i}{k \big(\tau- \frac{3}{2}(1+w)\tau_e \big)} \bigg) e^{- i k \tau} \nonumber \\
    &\overset{k \ll a_{\rm I} H_{\rm I} }{\approx}\frac{-i}{\sqrt{2k}} \frac{1}{k \big(\tau- \frac{3}{2}(1+w)\tau_e \big)} e^{- i k \tau} = \frac{i}{2\sqrt{k}}\frac{\ai(\tau) \hi}{k}e^{- i k \tau}\,,    \label{sol1}
\end{align}
where the approximation in the second line above ($k \!\ll\! a_{\rm I} H_{\rm I}$) implies the modes are well inside the intermediate wavelength ({\it purple}) region.
In this limit longitudinal modes grow linearly with the scale factor $\ai(\tau)$ until the mode momentum $k$ crosses the Compton wavelength, i.e. $k=\ai(\tau) m_X$ line. This evolution of the longitudinal modes of the vector DM is similar to a massless scalar.

In the second case with long-wavelength $k \! \ll \!\ai(\tau) m_X \! \ll\! \ai(\tau) \hi$ ({\it blue} region), the mode
equation \eqref{eq:doe} reduces to
\begin{align}
    \frac{d^2}{d\eta^2}\xtl + \bigg( \frac{m_X^2}{\hi^2 \eta^2} + \frac{ k^2\hi^2}{m_X^2}  \bigg)\xtl=0,     \label{eq:modeEb}
\end{align}
with a solution 
\begin{align}
    \xtl^{(b)}(\eta) = \sqrt{ \eta } \left[C_1^{(b)} J_{n}\left(\frac{H_{\rm I}}{m_X}k  \eta  \right) + C_2^{(b)} Y_{n}\left(\frac{H_{\rm I}}{m_X}k  \eta  \right)\right],
\end{align}
where $J_{n}(x)$ and $Y_n(x)$ are the Bessel functions of the first and second kind, respectively. The order of Bessel functions $n$ is 
\begin{align}
    n= \frac{1}{2}\sqrt{1-\frac{4 m_X^2}{H_{\rm I}^2}}\approx \frac{1}{2}.
\end{align}
Consequently, the mode $\xtl^{(b)}$ simplifies as follows
\begin{align}
    \xtl^{(b)}(\tau) =- C_1^{(b)}  \sqrt{\frac{2 m_X}{ \pi k H_{\rm I}}} \sin{\left(\frac{k}{\ai(\tau)m_X} \right) - C_2^{(b)}  \sqrt{\frac{2 m_X}{ \pi k H_{\rm I}}} \cos{\left( \frac{k}{\ai(\tau) m_X}\right)} }. \label{eq:xsolb}
\end{align}
We fix coefficients $C_1^{(b)}$ and  $C_2^{(b)}$ by requiring,
\begin{align}
    \xtl^{(b)}(\tau_2)&= \xtl^{(a)}(\tau_2),      &\xtl^{(b)\prime}(\tau_2)&= \xtl^{(a) \prime}(\tau_2),
\end{align}
where $\tau_2$ is defined by the condition $k\approx a_{\rm I}(\tau_2)m_X$, i.e.
\begin{align}
    \tau_2 = \frac{3}{2}\tau_e (1+w) - \tilde b\frac{m_X}{k H_{\rm I}},    \label{eq:tau2}
\end{align}
here $\tilde b$ is an auxiliary parameter introduced to improve the agreement between numerical and analytical solutions, in the following calculations $\tilde b=1.2$ has been adopted. For the coefficients we find,
\begin{align}
    C_1^{(b)}\!&= ~\sqrt{\frac{\pi}{H_{\rm I}m_X}} \frac{(i H_{\rm I}^2 +\tilde b H_{\rm I}m_X - i \tilde b^2 m_X^2)\cos{(\tilde b)}-\tilde b(i H_{\rm I}^2+\tilde b H_{\rm I}m_X)\sin{(\tilde b)}}{2\tilde b^2 m_X}e^{-ik \tau_2}, \\
    C_2^{(b)}\!&=\! - \sqrt{\frac{\pi}{H_{\rm I}m_X}} \frac{(i H_{\rm I}^2 +\tilde b H_{\rm I}m_X - i \tilde b^2 m_X^2)\sin{(\tilde b)}+\tilde b(i H_{\rm I}^2+\tilde b H_{\rm I}m_X)\cos{(\tilde b)}}{2 \tilde b^2 m_X}e^{-ik \tau_2}.
\end{align}
Expanding the above result in the limit $k \! \ll \!\ai(\tau) m_X \! \ll \!\ai(\tau) \hi$ we can approximate \eq{eq:xsolb} by the following constant solution
\beq
    \xtl^{(b)} \approx \frac{i}{\sqrt{2k}} \frac{H_{\rm I}}{m_X}\frac{(\sin{(\tilde b)}+ \tilde b\cos{(\tilde b)})}{\tilde b^2 } \,e^{-ik \big( \frac{3}{2}(1+w) \tau_e - \tilde b\frac{m_X}{k H_{\rm I}}\big)}\,.
        \label{sol2}
\eeq
The factor $(\sin{(\tilde b)}+ \tilde b\cos{(b)})/\tilde b^2\approx 0.95$ for $\tilde b=1.2$ and could be safely neglected. 
Note that in this case, with $k \! \ll \ai(\tau) m_X \! \ll \ai(\tau) \hi$, we have $\omega_L^2\!>\!0$~\eqref{eq:modeEb}. Hence there should be no tachyonic enhancement for $k$ values in this regime, as it is indeed confirmed by the above mode function being constant. Note however that in the long-wavelength region ({\it blue}) in \fig{fig:diagenergyscales} the constant mode function amplitude square $|\xtl^{(b)}|^2$ scales as $k^{-1}$ which is different from that of a massive scalar field which scales as $k^{-3/2}$.

\begin{figure}[t]
 \centering
    \includegraphics[width=0.5\textwidth]{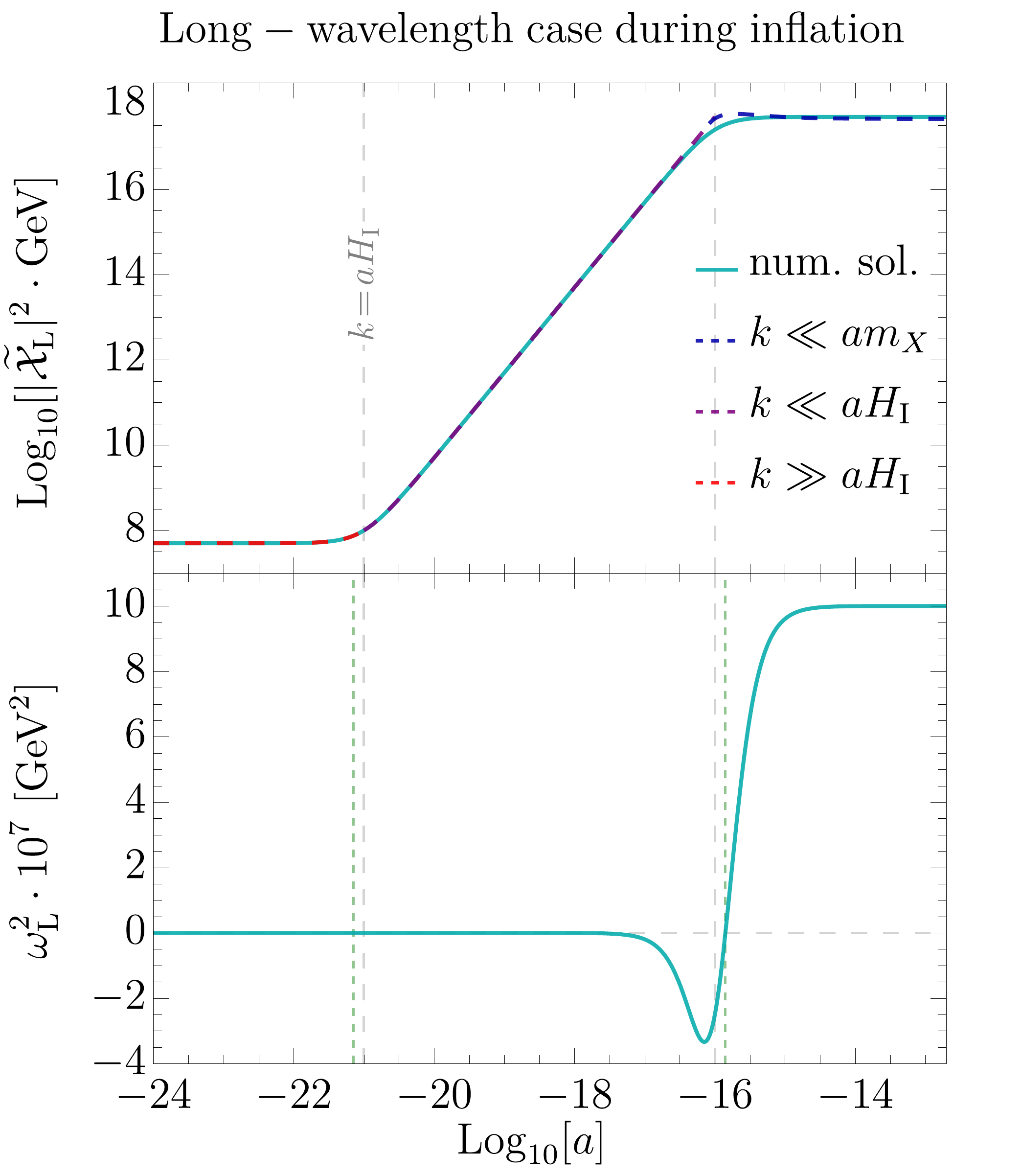}\!\!\!
    \includegraphics[width=0.5\textwidth]{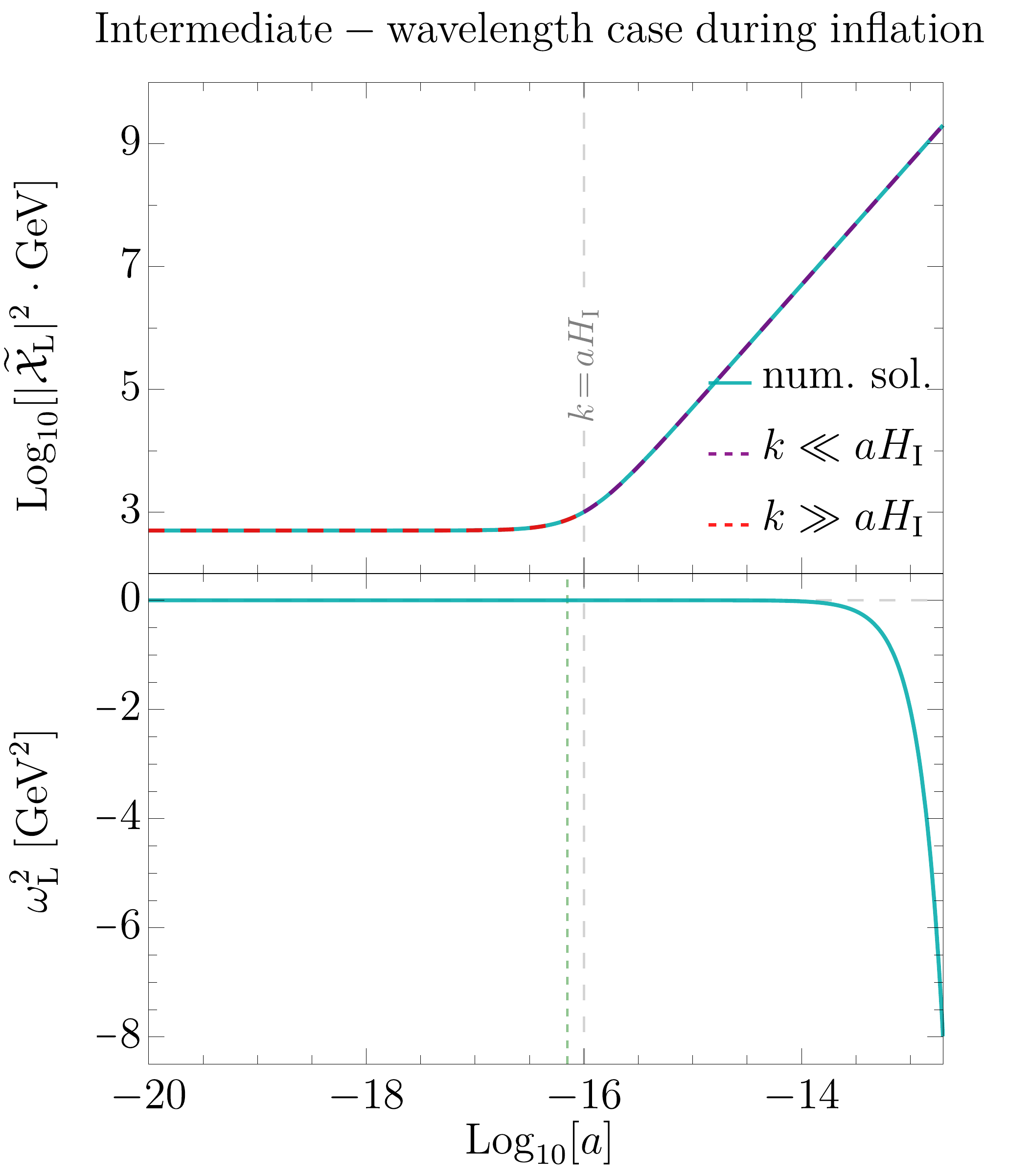}
    \caption{Evolution of the mode function $\lvert \widetilde{\cal X}_L \rvert^2$ and the frequency $\omega_L^2$ with the scale factor $a$ during inflation for $k \!\ll\! a_e m_X  \!\ll\! a_e \hi$ (left-panels) and $a_e m_X \!\ll\! k \!\ll\! a_e \hi$ (right-panels). In the upper panels cyan solid lines correspond to numerical solutions, while red, purple and blue dashed lines show analytical predictions specified in the main text by (\ref{eq:bunchdavies}), (\ref{sol1}) and (\ref{sol2}), respectively.  In the lower panel green dashed vertical lines indicate zeros of  $\omega_L^2$.   Here $m_X\!=\!10^8 \; \rm{GeV}$, $\hi\!=\!10^{13} \; \rm{GeV}$, $w\!=\!0$ and $k\!=\!10^{-8} \; \rm{GeV} $ ($k\!=\!10^{-3} \; \rm{GeV} $) for the left (right) panel. }
\label{fig:chi_and_omega}
\end{figure}

In \fig{fig:chi_and_omega} we show $\lvert\widetilde{\cal X}_L \rvert^2$ and the frequency $\omega_L^2$ as a function of the scale factor $a$ during inflation for $m_X=10^8\gev$, $\hi=10^{13} \gev$, and two values of the momentum $k=10^{-8} \; \rm{GeV} $ (left-panels) and $k=10^{-3} \; \rm{GeV} $ (right-panels) representing the two regimes $k \!\ll\! a_e m_X  \!\ll\! a_e \hi$ and $a_e m_X \!\ll\! k \!\ll\! a_e \hi$, respectively.  In the upper panels the approximate analytical solutions (\ref{eq:bunchdavies}), (\ref{sol1}) and (\ref{sol2}) are compared with exact numerical solutions. As it is seen the agreement is excellent. The vertical gray dashed lines show the value of the scale factor corresponding to the radius of the Hubble sphere during the inflation $a \!=\! k/\hi$ and $a\!=\!k/m_X$. At those points we have matched the approximate analytical solutions as discussed above.
In the lower panels we plot $\omega^2_L$ and show its zeros by vertical green dashed lines. Note nearly perfect correlation between enhancement (tachyonic) of $\lvert \widetilde{\cal X}_L \rvert^2$ and regions of negative $\omega^2_L$. This behavior has been anticipated.

Our goal is to find energy density of the longitudinal modes $\mathcal{\widetilde{X}}_L$ during inflation,
\begin{align}
    \langle \rho_{L}^{} \rangle = \frac{1}{4 \pi^2 a^4}\int dk k^2 &\bigg\{\lvert \mathcal{\widetilde{X}}'_L \rvert^2 - \big(\mathcal{\widetilde{X}}'_L \mathcal{\widetilde{X}}^*_L + \mathcal{\widetilde{X}}^{'*}_L \mathcal{\widetilde{X}}_L\big)  \frac{k^2 }{k^2 + a^2 m_X^2} a \hi   \nonumber\\ 
    &+ \bigg(\frac{k^4 }{(k^2 + a^2 m_X^2)^2} a^2 \hi^2 +  k^2 +  a^2 m_X^2 \bigg) \lvert \mathcal{\widetilde{X}}_L \rvert^2\bigg\}\,, \label{eq:energyden}
\end{align}
where $\langle \rho_L^{} \rangle$ denotes vacuum expectation value of the energy density calculated  with respect to the Bunch-Davies vacuum. 
In the case of the sub-horizon modes $\ai(\tau) \hi \ll k $ ({\it red} region in \fig{fig:diagenergyscales}) energy density per log momentum is given by
\beq
\frac{ d\langle \rho_L (a)\rangle}{ d \ln{k}} = \frac{k^4}{4 \pi ^2 a^4}+\frac{k^2 \left(\hi^2+m_X^2\right)}{8 \pi ^2 a^2}-\frac{\hi^2 m_X^2}{4 \pi ^2}+\frac{3 a^2 \hi^2 m_X^4}{8 \pi ^2 k^2} + {\cal O}\bigg(\Big(\frac{a m_X}{k} \Big)^2,\Big(\frac{a H_{\rm I}}{k}\Big)^4\bigg)\,.
\label{sub_hor}
\eeq
The first three terms are divergent in the limit $k \rightarrow \infty$ while integrating over $k$. However, since we are going to use a cut-off $\Lambda \!=\! a_e \hi$ for the integration,  that will eliminate contributions of those sub-horizon modes.

For modes with $\ai m_X \!\ll \!k \!\ll\! \ai \hi$ ({\it purple} region in \fig{fig:diagenergyscales}) the energy density per log momentum is
\beq
    \frac{d \langle \rho_L (a) \rangle }{d \ln{k}} = 
\frac{1}{2} \left(\frac{k}{a} \right)^2 \left( \frac{H_{\rm I}}{2 \pi} \right)^2    
\left\{ 1 + {\cal O}\bigg(\Big(\frac{m_X}{H_{\rm I}} \Big)^2,\Big(\frac{k}{a H_{\rm I}}\Big)^2\bigg)\right\}\,.
\label{k_mid}
\eeq
The modes with $k \!\ll\! \ai m_X \!\ll\! \ai \hi$ ({\it blue} region in \fig{fig:diagenergyscales}) contribute to the energy density per log momentum as,
\beq
    \frac{d \langle \rho_L (a) \rangle }{d \ln{k}} = 
\frac{1}{2} \left(\frac{k}{a} \right)^2 \left( \frac{H_{\rm I}}{2 \pi} \right)^2    
\bigg\{ 1 + {\cal O}\bigg(\Big(\frac{k}{a m_X} \Big)^2,\Big(\frac{k}{a m_X} \Big)^4 \cdot \Big(\frac{H_{\rm I}}{m_X}\Big)^2\bigg)\bigg\}.
\label{k_low}
\eeq
Therefore, at the end of inflation, $a_e\!\equiv\! a(\tau_e)$, the energy density per $\ln k$ of the longitudinal modes with $m_X \!\lesssim\!H_{\rm I}$ scales as,
\begin{align}
    \frac{d \langle \rho_L(a_e) \rangle }{d \ln{k}} \approx
    \begin{dcases}
    \frac{k^4}{4 \pi^2 a_e^4}, & \text{sub-horizon modes }  k\!\gg\!a_e H_{\rm I} ~ ({\rm \textcolor{Maroon!70}{red}}),\\
    \frac{1}{2} \left(\frac{k}{a_e} \right)^2 \left( \frac{H_{\rm I}}{2 \pi} \right)^2, & \text{super-horizon modes }  k \!\ll\!a_e H_{\rm I} ~ ({\rm \textcolor{purple!57}{purple},\textcolor{blue!70}{blue}}).\\
    \end{dcases}
\end{align}

\subsection{Longitudinal modes after inflation}
\label{sec:long_rh}
In this subsection our goal is to find solutions to the equation of motion~\eqref{eq:doe} for the longitudinal modes after inflation. In particular, we do not assume standard cosmological evolution, in which the inflationary period is followed by the $\rm{RD}$ epoch until the matter-radiation equality (mre). Instead, we allow the possibility for a non-standard cosmological evolution after the end of inflation by assuming a general equation of state $\eqref{eq:eosinf}$ such that the dominant energy density scales as $a^{-3(1+w)}$.
The period of non-standard cosmology, referred to as reheating, is then taken over by the standard cosmology with the RD universe.

After inflation, for $a\!>\! a_e$, the longitudinal mode frequency $\omega_L^2$~\eqref{eq:omegaL} takes the form
\beq
\omega_L^2 (\tau)= k^2 + a^2m_X^2 - \frac{k^2}{k^2+a^2m_X^2}\bigg(\frac{1-3w}{2}-\frac{3a^2 m_X^2}{k^2 +a^2 m_X^2} \bigg) a^2 H^2\,,
\eeq
where we have used the following relations 
\begin{align}
\left( \frac{a'}{a} \right)^{2}&= a^2 H^2,		&\frac{a^{\p\p}}{a}&= \frac{1-3w}{2}a^2 H^2\,.
\label{fried_eq}
\end{align}
After inflation the scale factor $a(\tau)$ and the Hubble rate $H(a)$ are given in Eqs.~\eqref{eq:scalefactor} and \eqref{eq:Ha}, respectively.
Note that the longitudinal mode frequency $\omega_L^2$ (\ref{eq:omegaL}) contains a term proportional to $a^{\prime \prime}$, which is discontinuous across the two transition points $\tau_e$ and $\tau_{\rm rh}$, therefore the longitudinal mode frequency also has a jump at these two points. 
Technically, discontinuity disappears by use of the Friedmann equations (\ref{fried_eq}), however, in some sense, it is still present due to the fact that we assume an instantaneous transition from the inflationary period to a non-standard cosmological epoch of reheating parameterized by $w$ and similarly for the transition from non-standard cosmological epoch to RD universe.
In a case of describing the inflation and reheating by a dynamical inflaton/reheaton field, the discontinuity would disappear as $w$ would be a continuous function of the inflaton/reheaton field. However, we would like to emphasize that the energy density \eqref{eq:energyden} contains only terms proportional to $a$, $a^{\prime}$ and $\xtl,$ $\xtl^{\prime}$. Since the scale factor and the Hubble rate are continuous across the transition points, by requiring the mode functions to match at $\tau_e$ and $\tau_{\rm rh}$:
\begin{align}
    \xtl^{(-)}(\tau_{e, \rm{rh}})=\xtl^{(+)}(\tau_{e, \rm{rh}}), \qquad\qquad  \xtl^{(-)\prime}(\tau_{e, \rm{rh}})=\xtl^{(+)\prime}(\tau_{e, \rm{rh}}), \label{eq:modereq}
\end{align}
where $\xtl^{(-/+)}$ denotes the {\it left/right} limits of the solutions to the modes equation, i.e. \(\displaystyle \xtl^{(-/+)}(\tau_{e,\rm{rh}})=\lim_{\tau \rightarrow \tau^{-/+}_{e, \rm{rh}}}\xtl(\tau)\), this ensures the energy density to be continuous. 

To find an approximate solution to the equation of motion for the longitudinal mode after inflation, we consider the following regimes in parameter space together with the corresponding frequency~$\omega_L^2$ values:
\begin{subequations}
\begin{empheq}[left={\displaystyle \omega_L^2\! \approx \!\empheqlbrace}]{align}
& k^2  & a(\tau)m_X, a(\tau) H(\tau) \ll k&~ ({\rm \textcolor{Maroon!70}{red}})     \label{eq:omegaL_a}\\
&\!\!\!-\!\!\bigg(\!\frac{1\!-\! 3 w}{2} \!-\!\frac{3a^2(\tau) m_X^2}{k^2} \!\bigg)a^2(\tau) H^2(\tau)\!\! & \!a(\tau)m_X \!\ll\! k\! \ll\! a(\tau) H(\tau) &~({\rm \textcolor{purple!57}{purple}})  \label{eq:omegaL_b}\\
&\frac{k^2}{a^2(\tau) m_X^2}\bigg(\frac{5+ 3 w}{2} \bigg)a^2 (\tau) H^2 (\tau) & \!k\! \ll\! a(\tau) m_X \!\ll\! a(\tau) H(\tau)& ~({\rm \textcolor{blue!70}{blue}})     \label{eq:omegaL_c} \\
&a^2(\tau) m_X^2 &  a(\tau)H(\tau), k \ll a(\tau) m_X&~ ({\rm \textcolor{OliveGreen!70}{green}})    \,, \label{eq:omegaL_d}
\end{empheq}
\end{subequations}
where we have indicated colors which, in \fig{fig:diagenergyscales}, mark regions where a given formula for the frequency applies.

In the following analysis we restrict our considerations to modes that went outside the horizon during inflation since only these modes receive tachyonic enhancement, i.e. $k\!<\!k_e\!=\!a_e \hi$.  
After the end of inflation, modes with mass $m_X \!\ll\! H_{\rm I}$ can enter two regions (see \fig{fig:diagenergyscales}): {\it purple} region, in which $am_X\!<\!k\!<\!aH$, and {\it blue} region, which describes modes with $k\!<\!am_X\!<\!aH_{\rm I}$. 
The modes with different $k$ values evolve differently and they all eventually enter the {\it green} area in which $a m_X\!\gg\! k, a H$ is the dominant energy scale.

Our aim is to find approximate analytical solutions for the mode equation in different mass and wavelength regions after the end of inflation. 
\subsubsection{Heavy vector dark matter: $H_{\rm rh}\!\leq\!m_X\!<\!H_{\rm I}$}

In the case of heavy vector DM we consider the following three regimes w.r.t.~the wavelength of the modes.  

\begin{figure}[t]
\centering\hspace{-7pt}
    \includegraphics[width=0.35\textwidth]{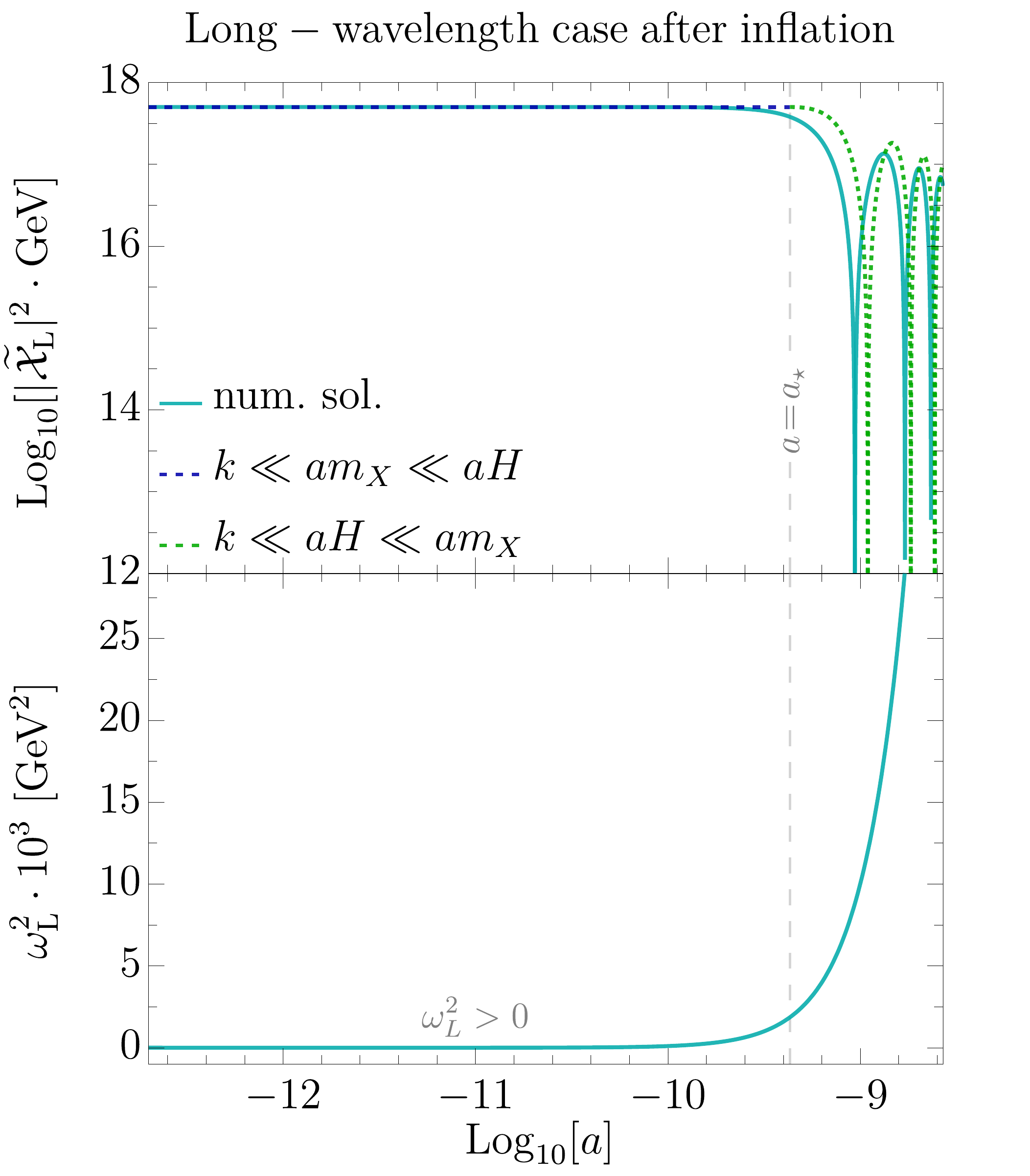}\hspace{-13pt}
    \includegraphics[width=0.35\textwidth]{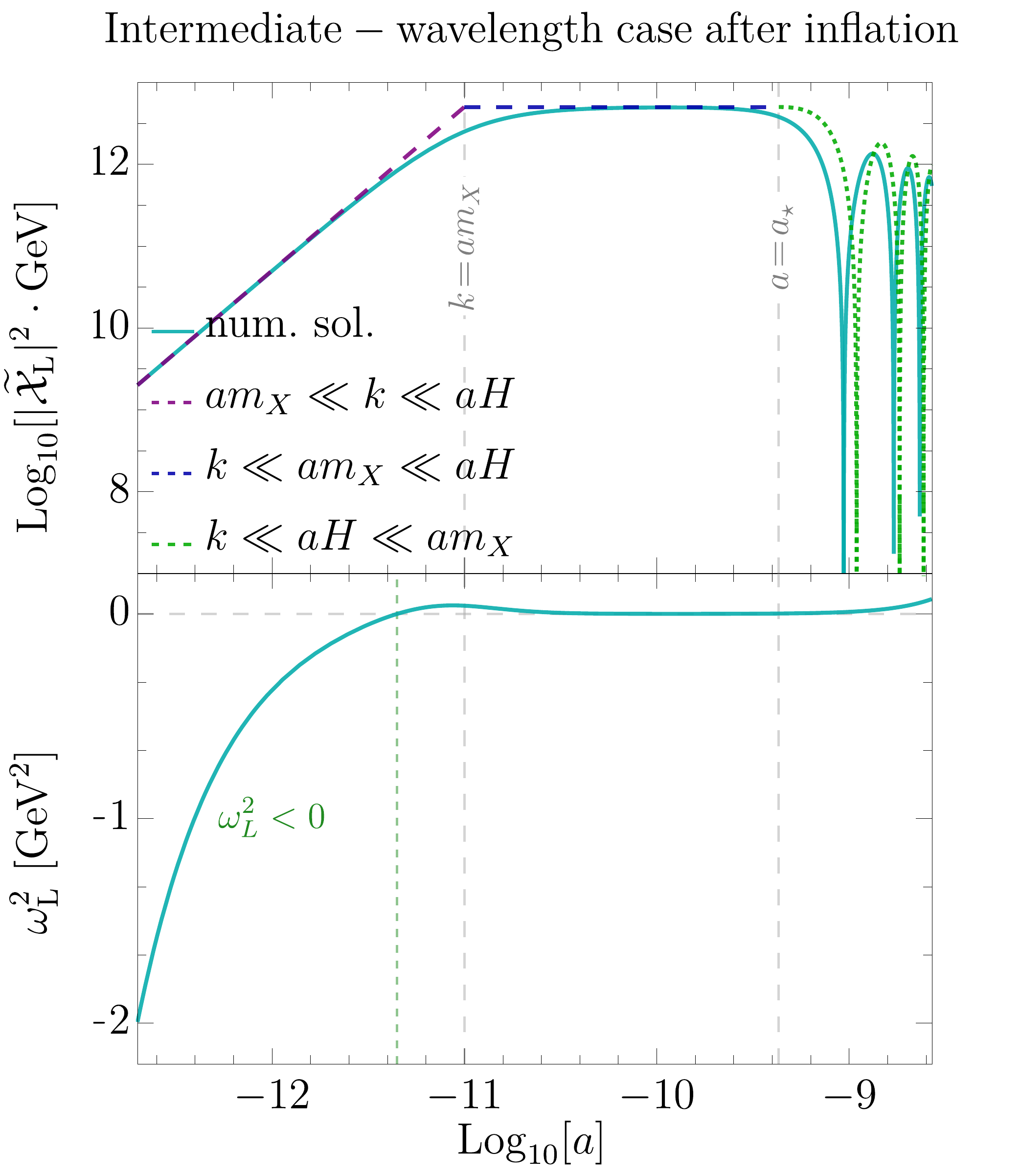}\hspace{-13pt}
    \includegraphics[width=0.35\textwidth]{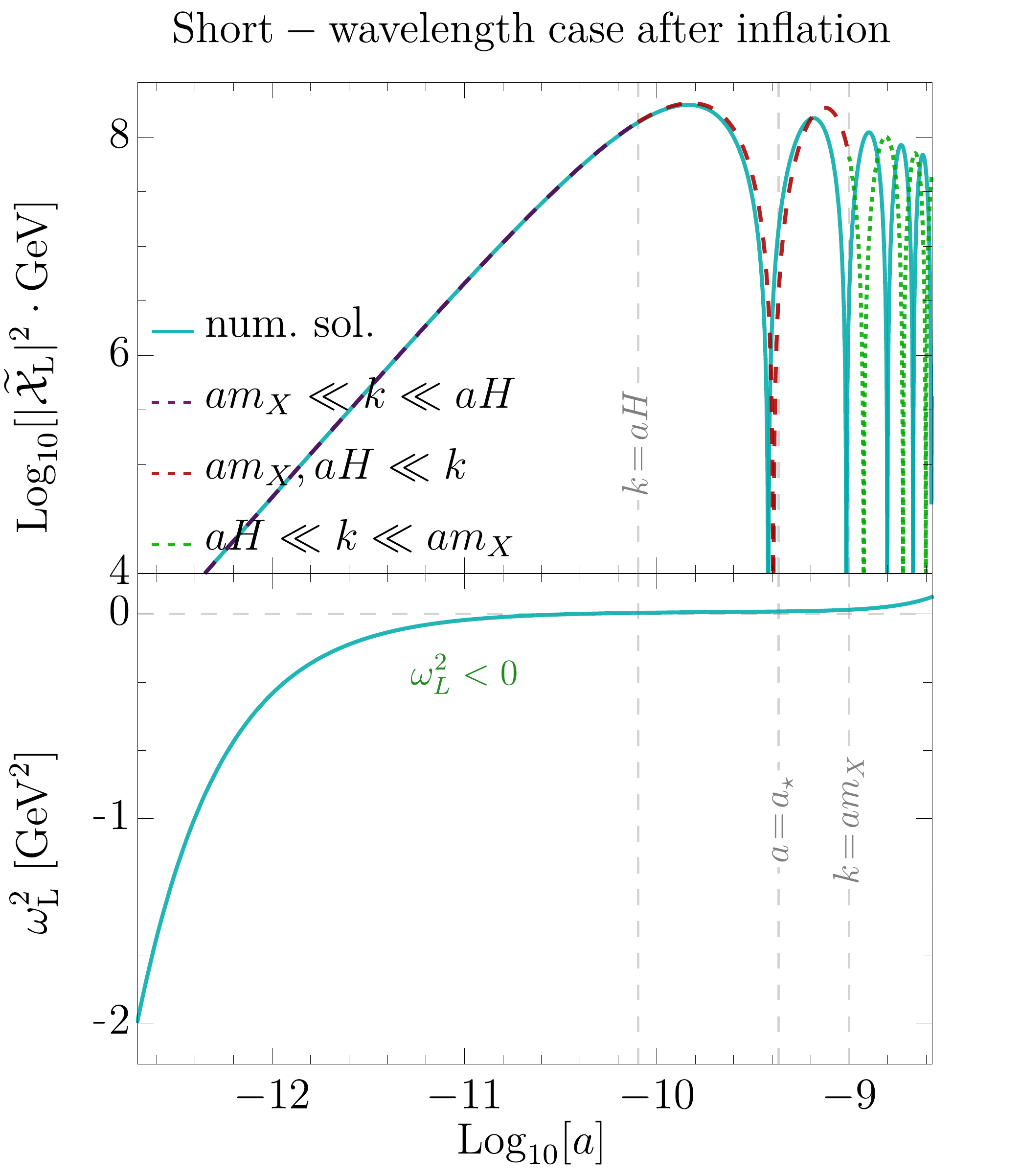}\hspace{-7pt}
    \caption{Evolution of the mode function $\lvert \widetilde{\mathcal{X}}_L \rvert^2$ (upper) and frequency $\omega_L^2$ (lower) with the scale factor $a$ after inflation for $k\!<\!k_m $ (left-panel), $k_m \!<\!k\!<\!k_{\star}$ (middle panel), and $k_{\star}\!<\!k\!<\!k_e$ (right-panel). The solid cyan curves correspond to the exact numerical solution, while colored dashed lines show analytical predictions. Here we take $m_X=10^8 \; \rm{GeV}$, $\hi=10^{13} \gev$, $w\!=\!0$ with $k=10^{-8}\gev, 10^{-3}\gev, 1\gev$ for the left, center, and right plots, respectively. }
    \label{fig:afterInflation}
\end{figure}

\paragraph{Long-wavelength case ($k\!\ll\! k_m \!\equiv \!a_em_X$):}~
After the end of inflation, modes with $k\!<\!k_m \!\equiv \!a_em_X $, for instance $k_1^{-1}$ line in \fig{fig:diagenergyscales}, pass through the {\it blue} region to the {\it green} area, where they eventually re-enter the horizon. 
At the early times, when the modes $k \!\ll\! a(\tau) m_X\! \ll\! a(\tau) H(\tau)$ are still in the blue region, the longitudinal mode frequency is negligible, i.e. $\omega_L^2 \rightarrow 0$. Consequently, in this regime, we obtain the following solution to the mode equation
\begin{align}
    \xtl^{(\rm{I})}(a) = C_1^{(\rm{I})} + C_2^{(\rm{I})}a^{\frac{1+3w}{2}}. \label{eq:solblue}
\end{align}
We fix the coefficients $C_1^{(\rm{I})}$ and $C_2^{(\rm{I})}$ using the continuity condition \eqref{eq:modereq} at the transition point $\tau=\tau_e$. In this case, we match $\xtl^{(\rm{I})}(a_e)$ with the solution $\xtl^{(b)}$~\eqref{sol2}, which implies
\begin{align*}
    C_1^{(\rm{I})}=\frac{i}{\sqrt{2k}} \frac{H_{\rm I}}{m_X} e^{-i k \Big(\frac{3}{2}\tau_e(1+w)- \tilde{b}\frac{m_X}{kH_{\rm I}} \Big)} , \qquad\qquad C_2^{(\rm{I})}=0.
\end{align*}
Therefore, we get a constant solution in this region
\begin{align}
    \xtl^{(\rm{I})}= \frac{i}{\sqrt{2k}}\frac{H_{\rm I}}{m_X} e^{-i k \left(\frac{3}{2}\tau_e(1+w)- \tilde{b}\frac{m_X}{kH_{\rm I}} \right)},
\end{align}
where we have neglected the factor $(\sin{(\tilde b)}+ \tilde b\cos{(b)})/\tilde b^2\!\approx\!1$ in \eqref{sol2}.
This approximate solution is shown in the left-panel of \fig{fig:afterInflation} as dashed blue line which matches well with the exact numerical solution presented as solid cyan curve. Again note that the amplitude of the modes in the {\it blue} region scales w.r.t. the wavelength as $\lvert \xtl^{(\rm{I})}\rvert^2\propto k^{-1}$.

Next, at $\tau=\tau_{\star}=\big(\tilde{b}m_X/H_{\rm I} \big)^{-\frac{1+3w}{3(1+w)}}$ such that $m_X\!\approx \!H(\tau_{\star})$, the modes enter the {\it green} region in the left-panel of \fig{fig:diagenergyscales}, in which the longitudinal frequency is well approximated by $\omega_L^2\approx a^2 m_X^2$. Note that in this regime $\omega_L^2$ changes slowly, i.e.
\beq
2 \pi \lvert \omega'_L \rvert \!\ll\!\omega_L^2 \label{eq:adiabaticcon}.
\eeq
Since in this regime $\omega_L^2 \approx a^2 m_X^2$, therefore the above condition implies
\begin{align}
    2 \pi H\ll m_X\,,
\end{align}
which usually (except a short period  just after $\tau = \tau_\star$) is satisfied in this region.
Hence, we can use the following approximation to the solution of the mode equation
\beq
\mathcal{\tilde{X}}^{(\rm{II})}_L(\tau)\! =\! \frac{C_1^{(\rm{II})}}{\sqrt{2 a(\tau) m_X}} \exp\!\Big(i \!\!\int_{\tau_{\star}}^\tau\! d\tau' \,a(\tau') m_X \Big)+ \frac{C_2^{(\rm{II})}}{\sqrt{2 a(\tau) m_X}} \exp\!\Big(\!\!-i \!\!\int_{\tau_{\star}}^\tau\! d\tau' \,a(\tau') m_X \Big). \label{eq:solgreen}
\eeq
The $\tau$  dependance could be easily converted into a dependance on the scale factor $a$. To find $C_1^{(\rm{II})}, C_2^{(\rm{II})}$ we should merge the above solution with $\xtl^{(\rm{I})}$ at $\tau= \tau_{\star}$. 
The approximate form of these coefficients is 
\begin{align}
C_{1,2}^{(\rm II)} = \frac{H_{\rm I}}{2}\sqrt{\frac{a(\tau_{\star})}{m_X k}} \bigg[\frac{\tilde{b}}{2} \bigg(\frac{ \tilde{b}m_X}{H_{\rm I}}\bigg)^{-\frac{2}{3 (w+1)}} \frac{a_e}{a(\tau_{\star})}\pm  i\bigg]\exp\!\!\Big[\!\!-\!ik\Big(\frac{3}{2}\tau_e(1+w)- \frac{\tilde{b}m_X}{kH_{\rm I}} \Big)\Big].
\end{align}
The form of solutions in this region are oscillatory decaying as shown in the left-panel of \fig{fig:afterInflation} (dotted green) which match well with the exact numerical solution (solid cyan).
\paragraph{Intermediate-wavelength case ($k_m \!<\!k\!<\!k_{\star}\!\equiv\! a(\tau_{\star})m_X$):} 
After the end of inflation modes with wavevector in the range $k_m \!<\! k\!< \!k_{\star}$, e.g. $k_2^{-1}$ line in \fig{fig:diagenergyscales}, pass through the three regions ({\it purple}, {\it blue}, {\it green}). Their post-inflationary evolution starts in the {\it purple} region, where $am_X \!\ll\! k\! \ll\! aH$ so that  $a^2 m_X^2/k^2 \ll 1$. This implies that the second term in the parenthesis in Eq.~\eqref{eq:omegaL_b} can be neglected. Hence the mode equation reduces to
\beq
\mathcal{\tilde{X}}_L^{\p\p (\rm{III})} - \frac{(1- 3w)}{2}a^2 H^2 \mathcal{\tilde{X}}_L^{(\rm{III})} =0 \label{eq:solv},
\eeq
with the solution
\begin{equation*}
\mathcal{\tilde{X}}_L^{ (\rm{III})}(a) = C_1^{(\rm{III})} a + C_2^{(\rm{III})} a^{-\frac{(1-3w)}{2}}.        \label{eq:solb}
\end{equation*}
We fix the coefficients $C_1^{(\rm{III})}$ and $C_2^{(\rm{III})}$ by merging the above solution at $\tau\!=\!\tau_e$ with $\xtl^{(a)}(\tau_e)$~\eqref{sol1}. However, as long as $w\!<\!1$ the second term in \eqref{eq:solv} is small compared to the first one. Since we confine ourselves to the range $-1/3\!<\!w\!<\!1$, therefore, in the subsequent calculations we neglect this part of the solution. Hence the above mode solution reduces to
\begin{align}
\xtl^{\rm{III}}(a) &= C_1^{(\rm{III})}a, \label{eq:solviolet}
&{\rm where}\quad    C_1^{\rm{(III)}}& = \frac{iH_{\rm{I}}}{\sqrt{2} k^{3/2}}\,e^{-i k \tau_e}\,.
\end{align}
So, in this regime (purple region) the longitudinal modes evolve linearly with the scale factor $a$. We show this approximate analytic solution as the dashed purple curve in the upper-middle panel of \fig{fig:afterInflation}.
 
Next, at $\tau\!=\!\tilde{\tau}_2= \tau_e \big( \tilde{b}\, a_em_X /k\big)^{-(1+3w)/2}$ such that $k\!= \!\tilde{b} a(\tilde{\tau}_2)m_X$, the mode crosses the Compton wavelength line $(a m_X)^{-1}$ and enter the {\it blue} area in \fig{fig:diagenergyscales} (left-panel). 
In this region, we have found an asymptotic analytical solution given by \eqref{eq:solblue}. 
In order to get coefficients parametrizing solution \eqref{eq:solblue} for $k_m \!<\!k\!<\!k_{\star}$ we merge solution \eqref{eq:solblue} with \eqref{eq:solviolet} at $\tau\!=\! \tilde{\tau}_2$. 
Note that for $w\!>\! -1/3$ the second term in Eq.~\eqref{eq:solblue} is growing. 
Contrary to the long-wavelength case, now the modes with $k\!\ll\! am_X \!\ll\! aH$ follow the {\it purple} region, in which mode functions increase in amplitude according to solution~\eqref{eq:solviolet}. 
One would naively expect that the growing term in~\eqref{eq:solblue} should dominate the solution following~\eqref{eq:solviolet}. 
However, exact numerical solution presented in upper-middle panel of \fig{fig:afterInflation} shows that even in this case the growing part of the solution~\eqref{eq:solblue} is not the dominant one. Consequently, by keeping only the constant part of Eq.~\eqref{eq:solblue} we get,
\begin{align}
 \xtl^{(\rm{IV})} = C_1^{(\rm IV)}=C_1^{(\rm{III})}\frac{k}{m_X} \,.     \label{eq:solblue2}
\end{align}
Again note that the mode amplitude in this {\it blue} region scales as, $ |\xtl^{(\rm{IV})}|^2\propto\!k^{-1}$.
This constant solution is shown in the upper-middle panel of \fig{fig:afterInflation} as the dashed blue line. 

Finally at $\tau = \tau_{\star}$, modes with intermediate wavelength evolve to the {\it green} region where approximate solution to mode equation of motion in given by \eqref{eq:solgreen}. Once again we fix coefficients by merging  in Eq.~\eqref{eq:solgreen} with the solution~\eqref{eq:solblue2} at $\tau = \tau_{\star}$ and get decaying oscillatory solutions shown as dotted green in the middle panel of \fig{fig:afterInflation}. 
The approximate solution in this region is in the form given by~\eqref{eq:solgreen} with the following coefficients:
\begin{align}
C_{1,2}^{\rm (V)}= \frac{C_{1}^{(\rm III)}}{\sqrt{2}}\sqrt{m_{X}a(\tau_{\star})}\bigg[ 1\mp i \frac{a_e}{a(\tau_{\star})}\frac{H_{\rm _I}}{2 m_X}\bigg(\frac{\tilde{b} m_X}{H_{\rm I}}\bigg)^{\frac{3w+1}{3 (w+1)}}\bigg].
\end{align}

\paragraph{Short-wavelength case ($k_{\star}\!<\!k\!<\!k_{\rm{e}}$):}
After inflation the modes with $k_\star\!<\!k\!<\!k_e$ also start their evolution in the {\it purple} region (\fig{fig:diagenergyscales} left-panel), but unlike the previous cases they do not evolve into {\it blue} region. 
Instead at $\tau=\tilde{\tau}_1$, they re-enter the horizon, i.e. the {\it red} area, in which $k$ is the dominant energy scale. 
In this case, it is convenient to consider those two regimes collectively, i.e. in {\it purple} and {\it red} regions, we neglect in $\omega_L^2$ terms proportional to $a^2m_X^2$ and arrive at the following equation of motion,
\begin{align}
    \xtl^{\prime \prime \rm{(VI)}} + \left(k^2 - \frac{1-3w}{2}a^2 H^2\right)\xtl^{\rm{(VI)}}=0,
\end{align}
with the solution
\begin{align}
    \xtl^{\rm{(VI)}}(a)&=\sqrt{\frac{2}{(1+3w)aH}} \bigg[C_1^{(\rm{VI})} J_{\nu}\Big(\!\tfrac{2k}{(1+3w)aH} \!\Big)+C_2^{(\rm{VI})} Y_{\nu}\Big(\!\tfrac{2k}{(1+3w)aH}\!\Big)\bigg], 	\label{ew:solvr}
\end{align}
where the order of Bessel functions is $\nu \!\equiv\! \frac{-3 (1-w)}{2 (1+3w)}$.
Merging the above solution \eqref{ew:solvr} with $\xtl^{(a)}(\tau_e)$ at the end of inflation we fix the value of $C_1^{(\rm{VI})}$ and $C_2^{(\rm{VI})}$:
\begin{align}
C_1^{\rm{(VI)}} &= \frac{i\pi \exp\!\Big(\frac{-2 i k}{(1+3 w)k_e}\Big)}{2\sqrt{2(1+3 w)k/k_e}}\bigg[\!Y_{\nu-1}\Big(\!\tfrac{2k}{(1+3 w)k_e} \!\Big)+i Y_{\nu}\Big(\!\tfrac{2k}{(1+3 w)k_e} \!\Big)\!\bigg], 		\\
C_2^{\rm{(VI)}} &= \frac{-i\pi \exp\!\Big(\!\frac{-2 i k}{(1+3 w)k_e}\!\Big)}{2\sqrt{2(1+3 w)k/k_e}}\bigg[\!J_{\nu-1}\Big(\!\tfrac{2k}{(1+3 w)k_e} \!\Big)+i J_{\nu}\Big(\!\tfrac{2k}{(1+3 w)k_e}\! \Big)\!\bigg],
\end{align}
where $k_e\!\equiv\!a_e H_{\rm I}$.
This analytic solution in the {\it purple} and {\it red} regions is shown in the right-panel of \fig{fig:afterInflation} as dashed line in their respective colors. 

Then at $\tau=\tilde{\tau}_3$ the modes enter the {\it green} region with the solution given by Eq.~\eqref{eq:solgreen}. Again after some straightforward but tedious calculation we find that both coefficients appearing in~\eqref{eq:solgreen} are non-zero and the solution is decaying oscillatory as shown in the right-panel of \fig{fig:afterInflation} (dotted green). 

In lower panels of \fig{fig:afterInflation} we present frequency~$\omega_L^2$ as a function of the scale factor~$a$ in order to verify the expected correlation between regions of strong increase of mode functions and negativity of $\omega_L^2$. The correlation is indeed confirmed in \fig{fig:afterInflation}.
Furthermore, we would like to emphasize that the solutions for mode functions and therefore for energy densities obtained here are continuous functions, that can be seen just from inspection of upper panels of Figs.~\ref{fig:chi_and_omega} and \ref{fig:afterInflation}. However a small discontinuity appears is plots of the frequency $\omega_L^2$ as a function of $a$, a careful reader may notice it comparing $\omega_L^2$ at $a_e\!\approx\!10^{-12}$ in lower panels of Figs.~\ref{fig:chi_and_omega} and \ref{fig:afterInflation}.

\subsubsection{Light vector dark matter: $m_X\!<\! H_{\rm rh}$}

The above approximate solutions are also valid for the lighter vector DM case. 
As shown in the right-panel of \fig{fig:diagenergyscales}, in this case, all modes are in the {\it purple} region at the beginning of reheating. 
Previously, we have demonstrated that in this region modes increase in amplitude according to the solution~\eqref{eq:solviolet}. This growth is eventually terminated at the point when modes go outside this region. 
For modes with $k \!>\!a_{\rm rh} m_X$ it happens during the RD epoch. In this case, one can simplify the above results using the fact that in this era $w=1/3$. 
However, for the modes with $k \!<\!a_{\rm rh} m_X$ the transition from the {\it purple} region to {\it blue} happens during the reheating phase and results are similar to the heavy vector DM case discussed above.
The modes $k_\star \!<\!k \!<\!k_{\rm rh}$, e.g. $k_2$ in the right-panel of \fig{fig:diagenergyscales}, cross the horizon (transition from the {\it purple} to {\it red} region) during the RD epoch. In this case the solution~\eqref{ew:solvr} is valid, however with the fixed value of equation of state~$w\!=\!1/3$. 
On the other hand, the evolution of the mode functions with $k_{\rm rh}\!<\!k \!<\!k_e$ during the reheating period is same as for the heavy DM regime discussed above.

\subsubsection{Energy density scaling}

Before closing this section, it is instructive to get the approximate analytic results for the energy density per $\ln k$ in each colored region of \fig{fig:diagenergyscales}. 
Employing the approximate analytic mode function solutions, we use Eq.~\eqref{eq:rhox_long} to get the energy density per log momentum after the end of inflation and its scaling with $a$ in different regions.
In the {\it red} region where $k$ is the dominant energy scale one can approximate the mode solution~\eqref{ew:solvr} as
\begin{align*}
    \xtl^{\rm{(VI)}}(k\!\gg\!aH) \!\approx \!\frac{3 i \Gamma(-\nu)(1-w) (1+3 w)^{\frac{1-3 w}{1+3 w}}}{4 \sqrt{\pi k}} \Big(\frac{k}{k_e}\Big)^{-\frac{3 (1+w)}{1+3 w}} \sin\!\Big(\frac{2 \pi }{1+3 w}\!\Big) \sin\! \left(\!\frac{2 k+\pi aH}{aH(1+3 w)}\!\right), 
\end{align*}
which implies that energy density redshifts as,
\beq
    \frac{d \langle \rho_L^{\rm (a)} \rangle}{d \ln{k}} 
    \approx \!\frac{9 \Gamma^2(-\nu)(1-w)^2 (1+3 w)^{\frac{2(1-3 w)}{1+3 w}}}{32 \pi^3} \frac{k^4}{ a^4}\Big(\frac{k}{k_e}\Big)^{-\frac{3 (1+w)}{1+3 w}}\propto k^{-\frac{2 (1-3w)}{1+3 w}}a^{-4}\,.
\eeq
In the regime with $am_X\!\ll\! k\!\ll\! aH$, i.e. {\it purple} region, we approximate solution to mode equation by~\eqref{eq:solviolet} and the energy density scales as
\beq
    \frac{d \langle \rho_L^{\rm (b)} \rangle}{ d \ln{k}} \approx  \frac{k^5}{4\pi^2 a^2} \lvert C_1^{(\rm{III})} \rvert^2 \propto k^{2}a^{-2} \,.
\eeq
Note that the same scaling would be obtained if we used~\eqref{ew:solvr} in the limit $k \!\ll\! aH$.
In the {\it blue} region $k\!\ll\! am_X\!\ll\!aH$, using Eq.~\eqref{eq:solblue} with $C_2^{\rm(I)} \!\approx\! 0$, we get, 
\beq
\frac{d \langle \rho_L^{\rm(c)} \rangle}{d \ln{k}} \approx \frac{k^3}{4 \pi^2 a^4}\bigg(\frac{k^4}{a^4 m_X^4} a^2 H^2 + k^2 + a^2m_X^2  \bigg)\lvert C_1^{\rm(I)}\rvert^2 \propto k^2 a^{-2}\,.
\eeq
This scaling of energy density of the longitudinal modes $\propto a^{-2}$ is a very unique feature of the vector DM, see also~\cite{Graham:2015rva}. For instance, the energy density of a massive scalar DM scales as constant in this {\it blue} region which leads to enhanced isocurvature perturbations at large scales, such large isocurvature perturbations are severely  constrained by the CMB data~\cite{Akrami:2018odb}. 
Hence, the damping of the energy density  $\propto a^{-2}$ observed here is a welcome property of the vector DM that implies suppression of unwanted isocurvature perturbations at the large scales. 
Finally, in the {\it green} region where $a m_X$ is the dominant energy scale we obtain the following scaling
\begin{align}
\frac{d \langle \rho_L^{\rm(d)} \rangle}{d \ln{k}} &\approx
\frac{am_X\,k^3}{4 \pi^2 a^4}\left( \lvert C_{1}^{(\rm{II})}  \rvert^2 + \lvert C_{2}^{(\rm{II})}  \rvert^2 \right)\propto k^2 a^{-3}\,,    \nonumber 
\end{align}
where we have neglected the oscillatory terms proportional to $\exp\big[{\pm i \int d \tau a(\tau) m_X}\big]$. Note that is this region the energy density behave as of the matter density.

\begin{figure}[t!]
\centering 
    \includegraphics[width=0.5\textwidth]{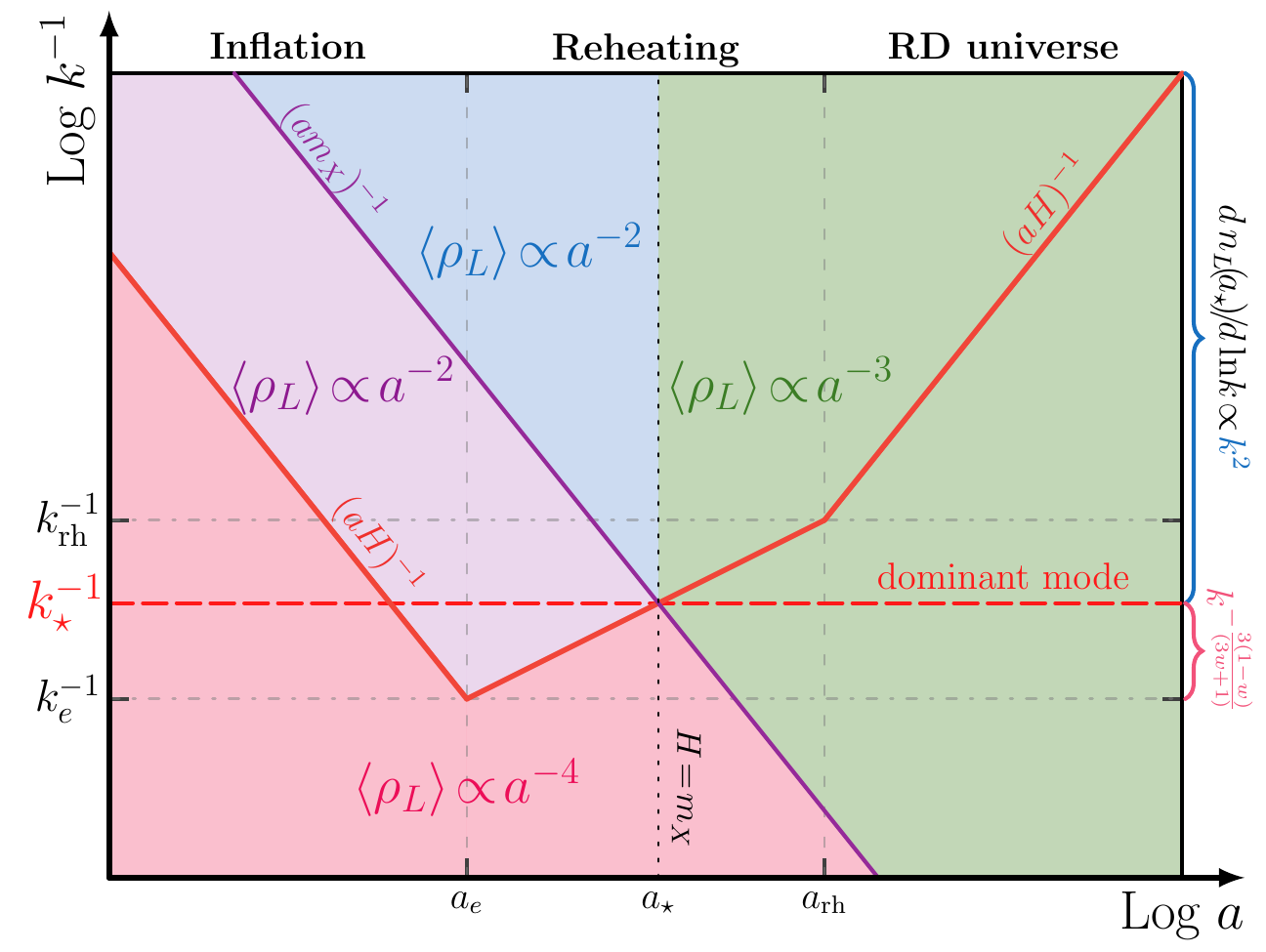}\!\!\!
    \includegraphics[width=0.5\textwidth]{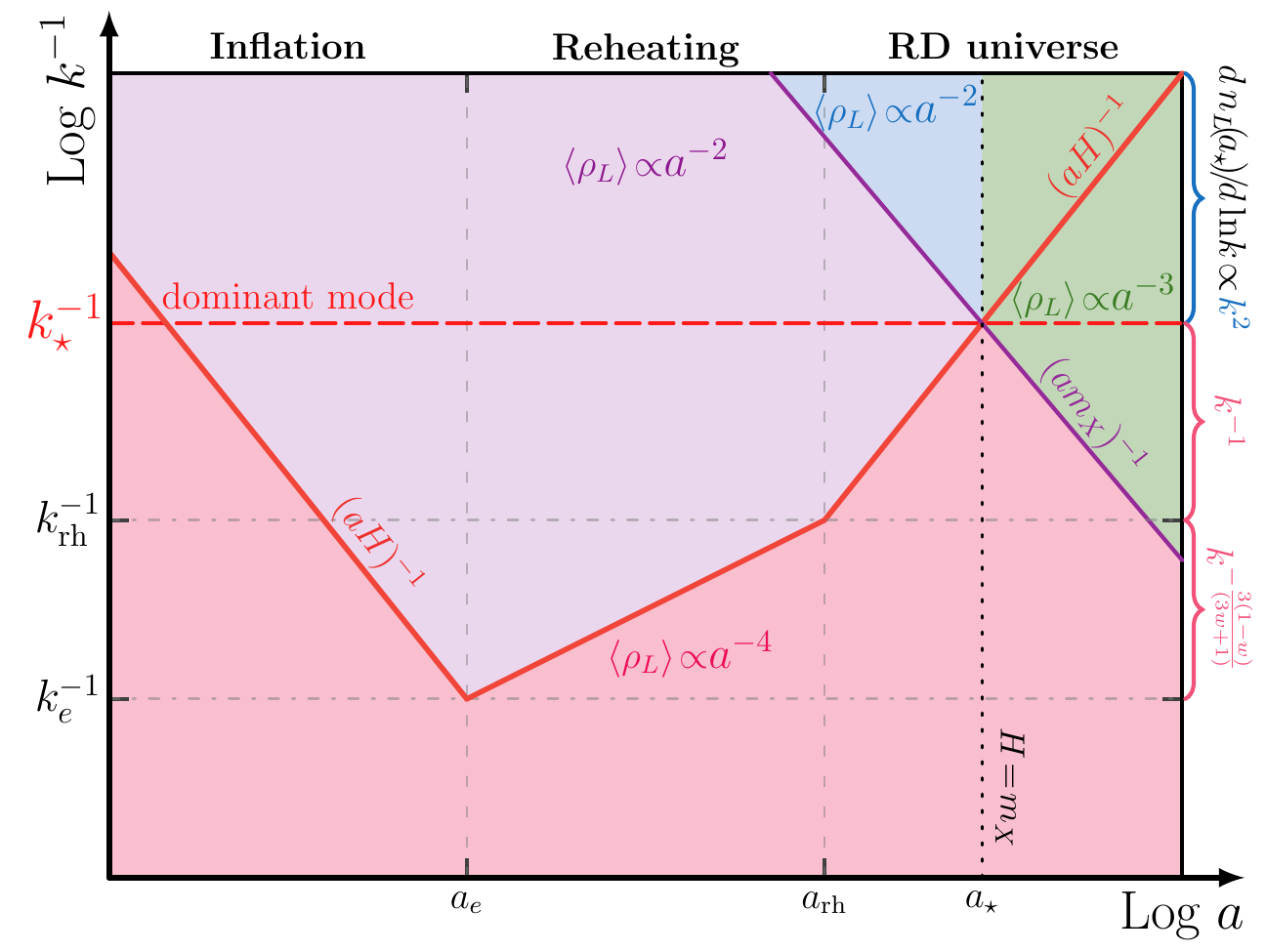}
    \caption{Scaling of the energy density per $\ln k$ as a function of the scale factor $a$ in various regions of the $(\log k^{-1},\log a)$ space  for heavy vector DM i.e $m_X\!>\! H_{\rm rh}$ (left diagram) and light vector DM $m_X\!<\!H_{\rm rh}$ (right diagram). The main contribution to the total energy density comes from the mode $k_{\star}\!\equiv\! a_{\star}m_X$. Here $k_e \equiv a_e H_e$ and $k_{\rm rh} \equiv a_{\rm rh} H_{\rm rh}$. Along the right sides of the panels we show scaling of number density per $\ln k$ at $a=a_\star$ w.r.t. mode momentum $k$.} \label{fig:drhodlnkA}
\end{figure}

The above red-shifting of the energy density and its scaling w.r.t. mode momentum $k$ can be summarized as,
\begin{empheq}[left={\displaystyle \frac{d \langle \rho_L \rangle}{d \ln{k}}\!\propto\!\empheqlbrace}]{align}
&k^{-\frac{2 (1-3w)}{1+3 w}} a^{-4}, & a(\tau)m_X, a(\tau) H(\tau) \ll k&~ ({\rm \textcolor{Maroon!70}{red}}),     \notag\\
&k^2 a^{-2}, & a(\tau) m_X, k \ll a(\tau) H(\tau)& ~({\rm \textcolor{purple!57}{purple},\textcolor{blue!70}{blue}}), \label{eq:drhodlnkA} \\
&k^2 a^{-3}, &  a(\tau)H(\tau), k \ll a(\tau) m_X&~ ({\rm \textcolor{OliveGreen!70}{green}})\,, \notag
\end{empheq}
where the colors represent different regions as in \fig{fig:drhodlnkA}. Note that the dependance of energy density per $\ln k$ on momentum $k$ is $w$-dependent for the modes with $k\!\gg\!a(\tau)m_X, a(\tau) H(\tau)$, i.e. the {\it red} region. 
However, if these modes are during the inflationary (de~Sitter) period then $w\!=\!-1$ and if they are in the RD epoch then $w\!=\!1/3$. Whereas, during the reheating phase the equation of state parameter $w$ takes the value in the range $w\!\in\!(\!-1/3,1)$.

\begin{figure}[t!]
    \centering
    \includegraphics[width=0.33\textwidth]{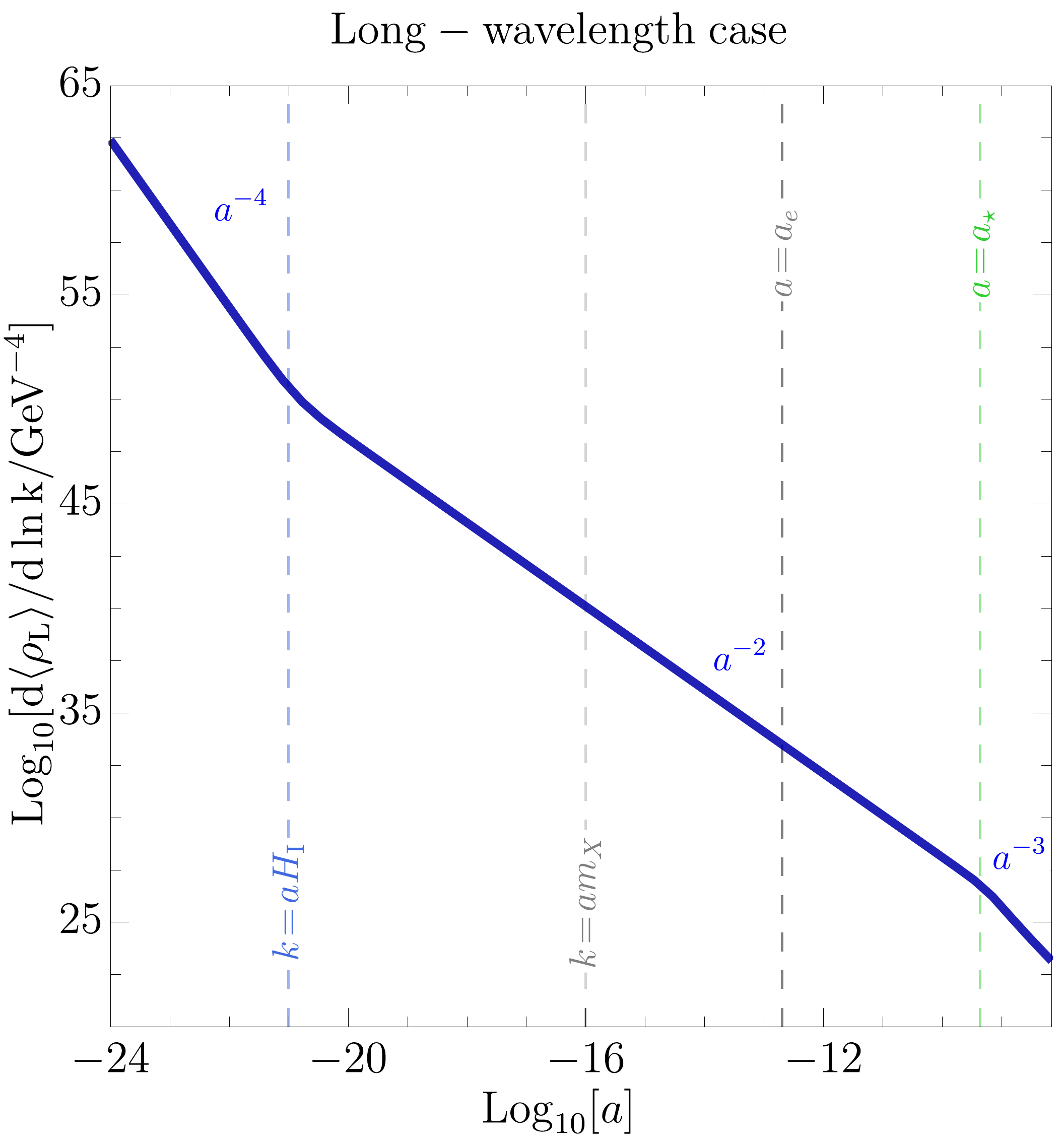}\!\!\!
    \includegraphics[width=0.33\textwidth]{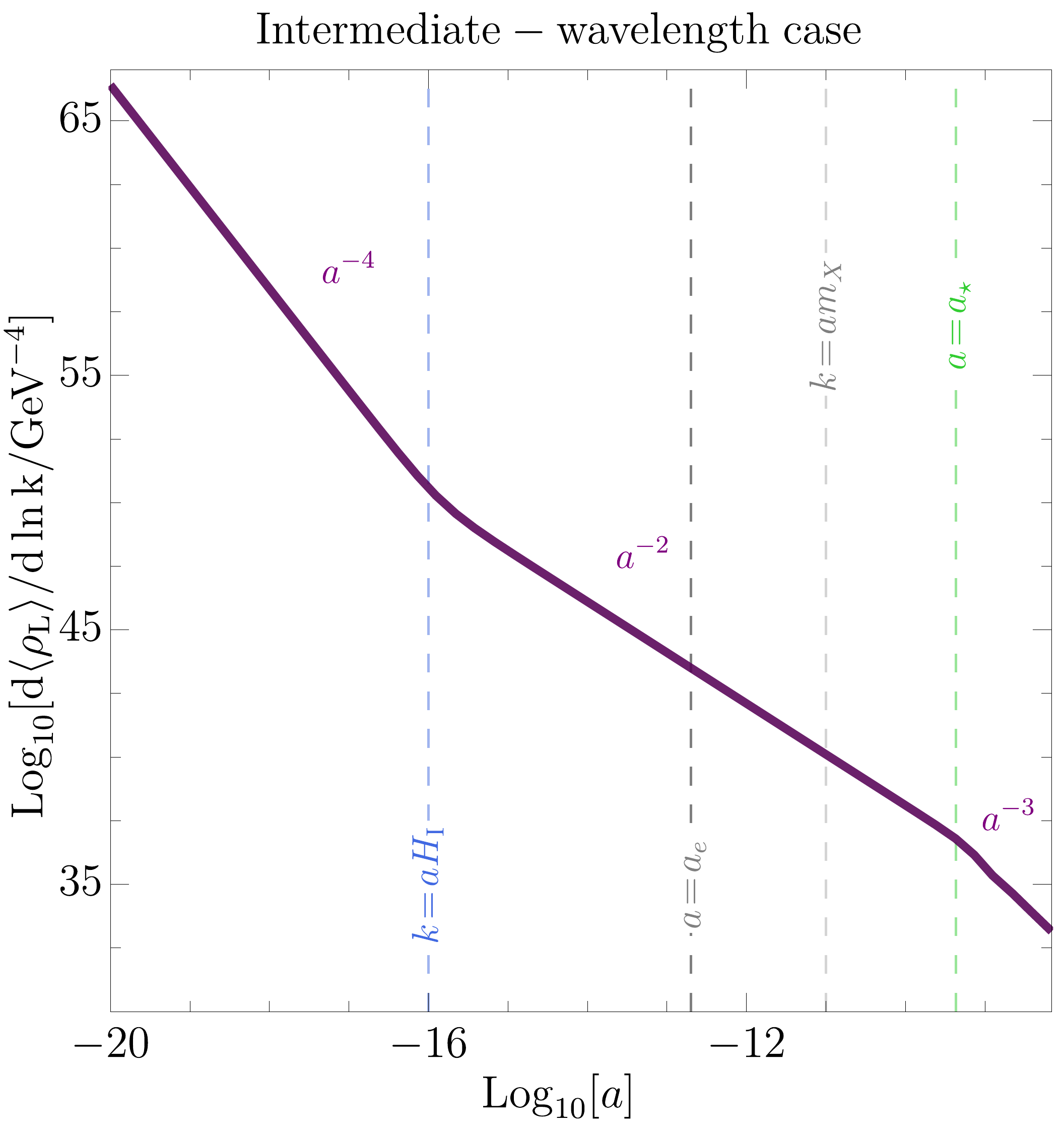}\!\!\!
    \includegraphics[width=0.33\textwidth]{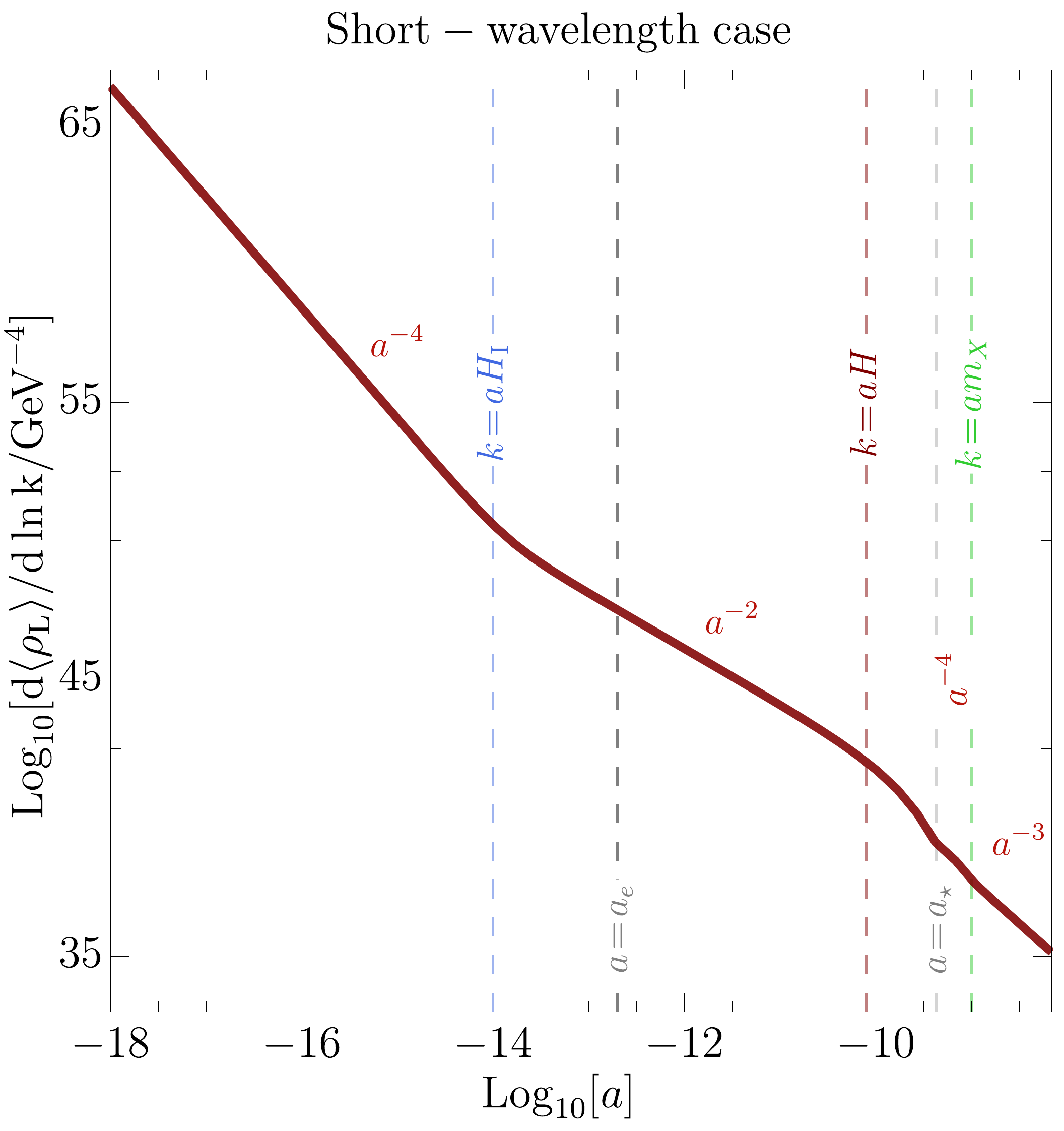}
    \caption{Evolution of the energy density per log momentum $d \langle \rho_L \rangle / (d \ln{k})$ with the scale factor $a$ for different choice of wavevector $k$: long-wavelength $k\!<\!k_m $ (left-panel), intermediate-wavelength $k_m\!<\!k \!<\!k_{\star}$ (middle-panel), and short-wavelength $k_{\star}\!<\!k\!<\!k_e$ (right-panel). The numerical analysis presented here agrees with the analytical predictions. Here we choose the parameter $m_X=10^8 \; \rm{GeV}$, $\hi=10^{13} \; \rm{GeV}$, $w=0$ and $k=10^{-8} \; \rm{GeV},10^{-3} \; \rm{GeV} $, and $4 \; \rm{GeV} $ for left, middle, and right plots, respectively.}
    \label{fig:drhodlnkN}
\end{figure}

To summarize the scaling of energy density w.r.t. the scale factor~$a$, we present the exact numerical results for different wavelength modes discussed above in \fig{fig:drhodlnkN}. We note that as the universe expands the redshift of the energy density varies for different modes and the approximate analytic scaling of the energy density~\eqref{eq:drhodlnkA} matches well with those of the exact numerical results. 
We see that at the end of inflation ($a\!=\!a_e$) the main contribution to the total energy density comes from modes with the shortest wavelength ($k\sim k_e$). However, after the end of inflation, those modes receive the strongest suppression proportional to $a^{-4}$. On the other hand, modes with longer wavelength initially have a smaller contribution to the total energy density, but their energy is also redshifted to a lesser extent. The intermediate-wavelength modes contribute the largest to the energy density as noted above.

\section{Relic abundance of vector dark matter}
\label{sec:relicDM}
The present energy/number density of DM particles could be expressed in terms of energy/number density at $H(a_\star)\!=\!m_X$. 
The number density of longitudinal modes, $\langle n_L \rangle$, is related to their energy density $\langle \rho_L \rangle$ as
\begin{align}
    d\,\langle n_L \rangle = \frac{d\,\langle \rho_L \rangle}{E_L},	\label{eq:nL_rhoL}
\end{align}
where $E_L$ denotes single-particle energy.
Consequently at $H(a_\star)\!=\!m_X$ the number density per $\ln k$ is computed as, 
\begin{align}
    \frac{d \langle n_L (a_\star) \rangle}{d \ln{k}} =\frac{1}{\sqrt{m_X^2 + k^2/a^2(\tau_\star)}} \frac{d \langle \rho_L(a_\star) \rangle }{d \ln{k}}.		\label{eq:dnL_drhoL_dlnk}
\end{align}
In the following, we evaluate the DM number density in two mass regimes, the heavy vector DM $H_{\rm rh}\!\leq\!m_X\!<\! H_{\rm I}$ and the light vector DM $m_X\!<\! H_{\rm rh}$.
Moreover, it should be stressed that we consider the evolution of number density for super-horizon modes with $k\!<\!k_e$, i.e. the modes which were outside of the horizon at the end of inflation.

Before presenting detailed numerical analysis it is instructive to estimate the number density per $\ln k$~\eqref{eq:dnL_drhoL_dlnk} of the longitudinal modes at $H(a_\star)\!=\!m_X$, from the energy density scaling observed in~\eq{eq:drhodlnkA}.
In \fig{fig:drhodlnkA}, we show the $d \langle n_L (a_\star) \rangle/ d \ln k$ scaling with $k$ for the longitudinal modes at $a_\star\!=\!a(\tau_\star)$. 
Note that the dominant number density is generated by the modes with comoving momentum $k_\star$ represented as dashed (red) line. 
After exiting the horizon, these dominant modes  pass through the {\it purple} region only until they reach $H(a_\star)=m_X$.
Focusing first on the heavy vector DM case (\fig{fig:drhodlnkA} left-panel), the modes with wavelengths longer that $k_\star^{-1}$, i.e. $k\!\lesssim\!k_\star$, would have to pass through the {\it blue} region where the energy density of these modes redshifts as $a^{-2}$. Therefore in this region the energy density associated with these longer wavelength modes $k\!\lesssim\!k_\star$ scales proportional to $k^2$. 
Hence the long wavelength longitudinal vector modes $k\!<\!k_\star$ lose their power proportional to $k^2$ and therefore are safe from generating dangerous isocurvature perturbations at the CMB scales. 
This result would remain same for the number density of modes in the {\it blue} region, i.e. $d \langle n_L (a_\star) \rangle/ d \ln k\propto k^2$. On the other hand, the shorter wavelength modes, $k\!>\!k_\star$, re-enter the horizon at $a\!<\!a_\star$ and follow the evolution in the {\it red} region (\fig{fig:drhodlnkA} left-panel) where the energy density of these modes redshifts as radiation i.e. $a^{-4}$. Since these shorter wavelength modes $k\!>\!k_\star$ re-enter the horizon during the reheating phase when the Hubble rate scales are $a^{-3(1+w)/2}$, therefore the number density of the modes scale as $k^{-\frac{3(1-w)}{(3w+1)}}$. 
Note that for $w\!<\!1$, the number density per $\ln k$ at $a_\star$ has a peak structure and the dominant modes are $k_\star$ where most of the number density is contained. This peak structure was first noted in~\cite{Graham:2015rva} for instantaneous reheating and RD universe after inflation. Here, we extended this observation for non-standard cosmology with equation of state $w\!<\!1$.

Similarly for the light DM case, $m_X\!<\! H_{\rm rh}$, we observe that the dominant modes (shown as dashed red line in the right-panel of \fig{fig:drhodlnkA}) are those with $k=k_\star$. 
The number density for longer wavelength modes ($k\!<\!k_\star$) scales as $k^2$ as they pass through the {\it blue} region to reach $a_\star=a(\tau_\star)$. 
However, in the light DM case, the modes with wavenumber $k_{\rm rh}\!<\!k\!<\!k_\star$ re-enter the horizon during the RD universe ($H(a)\!\propto\!a^{-2}$) and their energy density also redshifts like the radiation~$a^{-4}$. 
Therefore, the number density of the modes with $k_{\rm rh}\!<\!k\!<\!k_\star$ scales as $k^{-1}$. 
The modes with comoving momentum $k_{e}\!<\!k\!<\!k_{\rm rh}$ re-enter the horizon during the period of reheating, i.e. $H(a)\!\propto\!a^{-3(1+w)/2}$ and their energy density (in the {\it red} region) scales as $a^{-4}$, and the number density $d \langle n_L (a_\star) \rangle/ d \ln k$ of these modes scales as $k^{-\frac{3(1-w)}{(3w+1)}}$. 
Hence, for $w<1$, the number density per log momentum has a peak structure with maximum values around $k_\star$ modes. In the following we present detailed numerical analysis of this qualitative discussion by calculating the number density of the vector DM.

\subsubsection*{Number density of the heavy vector DM: \boldmath{$H_{\rm rh}\!\leq\!m_X \!<\! H_{I}$} }
\label{sec:density_mx_heavy}
For vector DM in the mass range $H_{\rm rh}\!\leq\!m_X \!<\! H_{I}$, we consider two cases for the momentum vector~$k$: 
\begin{enumerate}[label=(\alph*)]
\item $k_{\star}\!<\!k \!<\! k_e$,
\item  $k\!<\! k_{\star} $.
\end{enumerate}

The number density for modes with $k_{\star}\!<\!k\!<\! k_e$ can be written as
\beq
    \left.  \frac{d \langle n_L (a_{\star}) \rangle}{d \ln{k}} \right\vert_{k_{\star}<k<k_e} \approx \frac{a_{\star}}{k} \frac{d \langle \rho_L (a_{\star})}{d \ln{k}}=\frac{a_{\star}}{k} \frac{d \langle \rho_L (a_e) \rangle}{d \ln{k}}\left( \frac{a_e}{a_c}\right)^2 \left(\frac{a_c}{a_{\star}}
    \right)^4,	\label{eq:numden_m1a}
\eeq
where $a_c$ corresponds to the horizon crossing and it is defined as
\beq
 a_c = \frac{k}{H(a_c)}\,.
\eeq
Note that in this case the horizon crossing occurs during reheating, therefore, one can express $a_c$ as
\begin{align*}
   \left. a_c \right\vert_{k_{\star}<k<k_e} = a_e \left(\frac{H_
I a_c}{k} \right)^{\frac{2}{3(1+w)}},
\end{align*}
which implies
\begin{align}
\left. a_c \right\vert_{k_{\star}<k<k_e} = a_e^{\frac{3(1+w)}{1+3w}} \left(\frac{H_{\rm I}}{k} \right)^{\frac{2}{1+ 3w}}.
\end{align}
Consequently, in this limit we can rewrite Eq.~\eqref{eq:numden_m1a} as
\begin{align}
    \left. \frac{d \langle n_L (a_{\star}) \rangle}{d \ln{k}}\right\vert_{k_{\star}<k<k_e} \approx \frac{1}{8\pi^2}H_{\rm I}^{\frac{2\left(3 w^2+3 w+2\right)}{(1+w) (1+3w)}}m_X^{\frac{2}{1+w}}\left( \frac{a_e}{k}\right)^{\frac{3(1-w)}{(1+3w)}}.
\end{align}

In the other case, $k\!<\!k_{\star}$, the number density at $a=a_{\star}$ is given by
\beq
\left. \frac{d \langle n_L (a_{\star}) \rangle}{d \ln{k}} \right\vert_{k<k_{\star}} \approx \frac{1}{m_X} \frac{d \langle \rho_L (a_{\star})}{d \ln{k}} = \frac{1}{m_X} \frac{d \langle \rho_L (a_e)\rangle }{d \ln{k}} \left(\frac{a_e}{a_{\star}} \right)^2
= \frac{1}{8 \pi^2} H_{\rm I}^{\frac{2(1+3w)}{3(1+w)}}m_X^{\frac{1-3w}{3(1+w)}}\left(\frac{k}{a_e} \right)^2,
\eeq
where we have used the fact that in this limit $H(a)\!=\!m_X$ occurs during the reheating phase, i.e.
\begin{align}
    a_{\star} = a_e \left(\frac{H_{\rm I}}{m_X} \right)^{\frac{2}{3(1+w)}}.
\end{align}

\subsubsection*{Number density of the light vector DM: \boldmath{$m_X \!<\! H_{\rm rh}$}}
\label{sec:density_mx_light}
Now we focus on the case when the vector DM is lighter than the Hubble scale at the end of reheating, i.e. $m_X \!<\! H_{\rm rh}$. In this scenario, we calculate the DM number density for the following three cases:
\begin{enumerate}[label=(\alph*)]
    \item $ k_{\rm rh} < k< k_e$,
    \item $k_{\star} < k < k_{\rm rh}$,
    \item $k< k_{\star} $,
\end{enumerate}
where $k_e\equiv a_e H_{\rm I} $, $k_{\rm rh}\!\equiv\!a_{\rm rh} H_{\rm rh}$. The scale factor at the end of reheating $a_{\rm{rh}} \equiv a(\tau_{\rm{rh}})$ is given in terms of $a_e$ and $\gamma$ in \eq{eq:a_rh}. 

In the case (a), for large $k$ values, i.e. $k_{\rm rh}\!<\! k\!<\!k_e$, we get the number density as
\beq
    \frac{d \langle n_L (a_{\star}) \rangle}{d \ln{k}}\!\bigg\vert_{k_{\rm rh}< k<k_e}\!\! \approx \! \frac{a_{\star}}{k} \frac{\langle \rho_L(a_{\star}) \rangle}{d \ln{k}} \!=\! \frac{a_{\star}}{k}\frac{\langle \rho_L(a_e) \rangle}{d \ln{k}} \left(\!\frac{a_e}{a_c} \!\right)^2 \!\! \left(\!\frac{a_c}{a_{\rm rh}}\! \right)^4\!\! \left(\!\frac{a_{\rm rh}}{a_{\star}}\!\right)^4\!,
\eeq
where in this case the second horizon crossing occurs during the reheating phase at
\beq
a_c \Big\vert_{k_{\rm rh}< k<k_e}= a_e^{\frac{3(1+w)}{1+3w}} \left( \frac{H_{\rm I}}{k}\right)^{\frac{2}{1+3w}},
\label{eq:ac_sec}
\eeq
Hence the number density for the longitudinal modes for the case (a) can be rewritten as,
\begin{align}
   \left. \frac{d \langle n_L (a_{\star}) \rangle}{d \ln{k}}\right\vert_{k_{\rm rh}< k<k_e} & \approx  \frac{1}{8 \pi^2} m_X^{3/2} H_{\rm rh}^{\frac{1-3w}{2(1+w)}}H_{\rm I}^{\frac{2 \left(3 w^2+3 w+2\right)}{(1+w) (1+3w)}}\left( \frac{a_e}{k}\right)^{\frac{3(1-w)}{1+3w}}, \notag\\
   &= \frac{1}{8 \pi^2} m_X^{3/2}\gamma^{\frac{1-3w}{1+w}}H_{\rm I}^{\frac{3 (w+3)}{2 (1+3w)}}\left( \frac{a_e}{k}\right)^{\frac{3(1-w)}{1+3w}}.
\end{align}

In the case (b), i.e. $k_{\star} \!<\! k\!<\! k_{\rm rh}$, we get the number density for longitudinal modes as
\beq
   \frac{d \langle n_L (a_{\star}) \rangle}{d \ln{k}}\bigg\vert_{k_{\star}<k<k_{\rm rh}} \approx \frac{a_{\star}}{k}\frac{d \langle \rho_L (a_{\star})\rangle}{d \ln{k}} = \frac{a_{\star}}{k}\frac{d \langle \rho_L (\tau_e)\rangle}{d \ln{k}}\left(\frac{a_e}{a_{\rm rh}} \right)^2 \!\! \left(\!\frac{a_{\rm rh}}{a_c} \!\right)^2 \!\!\left(\!\frac{a_{c}}{a_{\star}} \!\right)^4\!,
\eeq
where again $a_c$ corresponds to value of the scale factor at the second horizon crossing and for light vector DM this happened during the RD epoch. Therefore,
\begin{align}
    \left. a_c \right\vert_{k_{\star}<k<k_{\rm rh}}&= a_e \left(\frac{H_{\rm I}}{H_{\rm rh}} \right)^{\frac{2}{3(1+w)}} \left( \frac{H_{\rm rh} a_c}{k}\right)^{1/2},
\end{align}
which implies
\begin{align}
    \left.a_c \right\vert_{k_{\star}<k<k_{\rm rh}} &= a_e^2 \left(\frac{H_{\rm I}}{H_{\rm rh}} \right)^{\frac{4}{3(1+w)}}\frac{H_{\rm rh}}{k}= a_e^2 \gamma^{-\frac{2(1-3w)}{3(1+w)}} \frac{H_{\rm I}}{k}.
\end{align}
Hence the number density in the case (b) is 
\beq
   \left. \frac{d \langle n_L (a_{\star}) \rangle}{d \ln{k}}\right\vert_{k_{\star}<k<k_{\rm rh}} \approx \frac{1}{8 \pi^2} m_X^{3/2} H_{\rm I}^{\frac{2(4+3w)}{3(1+w)}} H_{\rm rh}^{\frac{-1 +3w}{6(1+w)}}\left(\frac{a_e}{k}\right)
= \frac{1}{8 \pi^2}m_X^{3/2} H_{\rm I}^{5/2}\gamma^{\frac{-1+3w}{3(1+w)}} \left(\frac{a_e}{k} \right).
\eeq

Finally, for the longitudinal modes with $k< k_{\star}$ (c), we obtain the number density as
\begin{align}
    \left. \frac{d \langle n_L (a_{\star}) \rangle}{d \ln{k}}\right\vert_{k<k_{\star}} &\approx \frac{1}{m_X} \frac{\langle \rho_L(a_{\star})\rangle}{d \ln{k}} = \frac{1}{m_X} \frac{\langle \rho_L (a_e) \rangle}{d \ln{k}} \left(\frac{a_e}{a_{\rm rh}} \right)^2 \left(\frac{a_{\rm rh}}{a_{\star}} \right)^2,\\
    &= \frac{1}{8 \pi^2} H_{\rm I}^{\frac{2(1+3w)}{3(1+w)}}H_{\rm rh}^{\frac{1-3w}{3(1+w)}} \left(\frac{k}{a_e} \right)^2 = \frac{1}{8 \pi^2}H_{\rm I} \gamma^{\frac{2(1-3w)}{3(1+w)}} \left(\frac{k}{a_e} \right)^2\,, \non
\end{align}
where we have used the fact that in this case modes re-enter the horizon during the $\rm{RD}$ epoch, which implies
\beq
a_{\star}= a_e \left( \frac{H_{\rm I}}{H_{\rm rh}}\right)^{\frac{2}{3(1+w)}} \left(\frac{H_{\rm rh}}{m_X} \right)^{1/2} = a_e \gamma^{\frac{3 w-1}{3 (1+w)}} \left( \frac{ H_{\rm I}}{m_X}\right)^{1/2}.
\eeq

We can summarize our results for number density for the longitudinal modes at $a=a_\star$ in both mass regimes as follows:
\begin{itemize}
\item For heavy vector DM mass $H_{\rm rh}\!\leq\!m_X\!<\!H_{\rm I}$,
\begin{subequations}
\begin{empheq}[left={\displaystyle \hspace{-1cm}\frac{ d\langle n_L(a_{\star}) \rangle}{d\ln{k}}\!=\!\frac{1}{8\pi^2}\!\empheqlbrace}]{align}
& H_{\rm I}^{\frac{2\left(3 w^2+3 w+2\right)}{(1+w) (1+3w)}}m_X^{\frac{2}{1+w}}\left( \frac{a_e}{k}\right)^{\frac{3(1-w)}{(1+3w)}}, &&  k_{\star}\!<\!k\!<\!k_e,	\label{eq:nd_heavy_a}\\
& H_{\rm I}^{\frac{2(1+3w)}{3(1+w)}}m_X^{\frac{1-3w}{3(1+w)}}\left(\frac{k}{a_e} \right)^2,  &&  k<k_{\star} .	\label{eq:nd_heavy_b}
\end{empheq}
\end{subequations}    
\item  For light vector DM mass $m_X\!<\!H_{\rm rh}$,
\begin{subequations}
\begin{empheq}[left={\displaystyle \hspace{-1cm}\frac{ d\langle n_L(a_{\star}) \rangle}{d\ln{k}}\!=\!\frac{1}{8\pi^2}\!\empheqlbrace}]{align}
& m_X^{3/2}H_{\rm I}^{\frac{3 (3+w)}{2 (1+3w)}}\gamma^{\frac{1-3w}{1+w}}\left( \frac{a_e}{k}\right)^{\frac{3(1-w)}{1+3w}},
    &&  k_{\rm rh} \!<\!k\!<\!k_e,	\label{eq:nd_light_a}\\
&m_X^{3/2} H_{\rm I}^{5/2}\gamma^{\frac{-1+3w}{3(1+w)}} \left(\frac{a_e}{k} \right), && k_{\star}<k<k_{\rm rh},	\label{eq:nd_light_b}\\
& H_{\rm I} \gamma^{\frac{2(1-3w)}{3(1+w)}} \left(\frac{k}{a_e} \right)^2,  &&  k<k_{\star} .	\label{eq:nd_light_c}
\end{empheq}
\end{subequations} 
\end{itemize}

\begin{figure}[t!]
\centering 
    \includegraphics[width=0.5\textwidth]{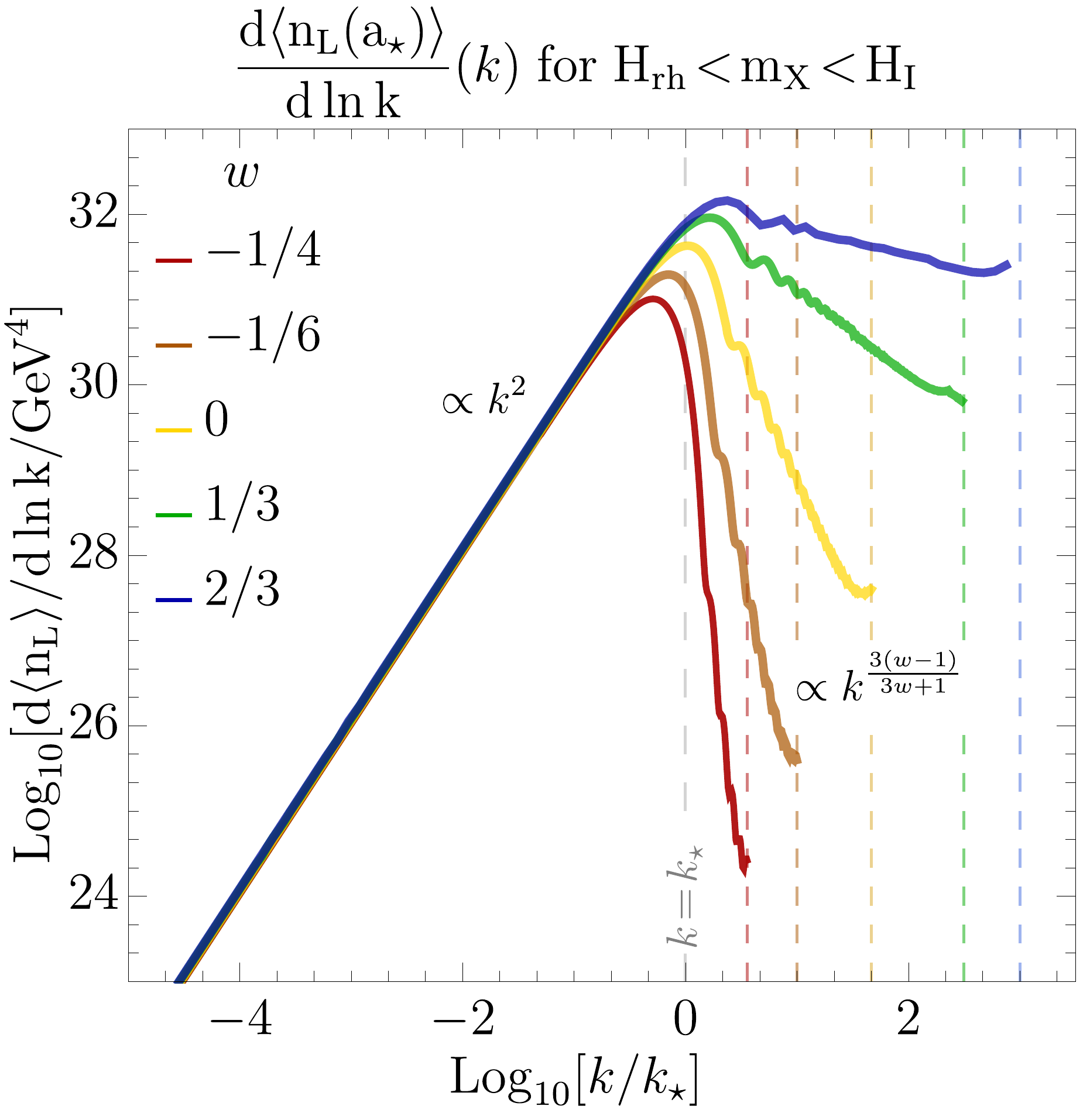}\!\!\!
    \includegraphics[width=0.5\textwidth]{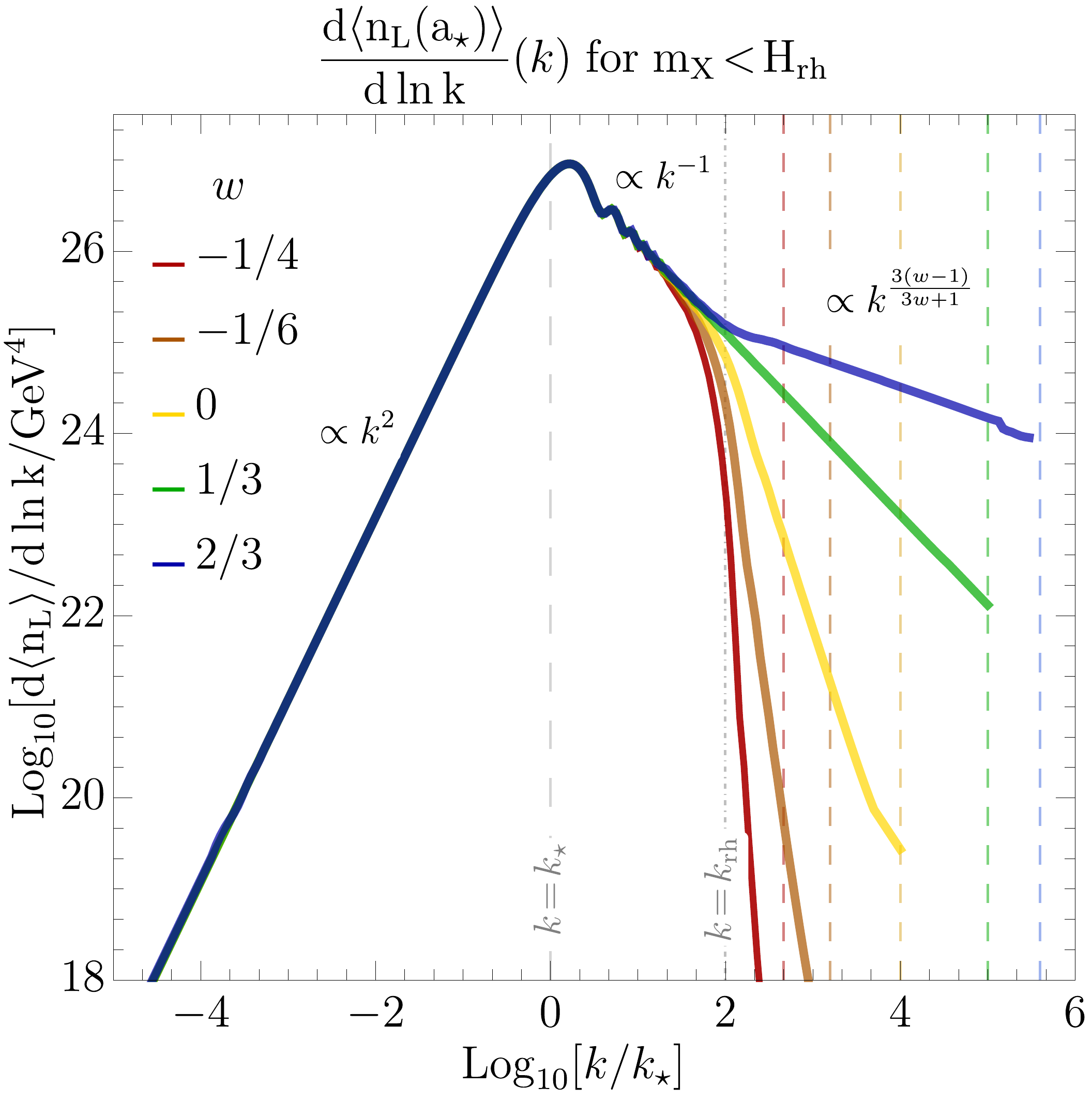}
    \caption{Expectation value of the number density per $\ln k$ as a function of wavevector $k$ at $a=a_{\star}$ for DM vector boson with mass in the range $H_{\rm rh}\!\leq\!m_X\!<\! H_{\rm I}$ (left-panel) and $m_X\!<\! H_{\rm rh}$ (right-panel). Different colors correspond to different values of equation of state parameter $w$. Grey dashed (dotted) lines indicate $k=k_{\star}(k_{\rm rh})$ while colored dashed lines refer to $k=k_e$ for corresponding value of $w$. For $k\!<\!k_{\star}$ the number density increases as $k^2$, while for $k_{\star}\!<\!k\!<\!k_e$ it is proportional to $k^{\frac{3(w-1)}{1+3w}}$. Here we take  $H_{\rm I}= 10^{13}  \gev$, $\gamma = 10^{-3}$,  $m_X=10^{8} \gev$ (left-panel), $m_X=10^{3} \gev$ (right-panel).}
    \label{fig:numberdensity}
\end{figure}

\begin{figure}[t!]
\centering 
    \includegraphics[width=0.5\textwidth]{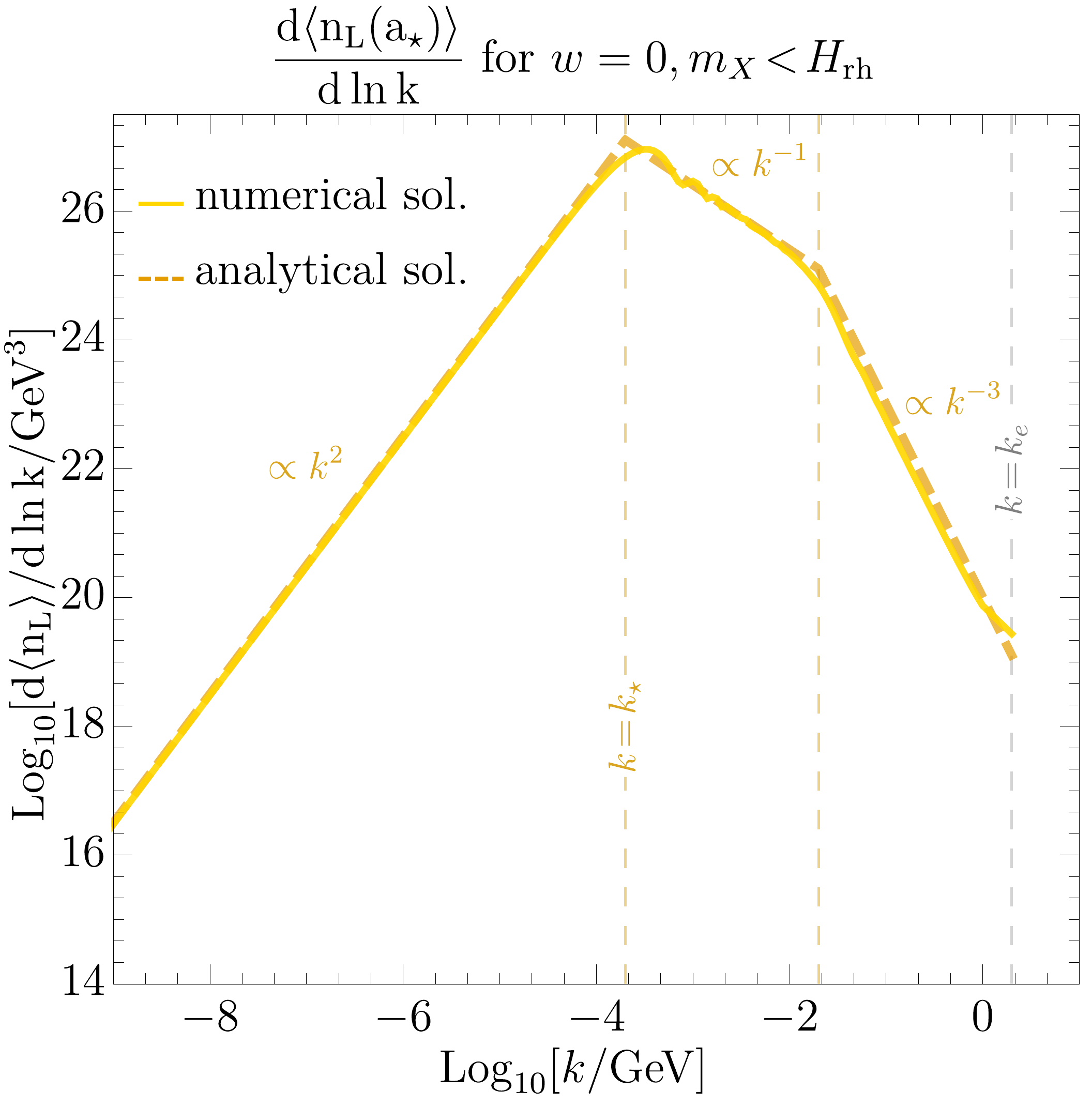}\!\!\!
    \includegraphics[width=0.5\textwidth]{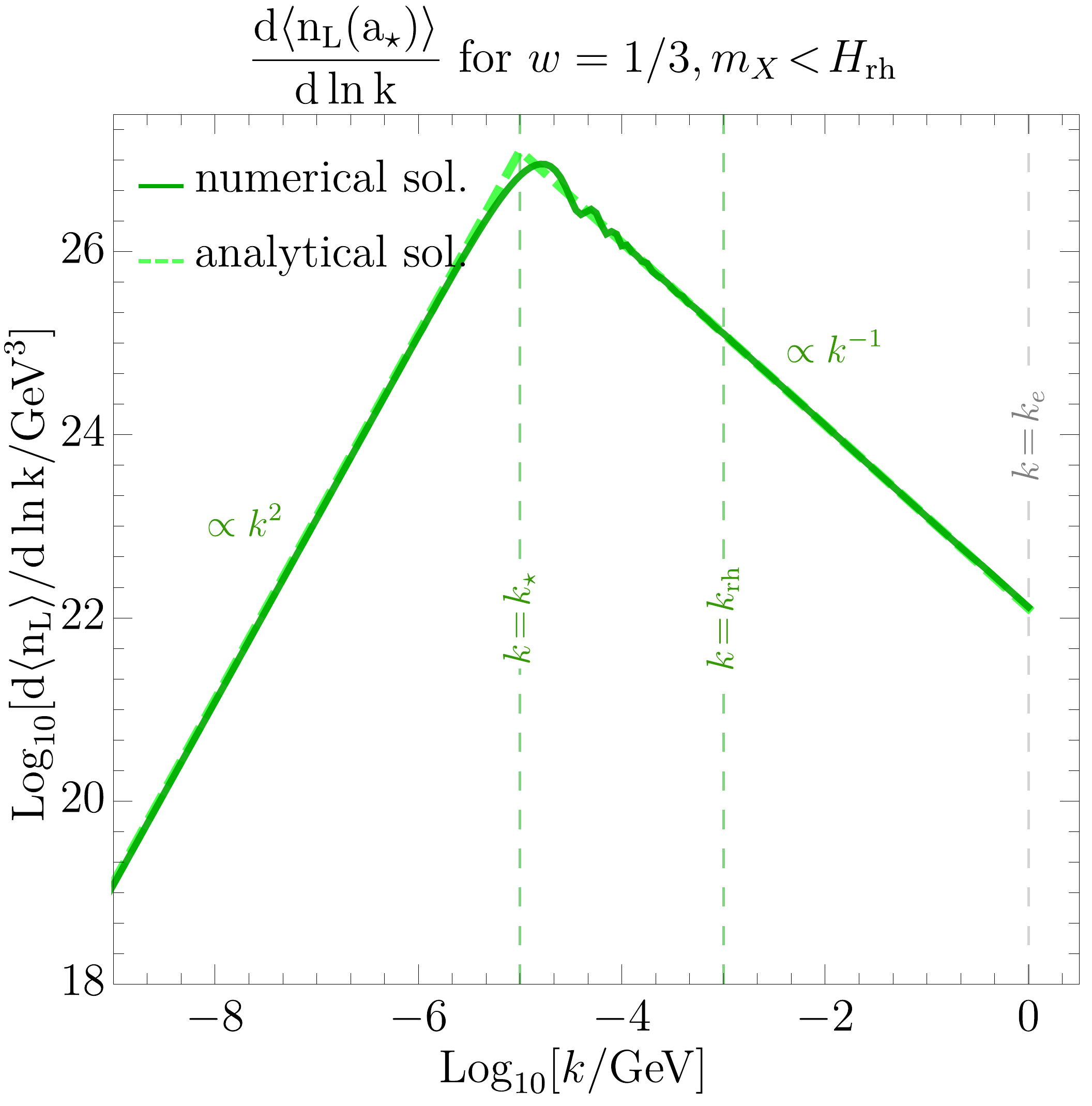}
    \caption{Comparison between numerical results (solid lines) and analytical predictions for the expectation value of number density per log momentum for $w=0$ (left-panel) and $w=1/3$ at $a=a_{\star}$. The same parameters as in Fig.~\ref{fig:numberdensity} have been adopted.}
\label{fig:numberdensity_num_an}
\end{figure}

In \fig{fig:numberdensity} we show results of exact numerical computation of the number density per $\ln k$ as a function of $k/k_\star$ for different values of $w$ at $a\!=\!a_\star$.
Note that in the above two vector DM mass regimes, the number density per log momentum has a peak structure if and only if $w \!\in\! (\!-\!1/3, 1)$. A similar peak structure was also observed in~\cite{Graham:2015rva} with standard cosmological history assuming instantaneous reheating and radiation dominated universe after the end of inflation. 
In this case, $d \langle n_L(a_\star) \rangle/d \ln{k}$ is dominated by modes with $k\! \sim\! k_{\star}$. 
In \fig{fig:numberdensity_num_an} we compare the exact numerical results (solid curves) with the approximate analytic form [\eqref{eq:nd_heavy_a}-\eqref{eq:nd_heavy_b}] (dashed curves) of the number density per $\ln k$ as a function of $k$ for $w=0,1/3$ at $a\!=\!a_\star$. As seen from the plots the exact numerical results agree very well with the approximate analytic solutions.

The total number density $\langle n_L (a_{\star}) \rangle$ for the case of heavy vector DM, $H_{\rm rh}\!\leq\!m_X\!<\!H_{\rm I}$, at the moment when $H(a_\star)\!=\!m_X$ is given by,
\begin{align}
\langle n_L^{\rm{hDM}} (a_{\star}) \rangle&= \frac{1}{8 \pi^2} \bigg[H_{\rm I}^{\frac{2(1+3w)}{3(1+w)}}m_X^{\frac{1-3w}{3(1+w)}} a_e^{-2} \int_0^{k_{\star}} k dk 	+  H_{\rm I}^{\frac{2\left(3 w^2+3 w+2\right)}{(1+w) (1+3w)}}m_X^{\frac{2}{1+w}}
    a_e^{\frac{3(1-w)}{(1+3w)}}\int_{k_{\star}}^{k_e} k^{\frac{-4}{1+3w}} dk\bigg],\nonumber \\
    &= \frac{1}{8 \pi^2} \bigg[\frac{1}{2}m_X H_{\rm I}^{2}+ \frac{1+3w}{3(w-1)}H_{\rm I}^{\frac{1+3w}{1+w}}m_X^{\frac{2}{1+w}}\bigg(1- \Big( \frac{H_{\rm I}}{m_X}\Big)^{\frac{1-w}{1+w}} \bigg)\bigg] , \nonumber \\
    &\approx \frac{1}{8 \pi^2}\left[\frac{1}{2}+\frac{1+3w}{3(1-w)}  \right]m_X H_{\rm I}^2, 	\label{eq:nhdm}
\end{align}
where in the last approximation we have assumed that $w\! \in\! (\!-1/3, 1)$ such that $\left(\! \frac{H_{\rm I}}{m_X}\!\right)^{\frac{1-w}{1+w}}\!\gg\!1$. 
Note that we have introduced a cut-off at $\Lambda = k_e$, so only modes that exit the horizon during inflation and re-enter during reheating contribute to the total number/energy density. Moreover, these sub-horizon modes, i.e. $\!k>k_e\!$, do not receive any tachyonic enhancement and hence their contribution would have been suppressed.
For the case of light vector DM, $m_X\!<\!H_{\rm rh}$, the number density is given by
\begin{align}
    \langle n_L^{\rm{lDM}} (a_{\star}) \rangle&= \frac{1}{8 \pi^2}\bigg[H_{\rm I} 
    \gamma^{\frac{2(1-3w)}{3(1+w)}}a_e^{-2}\int_{0}^{k_{\star}}k^{} dk  +m_X^{3/2}H_{\rm I} ^{5/2}\gamma^{\frac{-1+3w}{3(1+w)}}a_e \int_{k_\star }^{k_{\rm rh}} k^{-2} dk \nonumber \\
    &\qquad\lsp+ m_X^{3/2}\gamma^{\frac{1-3w}{1+w}}H_{\rm I}^{\frac{3(w+3)}{2(1+3w)}}a_e^{\frac{3(1-w)}{1+3w}}\int_{k_{\rm rh}}^{k_e}k^{\frac{-4}{1+3w}} dk\bigg], \nonumber \\
    &=\frac{1}{8 \pi^2}\bigg[\frac{1}{2}m_X H_{\rm I} ^2 - m_X^{3/2}H_{\rm I}^{3/2}\gamma^{-1} \bigg(1-\gamma \left(\frac{H_{\rm I}}{m_X} \right)^{1/2}\bigg) \nonumber \\
    &\qquad\lsp-\frac{1+3w}{3(1-w)}m_X^{3/2}\gamma^{\frac{1-3w}{1+w}}H_{\rm I}^{3/2}\bigg(1-\gamma^{\frac{2(w-1)}{1+w}} \bigg)\bigg], \nonumber \\
    &\approx \frac{1}{8 \pi^2}\bigg[\frac{3}{2} + \frac{1+3w}{3(1-w)}\sqrt\frac{m_X}{H_{\rm rh}}\,\bigg]m_X H_{\rm I}^2 ,		\label{eq:nldm}
\end{align}
where we have assumed that $w \!\in\! (\!-1/3, 1)$ such that in the last step we used $\gamma^{\frac{2(w-1)}{1+w}}\!\gg\!1$ approximation. 

\subsubsection*{Comments on isocurvature density perturbations}
As observed in the above analysis (see \fig{fig:numberdensity}), the most dominant modes are concentrated around the characteristic value of mode momentum $k_\star\!\equiv\!a_\star m_X$ for both the light ($m_X\!<\!H_{\rm rh}$) and the heavy ($H_{\rm rh}\!<\!m_X\!<\!H_{\rm I}$) DM mass regimes. The explicit value of $k_\star$ can be written as,
\beq
k_{\star}\!=\!a_{\rm mre} \sqrt{m_X H_{\rm rme}}
\begin{dcases}
\bigg(\frac{H_{\rm rh}}{m_X}\bigg)^{\!\frac{1-3w}{6(1+w)}}\,, & \qquad H_{\rm rh}\!\leq\!m_X\!<\!H_{\rm I}\,,\\
1\,, &\qquad H_{\rm mre}\!\leq\!m_X\!<\!H_{\rm rh}\,,
\end{dcases}	\label{eq:kstar}
\eeq
where $a_{\rm mre}$ and $H_{\rm mre}$ denote the scale factor and the Hubble rate at the matter-radiation equality, respectively.
As noted above, the number density per log momentum for the longitudinal vector DM drops as $k^2$ for the modes with momentum $k\!<\!k_\star$, 
i.e. for the long-wavelength modes. 
This result is independent of the vector DM mass regimes and the presence of the reheating phase. 
Since the vector DM is coupled to the SM (radiation) only through gravitational interactions, therefore the corresponding density fluctuations are of the isocurvature type.  
Hence the density perturbations of the vector DM with wavelengths of the size of CMB scale, i.e. $k\!\sim\!k_{\rm CMB}\!\approx\!0.05\,{\rm Mpc}^{-1}$ may lead to large isocurvature perturbations which are severely constrained by the Planck data~\cite{Akrami:2018odb}. 
It was first noted in Ref.~\cite{Graham:2015rva}, where an instantaneous reheating was assumed, that the power spectrum of longitudinal vector DM density fluctuations falls as $k^3$ for the long-wavelength modes $k\!<\!k_\star$. 
We also get the same scaling behavior $k^3$ for the long-wavelength modes $k\!<\!k_\star$ of the vector DM density perturbations (for both mass regimes), as the presence of non-standard reheating phase only affects the scaling behavior at the short-wavelengths $k\!>\!k_\star$.    
Hence, if the dominant modes $k_\star$ correspond to cosmological scales much smaller than the CMB scale, i.e. $k_\star\!\gg\!k_{\rm CMB}\!\approx\!0.05\,{\rm Mpc}^{-1}$, then thanks to $k^3$ suppression of the density perturbations, the longitudinal vector DM modes would not generate dangerous isocurvature perturbations. 

We can estimate the scale of the dominant modes $k_\star$ from \eqref{eq:kstar} as, 
\beq
k_{\star}\!\approx\!1400\,{\rm pc}^{-1}\,\sqrt{\frac{m_X}{10^{-14}\gev}}
\begin{dcases}
\bigg(\frac{H_{\rm rh}}{m_X}\bigg)^{\frac{1-3w}{6(1+w)}}\,, & \qquad H_{\rm rh}\!\leq\!m_X\!<\!H_{\rm I}\,,\\
1\,, &\qquad H_{\rm mre}\!\leq\!m_X\!<\!H_{\rm rh}\,.
\end{dcases}	\label{eq:kstar_approx}
\eeq
For the light DM mass scenario, as long as $m_X\!\geq\!10^{-14}\gev$, the vector DM would be safe from isocurvature modes since $1/k_\star$ is a tiny scale compared to the cosmological scales~\cite{Graham:2015rva}. 
However, for the heavy vector DM case ($H_{\rm rh}\!\leq\!m_X\!<\!H_{\rm I}$), the $w$ dependent extra factor is 
\beq
\bigg(\frac{H_{\rm rh}}{m_X}\bigg)^{\frac{1-3w}{6(1+w)}}=
\begin{dcases}
\bigg(\frac{m_X}{H_{\rm rh}}\bigg)^{[0,\,1/6)}\geq 1\,, & \qquad w\!=\![1/3,\,1)\,,\\
\bigg(\frac{H_{\rm rh}}{m_X}\bigg)^{(1/2,\,0)}<1\,, &\qquad w\!=\!(\!-\!1/3,\,1/3)\,.
\end{dcases}	\label{eq:kstar_approx}
\eeq 
Hence, in this case, the dominant mode momentum $k_\star\!\gg\! k_{\rm CMB}$ for $w\!=\![1/3,1)$ and  $m_X\!\geq\!10^{-14}\gev$, independent of the reheating scale $H_{\rm rh}$ (or reheating efficiency~$\gamma$). 
However, for the equation of state $w\!=\!(-1/3,1/3)$, the heavy DM dominant mode momentum $k_{\star}$ can be of the order of the CMB scale for $H_{\rm rh}\!\ll\!m_X$ and hence can generate dangerous isocurvature perturbations. In this case, we find the following condition on the reheating scale $H_{\rm rh}$ in order to safely avoid the isocurvature constraints with $k_\star\!\gtrsim\!1400\,{\rm pc}^{-1}\!\gg \!k_{\rm CMB}$,
\begin{align}
m_X\geq H_{\rm rh}&\geq 10^{-14}\gev\,\bigg(\frac{10^{-14}\gev}{m_X}\bigg)^{\frac{2(1+3w)}{(1-3w)}},	&{\rm with} \qquad	 w&=(\!-1/3,1/3).
\end{align}
The strongest lower bound on the reheating scale corresponds to the extreme case in our choice of parameters $w\!\simeq\!-1/3$ which implies $m_X\!\geq\!H_{\rm rh}\!\geq\!10^{-14}\gev$.
Hence, the heavy vector DM isocurvature perturbations are also highly suppressed at the cosmological scales for $m_X\!\geq\!H_{\rm rh}\!\geq\!10^{-14}\gev$ (corresponding to $T_{\rm rh}\gtrsim 100\gev$, see Eq.~\eqref{eq:Trh}), which is the case considered in this work.
Therefore, we conclude that the isocurvature modes are not problematic for the gravitationally produced vector DM in the parameter space of interest.

\subsubsection*{Relic abundance}
We calculate the present day relic abundance of the vector DM as 
\begin{align}
    \Omega_X h^2 = \frac{\rho_X}{\rho_c}h^2=\frac{m_X\, n_X(T_0)}{\rho_c} h^2,	\label{eq:omegah2}
\end{align}
where $\rho_c$ is  the critical density and $T_0$ refers to the present temperature. 
The present number density $n_X(T_0)$ is related to the number density $n_{\star}(T_{\star})\equiv \langle n_L(a_{\star}) \rangle$ at temperature~$T_{\star}$ such that $H(T_\star)=\!m_X$ as:
\begin{align}
    n_X(T_0)&= n_{\star}(T_{\star})\,\left(\frac{a_{\star}}{a_0} \right)^3 = n_{\star}(T_{\star})\, \frac{s_0}{s_{\rm rh}}\left(\frac{a_{\star}}{a_{\rm rh}} \right)^3,	\notag\\
  &=\frac{s_0}{s_{\rm rh}}\frac{H_{\rm rh}^2}{m_X^2}\begin{dcases}
\bigg(\frac{m_X}{H_{\rm rh}}\bigg)^{\!\frac{2w}{1+w}}\,\langle n_L^{\rm{hDM}} (a_{\star}) \rangle\,, & \qquad H_{\rm rh}\!\leq\!m_X\!<\!H_{\rm I},\\
\sqrt{\frac{m_X}{H_{\rm rh}}}\,\langle n_L^{\rm{lDM}} (a_{\star}) \rangle\,, &\qquad H_{\rm mre}\!\leq\!m_X\!<\!H_{\rm rh},
\end{dcases}	\label{eq:nxt0}
\end{align}
where in the last step we employed relation \eqref{eq:Ha} for the ratio $a_\star/a_{\rm rh}$, which is different for the two DM mass regimes. For the case of heavy vector DM with mass $H_{\rm rh}\!\leq\!m_X\!<\!H_{\rm{I}}$, the $H(a_\star)\!=\!m_X$ equality takes place during the reheating phase, while for the case of light DM with mass $m_X\!<\!H_{\rm rh}$, this condition is satisfied during the RD epoch. Above $s_0$ is the entropy density at present-day temperature $T_0$ and $s_{\rm rh}$ refers to the entropy density at the reheating temperature $T_{\rm rh}$, i.e.
\begin{align}
s_{\rm rh}&= \frac{4}{3}\frac{\rho_{\text{\tiny SM}}(a_{\rm rh})}{T_{\rm rh}}=
\frac{4 M_{\rm Pl}^2\, H_{\rm rh}^2}{T_{\rm rh}}\,,
\end{align}  
where the reheating temperature is given by,
\begin{align}
    T_{\rm rh}&= \left(\frac{90}{ \pi^2 g_{\star}(T_{\rm rh})} \right)^{1\!/\!4}\,\sqrt{\mpl\, H_{\rm rh}}\,.	\label{eq:Trh}
\end{align}

Finally the DM present-day relic abundance \eqref{eq:omegah2} can be calculated as,
\begin{align}
\Omega_X h^2&= \frac{s_0\, h^2}{4\mpl^2\,\rho_c} \frac{T_{\rm rh}}{m_X} 
\begin{dcases}
\bigg(\frac{m_X}{H_{\rm rh}}\bigg)^{\!\frac{2w}{1+w}}\,\langle n_L^{\rm{hDM}} (a_{\star}) \rangle , \qquad& H_{\rm rh}\!\leq\!m_X\!<\!H_{\rm I},\\
\sqrt{\frac{m_X}{H_{\rm rh}}}\,\langle n_L^{\rm{lDM}} (a_{\star}) \rangle, \qquad& m_X\!<\!H_{\rm rh},
\end{dcases}
\end{align}
where $s_0 \!=\! 2970~{\rm cm}^{-3}$ and $\rho_c\! =\! 1.054 \!\times\! 10^{-5} h^2 \gev \,{\rm cm}^{-3}$. Furthermore, we assume that $g_{\star}(T_{\rm rh}) \!\approx\! 106$, i.e. no extra relativistic d.o.f. beyond the SM. Employing the approximate results for $\langle n_L (a_{\star}) \rangle$ from Eqs.~\eqref{eq:nhdm}-\eqref{eq:nldm} and $H_{\rm rh}\!=\!\gamma^2 H_{\rm I}$ with $\gamma$ being the reheating efficiency, we get the vector DM relic abundance as,
\begin{align}
\Omega_X h^2&\approx 1.27\!\times\!10^{-22} \!\times\!\begin{dcases}
\left(\frac{1}{2}+\frac{1+3w}{3(1-w)}  \right) \bigg(\frac{m_X}{\gamma^2 H_{\rm I}}\bigg)^{\!\frac{2w}{1+w}}\,\gamma\,H_{\rm I}^{5/2} , \qquad& H_{\rm rh}\!\leq\!m_X\!<\!H_{\rm I},\\
\bigg(\frac{3}{2} + \frac{1+3w}{3(1-w)}\sqrt\frac{m_X}{\gamma^2H_{\rm I}}\bigg)\sqrt{m_X}\,H_{\rm I}^2 \,, \qquad& m_X\!<\!H_{\rm rh},
\end{dcases}	\notag\\
&\approx 0.12\!\times\!\begin{dcases}
\bigg(\frac{m_X}{0.33\gev}\bigg)^{\!\frac{2w}{1+w}}\bigg(\frac{H_{\rm I}}{3.3\!\times\!10^{10}\gev}\bigg)^{\!\frac{5+w}{2(1+w)}} \bigg(\frac{10^{-5}}{\gamma}\bigg)^{\!\frac{3w-1}{1+w}}, \qquad& H_{\rm rh}\!\leq\!m_X\!<\!H_{\rm I},\\
\bigg(\frac{m_X}{2\!\times\!10^{-14}\gev}\bigg)^{1/2}\bigg(\frac{H_{\rm I}}{6.6\!\times\!10^{13}\gev}\bigg)^2\,, \qquad& m_X\!<\!H_{\rm rh},
\end{dcases}	\label{eq:relic}
\end{align}
where $\Omega_X^{\rm obs} h^2 \!=\! 0.12\pm 0.0012$ is the observed DM relic abundance~\cite{Aghanim:2018eyx}.
In the last approximation for the heavy vector DM case we take $w\!=\!1/3$, however the dependance of the equation of state $w$ is power-law. 
Note for the heavy DM regime, the relic abundance is independent of the DM mass for the matter dominated ($w\!=\!0$) reheating phase.

Our result for light vector DM ($m_X\!<\!H_{\rm rh}$) relic abundance \eq{eq:relic} is consistent with Ref.~\cite{Graham:2015rva}, which employed instantaneous reheating followed by RD universe. 
Since the dominant modes for light vector DM cross the horizon during the RD era (see \fig{fig:drhodlnkA} right-panel), therefore the dependance of non-standard cosmology during the reheating phase is less significant and one gets the same results as for instantaneous reheating. 
In Ref.~\cite{Ema:2019yrd} a non-instantaneous phase of reheating with matter dominated ($w\!=\!0$) universe was considered. Our results for the case $w\!=\!0$ agree with Ref.~\cite{Ema:2019yrd}~\footnote{We thank the authors of Ref.~\cite{Ema:2019yrd} for pointing out a mistake in Eq.~\eqref{eq:nxt0} in an earlier version of the paper.}.

\begin{figure}[t!] 
    \centering
    \includegraphics[width=0.5\textwidth]{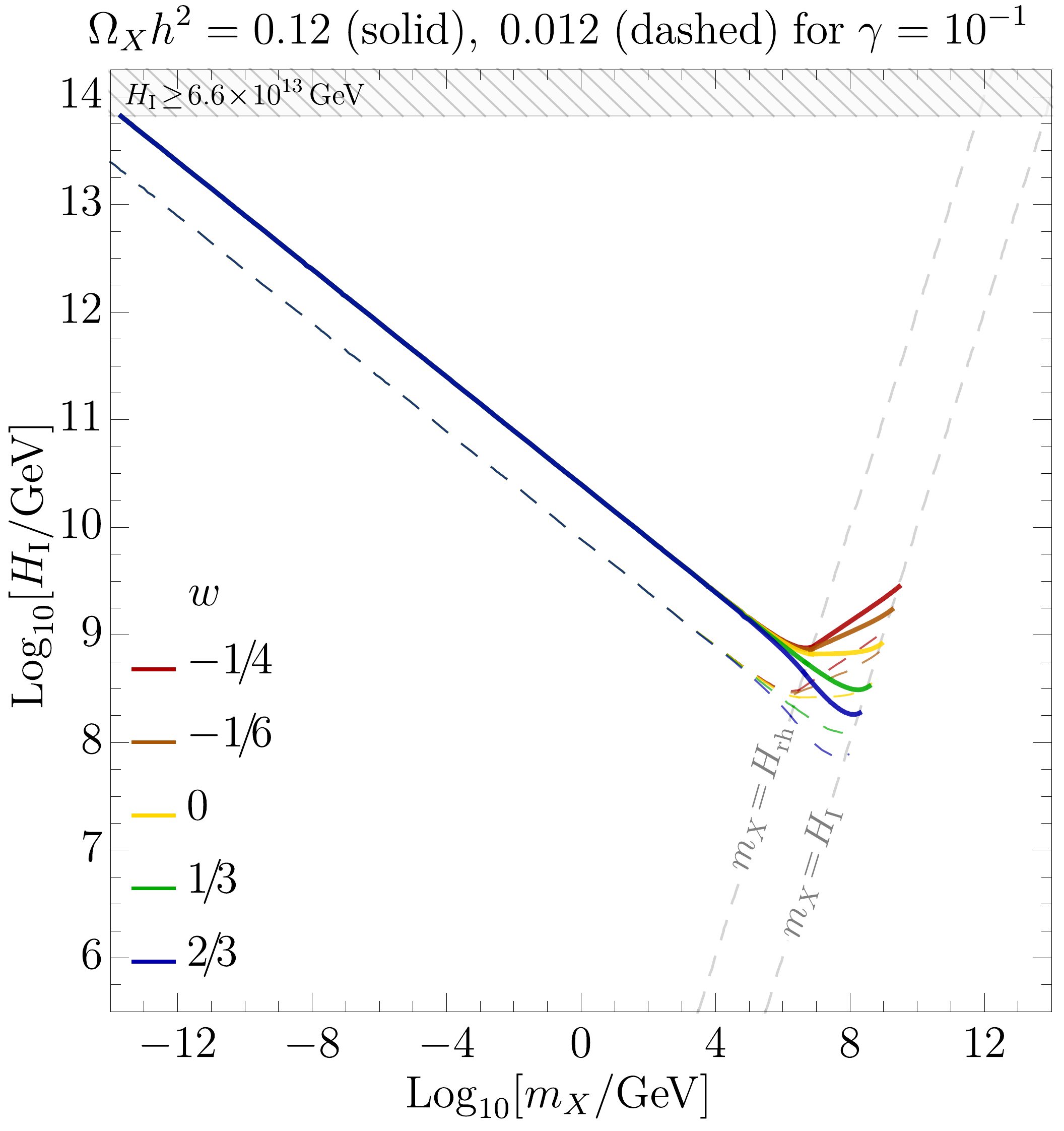}\!\!\!
    \includegraphics[width=0.5\textwidth]{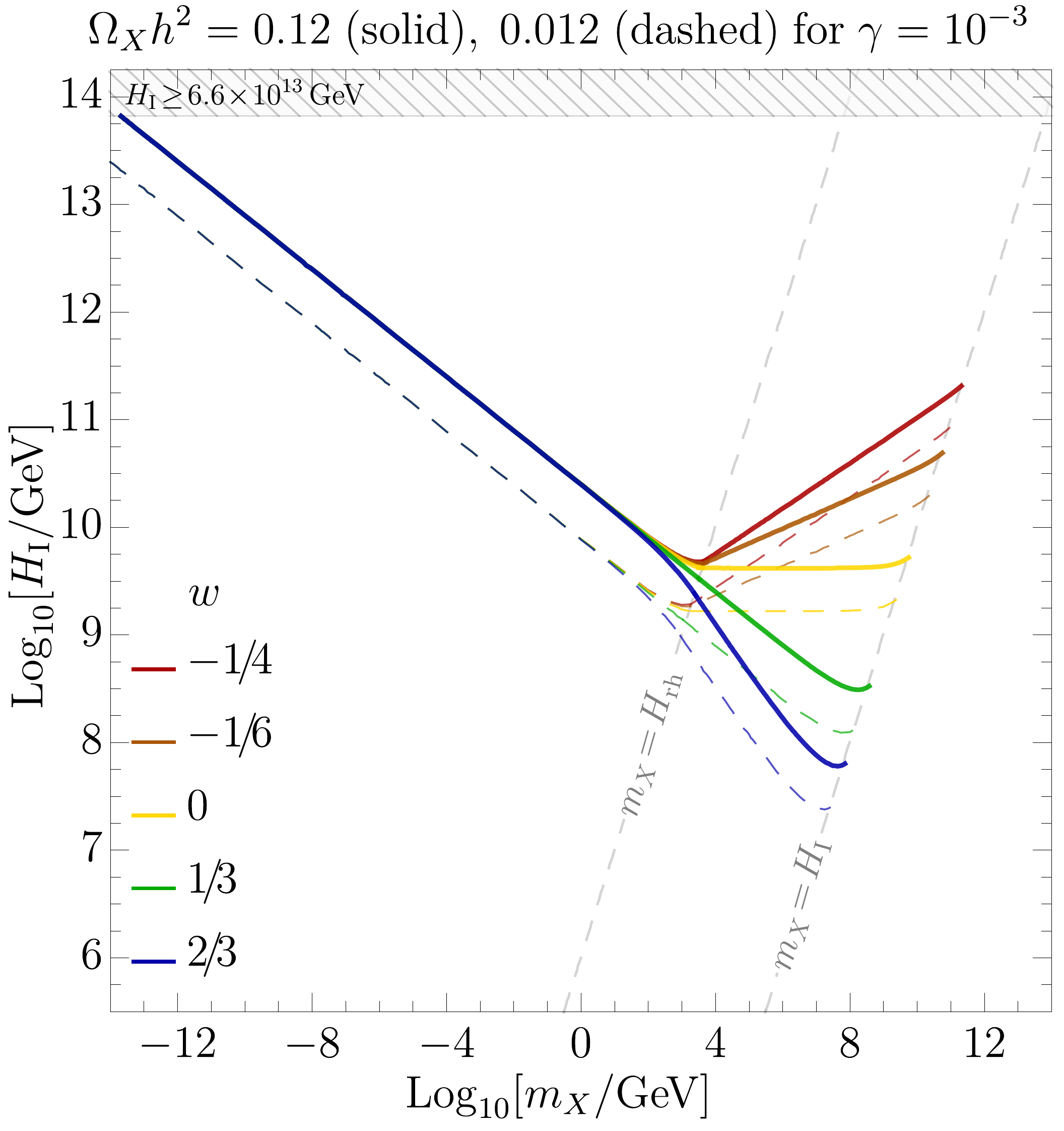}
    \includegraphics[width=0.5\textwidth]{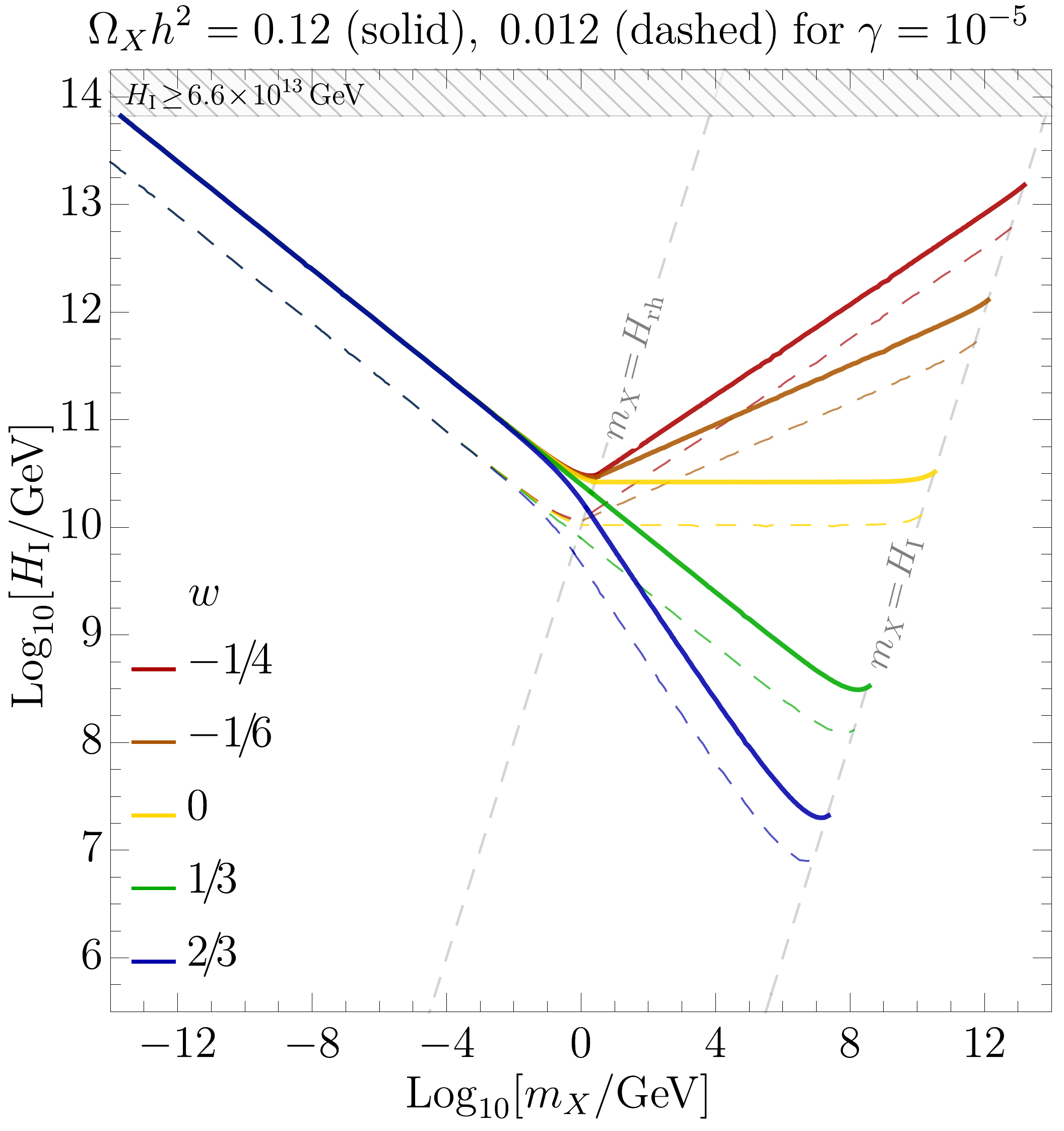}\!\!\!
    \includegraphics[width=0.5\textwidth]{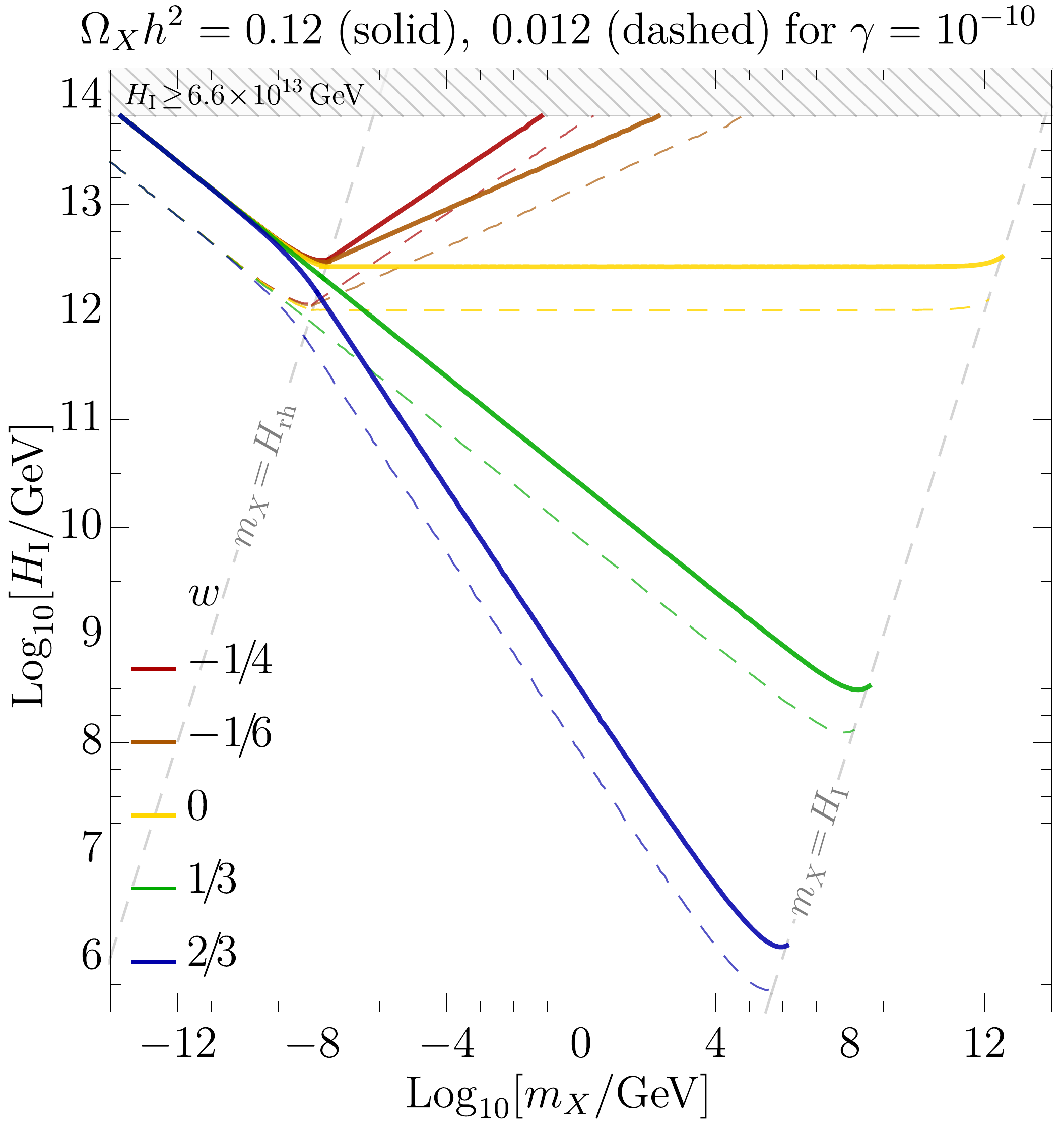}
    \caption{Relations between the Hubble rate during inflation $H_{\rm I}$ vs the vector DM mass $m_X$ that reproduces the observed relic abundance $\Omega^{\rm {obs}}_X$ (solid curves) and $10\%$ of the observed density (dashed curves) for different values of the equation of state parameter $w$. Each panel corresponds to a different reheating efficiency $\gamma \!=\! 10^{-1}, 10^{-3}, 10^{-5},10^{-10}$ as indicated in the title of each plot. The hatched gray region is excluded by the Planck satellite at 95\% C.L.~\cite{Akrami:2018odb}.}
    \label{fig:Hi_mX}
\end{figure}
In \fig{fig:Hi_mX} we present exact numerical results for the parameter space in the plane $\hi$ vs $m_X$ which leads to the production of observed DM relic abundance $\Omega^{\rm {obs}}_X h^2\!=\!0.12$ (solid curves) and 10\% of the observed relic abundance, i.e. $\Omega_X h^2\!=\!0.012$ (dashed curves), for four choices of the reheating efficiency $\gamma=10^{-1},10^{-3}, 10^{-5}$, and $10^{-10}$ with various values of the equation of state parameter~$w$ during the reheating phase. 
The hatched gray region corresponding to $H_{\rm I}\!\geq\!6.6\!\times\!10^{13}\gev$  is excluded by the Planck satellite at 95\%~C.L.~\cite{Akrami:2018odb}.
We note that for the light vector DM, $m_X\!<\!H_{\rm rh}$, the observed relic abundance is produced for~$m_X\!\approx\! 2\!\times\!10^{-14}\gev\big(6.6\!\times\!10^{13}\gev/H_{\rm I}\big)^4$.
However, for the heavy vector DM, $H_{\rm rh}\!\leq\!m_X\!<\! H_{\rm I}$, the presence of non-instantaneous reheating phase with general equation of state $w$ is of great significance as the DM relic abundance is power-law sensitive w.r.t. $w$ as the dependence scales as $m_X^{2w/(1+w)}$, $H^{(5+w)/(2+2w)}$, and $\gamma^{-(3w-1)/(1+w)}$.
Note that the effects of non-standard cosmological evolution during the reheating phase parametrized by $w$ are negligible for the light vector DM regime, as the DM mass gets smaller compared to the $H_{\rm rh}$. 
In \fig{fig:Hi_mX}, we show the DM mass equality to the reheating scale $H_{\rm rh}$ and the inflation scale $H_{\rm I}$ as dashed gray lines. 
For the reheating efficiency $\gamma=10^{-1}$ the purely gravitational production of vector DM can provide the observed relic abundance for vector DM mass $m_X\!\approx\![10^{-14},10^9]\gev$ for $H_{\rm I}\!\approx\![10^{14},10^8]\gev$. Whereas, for $\gamma=10^{-10}$ the gravitational vector DM can account for the observed DM relic for $m_X\!\approx\![10^{-14},10^{12}]\gev$ and $H_{\rm I}\!\approx\![10^{14},10^6]\gev$. 
Hence a wide range of parameter space can lead to the production of vector dark matter purely due to quantum fluctuation during the early universe.  

As a final remark, we address the question of detection of our purely gravitationally produced vector DM. 
It has been assumed that our vector DM candidate is absolutely stable due to the discrete $\mathbb{Z}_2$ symmetry, in addition it interacts with the SM via gravity only. Therefore detection of such DM in laboratory experiments is rather unlikely.
However, absolute stability of DM is not necessary as long as the lifetime of DM is larger than age of the Universe. 
In this case, one may allow a small mixing of vector DM with the SM photon via the dark $U(1)_X$ kinetic mixing with the SM hypercharge $U(1)_Y$, i.e. $\frac{1}{2}\epsilon B^{\mu\nu}X_{\mu\nu}$, where $B_{\mu\nu}$ is the field strength tensor for the SM $U(1)_Y$ gauge boson $B_\mu$. For very small values of the mixing parameter $\epsilon$ the vector DM can be stable at the time scales of the age of the Universe and there is possibility of direct detection, see~\cite{Graham:2015rva} and references therein.

\section{Conclusions}
\label{sec:con}
The aim of this work was to investigate the possibility of gravitational production of an Abelian vector dark matter due to rapidly expanding early universe. 
In this scenario the SM is extended by a $U(1)_X$ gauge group equipped with a stabilizing $\z2$ symmetry, so that the corresponding vector boson is a DM candidate. 
It has been assumed that the dark sector communicates with the SM only through gravitational interactions described by the General Relativity. 
We have focused here on the possibility of generating vacuum expectation value of energy density for the longitudinal component of vector DM in the presence of time dependent FLRW metric in the early universe. 
We have shown in detail how does the canonical quantization of the vector field in this varying gravitational background imply the tachyonic enhancement of some momentum modes of the field. 
In all cases approximate solutions of the mode equation have been found and verified against exact numerical solutions. 

We have assumed the period of inflation described effectively by de~Sitter geometry, however for the following period of reheating we have adopted a generic equation of state with its parameter varying in the window $-1/3 \!< \!w \!< \!1$.
That way we have effectively taken into account possibilities of unknown dynamics modeled by some unspecified inflation scenario followed by an extended period of reheating with non-standard early universe cosmology.
It has been shown that the spectrum of dark vectors produced that way is centered around a characteristic comoving momentum $k_\star$ that is determined in terms of the mass of the vector, the Hubble parameter during inflation $H_{\rm I}$, the equation of state parameter $w$ and the efficiency of reheating $\gamma$. 
The ultimate result of this work was to calculate the present total vector-dark-matter abundance produced purely gravitationally. 
Regions in the parameter space consistent with the Planck measurement of $\Omega_{\rm DM}$ have been determined justifying the gravitational production as a viable mechanism for vector dark matter production for a wide range of DM masses. 
In particular, we found non-trivial dependance of the relic abundance on the vector DM mass $m_X$ and the Hubble parameter during inflation~$H_{\rm I}$. 
This dependance of the relic abundance on $m_X$ and $H_{\rm I}$ is different for the heavy DM $H_{\rm rh}\!<\!m_X\!<\! H_{\rm I}$ and light DM $m_X\!<\! H_{\rm rh}$ regimes.
The results obtained in this paper are applicable within various possible models of inflation/reheating with non-standard cosmology parametrized by corresponding equation of state.  
  
\section*{Acknowledgments}

The authors acknowledge support by the National Science Centre (Poland), under the research project no~2017/25/B/ST2/00191.
AA is supported by FWO under the EOS-be.h project no. 30820817.

\appendix

\section{Quantization of the vector field in a curved background}
\label{app:quanta}
Here we collect essential details of the canonical quantization of a vector DM in a curved background. 
The action for an Abelian vector field reads
\begin{align}
    S_{\rm DM} = \int d^{4}x \sqrt{-g} \left[ - \frac{1}{4} g^{\mu \alpha} g^{\nu \beta} X_{\mu \nu} X_{\alpha \beta} + \frac{1}{2} m_X^2 g^{\mu \nu} X_{\mu} X_{\nu}  \right],
\end{align}
where $X_{\mu \nu} \equiv \partial_{\mu}X_{\nu}-\partial_{\nu}X_{\mu}$ denotes the field strength and $m_X$ is the vector boson mass.
The background metric is in the FLRW form~\eqref{eq:metric}. 
The above action results in the following equations of motion,
\begin{align}
     \vec{\nabla} \cdot \dot{\vec{X}} - \nabla^2 X_0 + m_X^2 a^2 X_0 &= 0\,, \label{et2}\\
    \Ddot{\vec{X}} + H \dot{\vec{X}} - \frac{1}{a^2} \nabla^2 \vec{X} + m_X^2 \vec{X}&= - 2 H \vec\nabla X_0\,,	 \label{es2}
\end{align}
where $\vec \nabla\equiv \partial/(\partial x^i)$, $\nabla^2\equiv \partial^2/(\partial x_i\partial x^i)$, and $H\equiv \dot a/a$ is the Hubble parameter.
It is convenient to adopt the Fourier transform 
\begin{align}
    X_{\mu}(t, \vec{x}) = \int \frac{d^3 k}{(2\pi)^{3/2}}  \mathcal{X}_{\mu}(t, \vec{k}) e^{i \vec{k} \cdot \vec{x}}, \label{fdec}
\end{align}
where the reality of the $X_{\mu}(t, \vec{x})$ field implies $\mathcal{X}_{\mu}(t, \vec{k})=\mathcal{X}^*_{\mu}(t, -\vec{k})$.
Inserting this decomposition into Eqs.~(\ref{et2}-\ref{es2}), we get,
\begin{align}
    \mathcal{X}_0 &= \frac{- i \vec{k} \cdot \partial_t{\vec{\cal X}}}{k^2 + a^2 m_X^2}, 	\label{x0mode}\\
    \partial_t^2{\vec{\cal X}}+ H \partial_t {\vec{\cal X}} + \bigg(\frac{k^2}{a^2} + m_X^2 \bigg){\vec{\cal X}} &= - 2 H\vec{k} \,\frac{\vec{k} \cdot \partial_t{\vec{\cal X}}}{k^2 + m_X^2 a^2}. \label{xvecmode}
\end{align}
Note that the $X_0$ is an auxiliary (unphysical) field and has no dynamics associated with it. However, we have used it in obtaining the above dynamical equation for the ${\vec{\cal X}}$ components.

The three components of $\mathcal{\vec{X}}$ field can be decomposed in a basis of helicity states, i.e.
\begin{align*}
    \mathcal{\vec{X}}(t, \vec{k})= \sum_{\lambda=\pm, L}\vec{\epsilon}_{\lambda}(\vec{k})\mathcal{X}_{\lambda}(t, \vec{k}),
\end{align*}
where $\mathcal{X}_{\pm}$ and $\mathcal{X}_L$ denote two transversely-polarized modes and a single longitudinally-polarized mode, respectively. By choosing the reference frame such that the vector $\vec{k}$ points the $z$-direction the explicit forms of the polarization vectors can be adopted as follows
\begin{align*}
    \vec{\epsilon}_L(k_z)=
\left(\begin{array}{c}
0 \\
0 \\
1
\end{array} \right), \quad     \vec{\epsilon}_{\pm}(k_z)=\mp \frac{1}{\sqrt{2}}
\left(\begin{array}{c}
1 \\
\pm i \\
0
\end{array} \right),
\end{align*}
which implies
\begin{align*}
    \vec{k} \cdot \mathcal{\vec{X}}(t, \vec{k}) =  k\, \mathcal{X}_L(t, \vec{k}).
\end{align*}
This allows us to rewrite \eq{xvecmode} in terms of the conformal time ($dt=a(\tau) d\tau$) as
\begin{align}
 \mathcal{X}^{\prime \prime}_{\pm} + \left( k^2 + a^2 m_X^2\right) \mathcal{X}_{\pm} = 0, \label{eomtm}\\
 \mathcal{X}^{\prime\prime}_{L} + \frac{2 k^2}{k^2 + a^2 m_X^2}\frac{a^{\prime}}{a}\mathcal{X}^{\prime}_L+  (k^2 + a^2 m_X^2)\mathcal{X}_L =0. \label{eomlc}
 \end{align}
Note that Eq.~\eqref{eomtm} is the harmonic oscillator equation with time-dependent frequency, $\omega_{\pm}^2 = k^2 + a^2 m_X^2$. It is worthwhile to mention that in this case $\omega_{\pm}^2$ is always positive.
However, for the longitudinal mode, $\mathcal{X}_L$, it is convenient to perform a field redefinition 
\begin{align}
    \mathcal{X}_L = \frac{\sqrt{k^2 +  a^2 m_X^2 }}{a m_X} \widetilde{\mathcal{X}}_L, \label{rfield}
\end{align}
that allows to rewrite \eq{eomlc} in the desired form of an oscillator equation.
We obtain the equation of motion for $\widetilde{\cal X}_L$ in the following form
\begin{align}
\widetilde{\mathcal{X}}^{\prime\prime}_L + \omega^2_L(\tau)\widetilde{\mathcal{X}}_L=0, \label{tdoe}
 \end{align} 
with the frequency
\begin{align}
 \omega^2_L(\tau)\equiv k^2 + m_X^2a^2 - \frac{k^2}{k^2 + m_X^2 a^2}\frac{a^{\prime\prime}}{a} + 3\frac{k^2 m_X^2 a^{\prime 2}}{(k^2 + m_X^2 a^2)^2}\,.
\end{align}

Let us here recall some basic mathematical facts about time-dependent oscillator equation. Such equations have a two-dimensional space of solutions, spanned by $\Big\{\widetilde{\mathcal{X}}_{L}^{(1)}, \widetilde{\mathcal{X}}_{L}^{(2)}\Big\}$ and $\Big\{\mathcal{X}_{\pm}^{(1)}, \mathcal{X}_{\pm}^{(2)}\Big\}$, respectively.
The general solutions are given by
\begin{align}
    \widetilde{\cal X}_L(\tau, \vec{k})&= a_{\vec{k}}^{-}\mathcal{\widetilde{X}}_{L}(\tau, \vec{k}) + a_{-\vec{k}}^{+}\mathcal{\widetilde{X}}_{L}^*(\tau, -\vec{k}). \label{xl1} \\
     {\cal X}_{\pm}(\tau, \vec{k})&= b_{\vec{k}, \pm}^{-}\mathcal{{X}}_{\pm}(\tau, \vec{k}) + b_{-\vec{k}}^+\mathcal{{X}}_{\pm}^*(\tau, -\vec{k}) \label{xpm1},
\end{align}
where $a^{\pm}_{\vec{k}},b_{\vec{k}}^{\pm} $ are complex time-independent constants and 
\begin{align}
    \mathcal{\widetilde{X}}_L &\equiv \widetilde{\mathcal{X}}_L^{(1)} + i \widetilde{\mathcal{X}}_L^{(2)}, 
    &\mathcal{X}_{\pm} &\equiv \mathcal{X}_{\pm}^{(1)} + i \mathcal{X}_{\pm}^{(2)}.
\end{align}
Using Eqs.~(\ref{fdec}, \ref{xl1}, \ref{xpm1}) we get
\begin{align*}
   \widetilde{X}_{L}(\tau, \vec{x})&= \int \frac{d^3 k}{(2 \pi)^{3/2}} \left\{\epsilon_L(\vec{k})a^-_{\vec{k}} \mathcal{\widetilde{X}}_{L}(\tau, \vec{k}) e^{i \vec{k} \cdot \vec{x}} 
	+  \vec{\epsilon}_L(\vec{k})a^{+}_{\vec{k}}\mathcal{\widetilde{X}}_{L}^*(\tau, \vec{k})e^{-i \vec{k} \cdot \vec{x}} \right\},		\\
    {X}_{\pm}(\tau, \vec{x})&= \int \frac{d^3 k}{(2 \pi)^{3/2}} \left\{\vec{\epsilon}_{\pm}(\vec{k})b^-_{\vec{k},\pm} \mathcal{{X}}_{\pm}(\tau, \vec{k}) e^{i \vec{k} \cdot \vec{x}} 
		+ \vec{\epsilon}_\pm^{\phantom{\pm}*}(\vec{k}) b^{+}_{\vec{k}, \pm}\mathcal{{X}}_{\pm}^*(\tau, \vec{k})e^{-i \vec{k} \cdot \vec{x}} \right\}\,.
\end{align*}

Next, we quantize the theory imposing equal-time commutation relations,
\begin{align}
    [\hat{\widetilde{X}}_L(\tau, \vec{x}), \Hat{\Pi}_{L}(\tau, \vec{y})] &= i  \delta^{(3)}(\vec{x} -\vec{y}), 
    & [\Hat{X}_{\pm}(\tau, \vec{x}), \Hat{\Pi}_{\pm}(\tau, \vec{y})] &= i  \delta^{(3)}(\vec{x} -\vec{y}), \label{crpm}
\end{align}
where we have promoted ${X}_{L(\pm)}$ to the quantum field operators $\hat{X}_{L( \pm)}$
\begin{align}
      \hat{\widetilde{X}}_{L}(\tau, \vec{x})&= \int \frac{d^3 k}{(2 \pi)^{3/2}} 
			\left\{\vec{\epsilon}_L(\vec{k})\hat{a}_{\vec{k}} \mathcal{\widetilde{X}}_{L}(\tau, \vec{k}) e^{i \vec{k} \cdot \vec{x}} 
			+  \vec{\epsilon}_L(\vec{k}) \hat{a}^{\dagger}_{\vec{k}}\mathcal{\widetilde{X}}_{L}^*(\tau, \vec{k})e^{-i \vec{k} \cdot \vec{x}} \right\}, \label{qol}\\
    \hat{{X}}_{\pm}(\tau, \vec{x})&= \int \frac{d^3 k}{(2 \pi)^{3/2}} 
		\left\{\vec{\epsilon}_{\pm}(\vec{k})\hat{b}_{\vec{k}, \pm} \mathcal{{X}}_{\pm}(\tau, \vec{k}) e^{i \vec{k} \cdot \vec{x}} 
		+ \vec{\epsilon}_{\pm}^{\phantom{\pm}*}(\vec{k}) \hat{b}^{\dagger}_{\vec{k}, \pm}\mathcal{{X}}_{\pm}^*(\tau, \vec{k})e^{-i \vec{k} \cdot \vec{x}} \right\}.\label{qopm}
\end{align}
where $\hat{a}_{\vec{k}}^{\dagger}(\hat{a}_{\vec{k}})$ and $\hat{b}_{\vec{k}}^{\dagger}(\hat{b}_{\vec{k}})$ are the creation (annihilation) operators for the $\hat{\widetilde{X}}_L$ and $\hat{X}_{\pm}$, respectively.
The canonical momenta are defined as
\begin{align}
    \hat{{\Pi}}_L(\tau, \vec{x}) &= \int \frac{d^3 k}{(2 \pi)^{3/2}} \left\{\vec{\epsilon}_L(\vec{k})\hat{a}_{\vec{k}}\mathcal{\widetilde{X}}^{\prime}_L(\tau, \vec{k}) e^{i \vec{k} \cdot \vec{x}} 
		+ \vec{\epsilon}_L(\vec{k})\hat{a}^{\dagger}_{\vec{k}}\mathcal{\widetilde{X}}^{\prime *}_L(\tau, \vec{k}) e^{-i \vec{k} \cdot \vec{x}}  \right\}, \\
    \hat{{\Pi}}_{\pm}(\tau, \vec{x}) &= \int \frac{d^3 k}{(2 \pi)^{3/2}} \left\{\vec{\epsilon}_{\pm}(\vec{k})\hat{b}_{\vec{k}, \pm}\mathcal{\widetilde{X}}^{\prime}_{\pm}(\tau, \vec{k}) e^{i \vec{k} \cdot \vec{x}} 
		+ \vec{\epsilon}_{\pm}^{\phantom{\pm}*}(\vec{k})\hat{b}^{\dagger}_{\vec{k}, \lambda}\mathcal{\widetilde{X}}^{\prime *}_{\pm}(\tau, \vec{k}) e^{-i \vec{k} \cdot \vec{x}}  \right\}, 
\end{align}
It is easy to check that conditions (\ref{crpm}) imply
\begin{align}
[\hat{a}_{\vec{k}}, \hat{a}^{\dagger}_{\vec{k^{\prime}}}] = \delta^{(3)}(\vec{k} -\vec{k^{\prime}}), \lsp
    [\hat{b}_{\vec{k},\lambda}, \hat{b}^{\dagger}_{\vec{k^{\prime}},\lambda^{\prime}}] = \delta_{\lambda \lambda^{\prime}}\delta^{(3)}(\vec{k} -\vec{k^{\prime}}),
\end{align}
if the Wronskian:
\begin{align*}
    W[v,v^{*}] \equiv v^{\prime}v^* - v^{\prime *}v
\end{align*}
is time-independent and normalized as follows 
\begin{align}
    W[\mathcal{\widetilde{X}}_{L},\mathcal{\widetilde{X}}_{L}^*]=W[\mathcal{{X}}_{\pm},\mathcal{{X}}_{\pm}^*]=-i. \label{wrcon}
\end{align}

To solve equations of motion for the two transversely-polarized mode \eqref{eomtm} and the single longitudinally-polarized mode \eqref{tdoe} we impose the Bunch-Davies initial conditions:
\begin{align}
    \lim_{\tau \rightarrow - \infty} \mathcal{\widetilde{X}}_L(\tau, \vec{k}) = \frac{1}{\sqrt{2k}} e^{-i k \tau}\,, \qquad\qquad 	\lim_{\tau \rightarrow - \infty} \mathcal{{X}}_{\pm}(\tau, \vec{k}) = \frac{1}{ \sqrt{2k}} e^{-i k \tau}\,.\label{bdav}
\end{align}  
This boundary condition together with \eqref{wrcon} completely fixes the mode functions $\mathcal{\widetilde{X}}_{L,\pm}$. 

\bibliography{bib_gdm}
\bibliographystyle{JHEP}


\end{document}